\documentclass[a4paper,12pt]{article}

\usepackage{amssymb,amsmath,mathbbol,mathrsfs}
\usepackage[usenames,dvipsnames]{color}
\usepackage{hyperref}
\usepackage{xcolor}
\usepackage{stmaryrd}
\usepackage{authblk}
\usepackage{framed}
\usepackage{empheq}
\usepackage{slashed}
\usepackage{hyphenat}
\usepackage{cite}

\usepackage{marginnote}    

\usepackage[left=.8in,right=.8in,top=.8in,bottom=.8in]{geometry}                  

\usepackage{sectsty}
\allsectionsfont{\sffamily\mdseries\upshape} 
\usepackage{tocloft}

\makeatletter
\renewenvironment{abstract}{%
    \if@twocolumn
      \section*{\abstractname}%
    \else 
      \begin{center}%
        {\bfseries\sffamily\abstractname\vspace{\z@}}
      \end{center}%
      \quotation
    \fi}
    {\if@twocolumn\else\endquotation\fi}
\makeatother

\numberwithin{equation}{section}
5

\hypersetup{
	colorlinks=true,         
	linkcolor=MidnightBlue,          
	citecolor=BrickRed,        
	urlcolor=MidnightBlue            
}

\newcommand{\BIBand}{and}

\newcommand{\be}{\begin{equation}}
\newcommand{\ee}{\end{equation}}

\newcommand{\fg}{{\mathfrak{g}}}

\newcommand{\fLie}{{\mathbb{L}}}
\newcommand{\fI}{{\mathbb{i}}}
\newcommand{\F}{{\Phi}}
\renewcommand{\d}{{\mathrm{d}}}
\newcommand{\D}{{\mathrm{D}}}

\newcommand{\SU}{{\mathrm{SU}}}
\newcommand{\YM}{{\text{YM}}}
\newcommand{\G}{{\mathcal{G}}}
\newcommand{\Ad}{{\mathrm{Ad}}}
\newcommand{\pp}{{\partial}}
\newcommand{\fvf}{{\mathbb{X}}}
\newcommand{\lvf}{{\xi}}

\newcommand{\fF}{{\mathbb{F}}}
\newcommand{\FYM}{{\Phi_\text{YM}}}

\newcommand{\fG}{{\mathrm{Lie}(\G)}}
\newcommand{\fX}{{\mathfrak{X}}}
\newcommand{\C}{{\mathbb{C}}}

\renewcommand{\bar}{\overline}
\newcommand{\dd}{{\mathbb{d}}}

\renewcommand{\hat}{\widehat}

\newcommand{\lbr}{\llbracket}
\newcommand{\rbr}{\rrbracket}

\newcommand{\cint}{{\int\kern-.87em{<}}}
\newcommand{\sint}{{\int\kern-.75em{\sim}}}
\newcommand{\fint}{{\int\kern-1.00em{\int}}}

\newcommand{\bb}{\mathbb}

\newcommand{\tr}{\text{tr}}

\renewcommand{\#}{\sharp}

\let\oldmarginpar\marginpar
\renewcommand\marginpar[1]{\oldmarginpar{\color{red}\raggedright\footnotesize #1}}

\newcommand{\aldo}{\color{black}}
\newcommand{\florian}{\color{black}}
\newcommand{\henrique}{\color{black}}

\newcommand{\new}{\color{black}}

\title{\sffamily A unified geometric framework for boundary charges and dressings: non-Abelian theory and matter}
\author[1,2]{\sffamily Henrique Gomes\thanks{gomes.ha@gmail.com}}
\author[1]{\sffamily Florian Hopfm\"uller\thanks{fhopfmueller@perimeterinstitute.ca}}
\author[1]{\sffamily Aldo Riello\thanks{ariello@perimeterinstitute.ca}}
\affil[1]{\small Perimeter Institute for Theoretical Physics\break 31 Caroline St. N., Waterloo, ON N2L2Y5, Canada}
\affil[2]{\small Trinity College, Cambridge University\break Cambridge CB2 1TQ, England}

\begin{document}
\maketitle

\abstract{
	\noindent  Boundaries in gauge theories  are a delicate issue. Arbitrary boundary choices enter the calculation of charges via Noether's second theorem, obstructing the assignment of unambiguous physical charges to local gauge symmetries. 
 Replacing the arbitrary boundary choice with new degrees of freedom suggests itself. 
 But, concretely, such boundary degrees of freedom are spurious---i.e. they are not part of the original field content of the theory---and have to disappear upon gluing. How should we fit them into what we know about field-theory? 
We resolve these issues in a unified and geometric manner, by introducing a connection 1-form, $\varpi$,  in the field-space of Yang--Mills theory. Using this geometric tool, a modified version of symplectic geometry---here called `horizontal'---is possible.  Independently of boundary conditions, this formalism bestows to each region a physical notion of charge: the horizontal Noether charge. The horizontal gauge charges always vanish, while global charges still arise for reducible configurations characterized by global symmetries. The field-content itself is used as a reference frame to distinguish `gauge' and `physical'; no new degrees of freedom, such as group-valued edge modes, are required. Different choices of reference fields give different $\varpi$'s, which are cousins of gauge-fixings like the Higgs-unitary and Coulomb gauges. But the formalism extends well beyond gauge-fixings, for instance by avoiding the Gribov problem. For one choice of $\varpi$, would-be Goldstone modes arising from the condensation of matter degrees of freedom play precisely the role of the known group-valued edge modes, but here they arise as preferred coordinates in field space, rather than new fields. For another choice, in the Abelian case, $\varpi$ recovers the Dirac dressing of the electron. }

\newpage
{\hypersetup{	linkcolor=black, }\tableofcontents}

\begin{center}
	\rule{8cm}{0.4pt}
\end{center}


\section{Introduction}

In the covariant symplectic formalism for field theories, it is standard to require  the symplectic flow of the symmetry to be generated by functionally differentiable charges. 
For gauge theories, in the presence of boundaries, if no supplementary conditions are introduced a somewhat surprising (and well-documented) feature arises: the constructed symplectic charges may differ from the constraints by a boundary term, and hence will not vanish on the constraint surface. 

In some discussions, this is taken to mean that these boundary charges carry information about physical, rather than `pure gauge', symmetries; in other discussions more stringent boundary conditions are imposed, eliminating the \lq{}pure gauge\rq{} charges.

In view of these ambiguities,  the presence of (spatial) boundaries often seems to beg for the introduction of edge modes; they can be used both to cancel unwanted  charges---which may arise due to overly weak boundary conditions on the gauge degrees of freedom,---or to reinstate gauge symmetries---which may have been lost due to overly strong ones. 

Here, we keep track of gauge degrees of freedom in a purely relational way, without requiring extra degrees of freedom or boundary conditions, nor adding new boundary terms to the action. Different regions, with their Noether charges and unrestricted gauge-invariance, can be treated independently and composed, with a treatment applicable  to  all boundary conditions. 
We accomplish this by working directly on the space of fields, which already contains each and every boundary condition. Barring some (important) obstructions---posed by the existence of field configurations with global symmetries---field-space can generically be understood as a principal fiber bundle, wherein we introduce a connection-form, $\varpi$ (read \textsc{Var-Pie}).  In this paper we have focused on the field-space of Yang-Mills theory with scalar or fermionic matter, although the formalism could well turn out to encompass more general theories, such as general relativity and BF theory.

Using a gauge-covariant notion of  functional variation in this field-space,  charges associated to pure gauge-transformations are appropriately screened, without additional assumptions on the boundary conditions. This screening happens generically in field-space. However,  backgrounds which have  global symmetries obstruct the principal bundle description, which leads  the gauge-covariant derivative to fail in the directions of global symmetry---$\varpi$ is blind to such directions. In those circumstances, global symmetries are not screened; instead, they give non-trivial Noether charges.

The key to our results {is}  the possibility of describing important relations within and between subsystems  through the use of connection-forms.   Connection-forms encode field-variations in terms of the fields themselves---they split variations into physical and gauge with respect to the field content itself. They can be non-local but are always regional, meaning they can be consistently defined intrinsically in sub-regions of space.  Such a relational description in terms of how fields and regions couple to each other has been heuristically deemed by many to be  the defining feature of gauge theories (see e.g. \cite{Rovelli:2013fga, MachsBucket}).  As far as we know,  any relational splitting of field variations which is to be fully compatible with gauge symmetry can be encoded in some connection-form.

Connection-forms, in turn, call for Wilson lines. In some circumstances, Wilson lines allow for the construction of classical  dressings for charged fields; i.e. they attach certain Lie-group valued functions to the charged fields, rendering them gauge-invariant. These relations---which do not always exist---provide a direct link between boundary charges, dressings, and gauge invariance, unified through the concept of field-space covariance.

\subsection{Summary of results and roadmap}

Throughout the paper, we have included the \lq{}take-home\rq{} message of each section in the \lq{}{\textbf{Remarks on Section X}\rq{}  paragraphs.  In this subsection, we provide a concise summary of the main points, jointly with a guide to the paper.

	The first part of the paper,  sections \ref{sec:fieldspace_geo}-\ref{sec:Singerconnection}, is focused mostly on the mathematical aspects of our work. That being said, section \ref{sec:Singerconnection} has more physical content, and makes the transition to Part II (sections \ref{sec:Noether_hor}-\ref{sec:dressing}), wherein we discuss all the physical applications we have investigated so far.

	\paragraph*{Part I -- Mathematical theory}
	~ \\
	
	The fields we will be concretely working with are: the gauge potential, scalar and spinorial matter fields. These are introduced in section \ref{sec:fieldspace_geo}, where we also review the \textit{principal fiber bundle} (PFB) structure of the  field-space of Yang-Mills theory, and spell out the aforementioned obstructions to such a structure presented by backgrounds with global symmetries \cite{Ebin, Palais, YangMillsSlice, kondracki1983, Mitter:1979un, Singer:1978dk, Singer:1981xw,  isenberg1982slice}.  A PFB structure does \textit{not} imply a global product space structure. Indeed, the lack of such product structure is  caused by another famous obstruction: the Gribov problem \cite{Gribov:1977wm, Singer:1978dk, Singer:1981xw, Vandersickel:2012tz}.
	
	The most important mathematical object in  PFB\rq{}s, and in this work, is the connection 1-form $\varpi$. In section \ref{sec:general_YM} we introduce a general notion of  $\varpi$ for the field-space of Yang--Mills theory (YM), and discuss properties of field-space gauge-covariance and its relation to the geometrical notion of {\it horizontality} \cite{kobayashivol1}. 
	This is another fundamental concept in our constructions: a connection-form defines an infinitesimal transverse plane to the gauge-orbits---termed horizontal---\textit{in a gauge-covariant manner} (therefore its relation to `geometric' BRST \cite{Gomes:2016mwl, Thierry-MiegJMP, Bonora1983}). Horizontality is such an important concept because it will provide a notion of \textit{physical}. Namely, horizontal will mean \textit{physical with respect to a given \lq{}observer\rq{}, or field}.

	In principle, one could add new degrees of freedom---extra fields---to play the role of such observers, or frames of reference. But the addition of extra-fields is not \textit{necessary} for the construction of $\varpi$; indeed it goes against the relational foundation of this work---which only relies on the pre-existing physical fields. Nonetheless, we present a simple example in which a  $\varpi$ with the correct  properties \textit{is} constructed with the aid of extra fields. This construction recovers previously studied  edge-modes, introduced to deal with gauge-invariance in the presence of boundaries \cite{Donnelly:2014fua, Wadia, Regge:1974zd, Henneaux:2018gfi, Carlip:2004mn, Balachandran:1994up} and sometimes identified precisely with observer (or frame) degrees of freedom \cite{Donnelly:2016auv}.

	Apart from this brief excursion, in all cases studied here we have induced connection-forms from \textit{supermetrics} on field-space, that is, directly from the geometry of field-space itself. Interestingly, the geometry of field-space can be severely restricted by the assumption that the field-space supermetric is ultralocal \cite{Singer:1978dk, Singer:1981xw, Narasimhan:1979kf, Babelon:1979wd, Babelon:1980uj, Mitter:1979un,AsoreyMitter,DeWitt_Book}. This is discussed in section \ref{sec:dw_connection}.
	
	One main difference between the simple connection-forms induced from extra fields and the more physically motivated ones---{relationally defined via} the intrinsic geometry of field-space---is that the latter may have associated \textit{curvature}.  Curvature may arise either from field-space dependence of the supermetric itself (as in general relativity) or from field-space dependence of the gauge-group action on field-space (as in non-Abelian gauge theories and  general relativity).
	
	Interestingly, curvature also plays a role in the relation between our formalism and gauge fixings \cite{Singer:1978dk, Singer:1981xw}. Namely, one can always relate, at a perturbative level, a choice of a connection-form to a gauge-fixing. The main idea of this relation is that an infinitesimal horizontal plane can be locally extended in field-space in an affine manner. In the Abelian Yang-Mills case for example, this analogy holds also globally. In more general cases, a connection-form will have associated curvature; then, it can still be global, but will not be integrable, and therefore not equivalent to any choice of gauge-fixing.
	
	By this account, a $\varpi$ constructed from the geometry of the {space of} YM gauge-potentials  will have broadly different properties depending on whether the field is Abelian or not. In the Abelian case, there is no Gribov problem and the PFB is trivial; in the non-Abelian case, there is a Gribov obstruction and the bundle  does not admit a global section. In the Abelian case, there is no curvature, while in the non-Abelian there is. In either case,  our connection-form  requires  neither a gauge-fixing section nor an explicit parametrization of the base  manifold--- representing the physical degrees of freedom---and therefore the Gribov obstruction plays no role in the use of $\varpi$. The corresponding topological information of field-space is taken up by the field-space curvature of $\varpi$, which may therefore start playing an important role in non-perturbative aspects of the path integral (which we plan to investigate in the near future). The specific connection-forms and their curvatures for the ultralocal geometry in the space of YM vector potentials are calculated in section \ref{sec:Singerconnection}. This choice of $\varpi$ is termed the Singer-DeWitt (SdW) connection. 
	
	Due to the nature of the gauge-potentials and of the gauge action on them, SdW connections are nonlocal functionals.
	Already shifting to the physical applications,  section  \ref{sec:Singerconnection} then discusses the relation between  horizontality in field-space and the compatibility of this non-locality with the restriction to regions. 
	If we associate a field-space to each physical region, it  will be equipped with its own regional notion of horizontality, which is nonlocal for the $\varpi$ based on the YM potential.  The most important remark on this issue is that---as a consequence of nonlocality---{\it regional restrictions  do not commute with horizontal projections}. Nonetheless, the composition of regions and their physical charges is well-defined and consistent; we therefore say such connections are nonlocal but \textit{regional}.

	\paragraph*{Part II -- Physical applications}
	~ \\ 
	
	In this paper, we also explore the roles which simple geometric connection\hyp{}forms have for physical theories. We start this exploration with the $\varpi$ based on the space of YM vector potentials, in section \ref{sec:Noether_hor}, where we apply it to the covariant symplectic treatment of  Yang--Mills theory  proposed by two of us in \cite{Gomes:2016mwl, Gomes:2018shn}.   The difference between the standard symplectic potential and our modified (horizontal) one is given by a boundary term.
	
	Again we find that in the absence of boundaries our formalism reduces to standard treatments.  But this almost-everywhere coincidence  should not {distract from} the importance of their difference; charge currents require boundaries, and we find that for generic  backgrounds all the appropriately modified, so-called \textit{horizontal} Noether currents are screened in the non-Abelian case.  For backgrounds which possess global symmetries, on the other hand, the SdW connection picks out the global charges as the only physical ones. For the Abelian case (electrodynamics), the horizontal Noether current is always precisely the total current density of electrons. Importantly, while standard  Noether currents can be associated to any gauge-parameter \cite{Lee:1990nz}, our horizontal ones are non-trivial only for (at most) a finite number of parameters. Horizontal charges are related to objectively conserved physical quantities  in the sense of the first Noether theorem. 
	Nonetheless, in spite of these differences, the derived symplectic 2-form of our formalism is still closed, and therefore equips the field-space of the theory with a well-defined symplectic structure.  
	
	Each field-space sector---be it of matter or of gauge potentials---carries its own geometry. Nonetheless, as long as we have the same gauge group acting on each, covariantization of one sector implies covariantization of all. Some of the most interesting outcomes of our work come from the  study of the natural (i.e. ultralocal) field-space geometry for the matter sector of the theory, pursued in  section \ref{sec:matter}.  
	
	Since gauge transformations of matter-fields do not involve derivatives, their associated connections are not only regional, but indeed completely local in spacetime, and are moreover  locally flat in field-space. However, as we will see shortly, they fail to be defined everywhere in field-space. The  connection $\varpi$ emerging from this procedure---termed {\it Higgs connection}---{is flat and can therefore} be put in correspondence {to a gauge-fixing. This is indeed} the celebrated \lq{}Higgs unitary (partial) gauge\rq{} \cite{WeinbergQFT2, Weinberg:1973ew}; more connections  to the Higgs are coming. This type of locality is one of the main differences between matter-induced relationalism and  vector-potential relationalism, but there are others.
	
	It turns out no smooth {Higgs connection} will exist for background-fields which vanish at any given point in space. In other words, { the Higgs connection}  only exists for symmetry-broken configurations, where the vacuum expectation value of the field does not vanish anywhere. Such configurations are known as {condensates}. This example provides an intuitive physical meaning to the screening of charges by the {Higgs connection}, as due to a proliferation of charged particles formed by the condensate.
	
	To realize the depth of the analogy, we first note that, in the same way as for the Yang--Mills vector potential, the {Higgs} connection will be \lq{}blind\rq{} to the directions of global symmetry, at backgrounds for which those exist. This merely states that if {the reference matter} field is insensitive to some  transformations, they cannot be measured in relation to it.  
	In this case, as a consequence of the locality of the gauge action on matter, and therefore in contrast to what happens in the case of the YM potentials and the SdW connection, the `global' symmetries---if they exist---factorize at every point of spacetime, thereby forming an infinite-dimensional group.
	
	In the Higgs-condensate analogy, this infinite-dimensional group is associated to the Goldstone modes of the symmetry-breaking, i.e. to the `residual' massless gauge vector bosons. {These directions cannot be dressed by the corresponding Higgs connection.}  Therefore they will still have associated non-trivial Noether charges and correspond to long-range interactions; these components \textit{are not} screened by the  Higgs connection. The other directions---those which {act nontrivially on the reference field}---acquire a non-vanishing mass. As a consequence, the interaction such field components mediate is not long-ranged, which of course affects the Gauss law in this broken phase. In other words, these components \textit{are} screened by the $\varpi$-condensate. All of these facts are automatically taken into account by our horizontal Noether charge. These points are discussed {in greater detail} in section \ref{sec:matter}, as well as the relation of the Higgs connection to physical reference frames.

	A last example of connection-form is presented in section \ref{sec:EC}. 
	There, we study the Higgs connection associated to the Lorentz symmetry in the vielbein (Einstein--Cartan) formulation of general relativity. 
	The reference field for this Higgs connection is the vielbein field, whose expectation value cannot vanish in a nondegenerate spacetime.
	The interest of this construction lies in the interplay of Lorentz and diffeomorphism symmetry in vielbein general relativity,  which has raised puzzles on the status of the Hamiltonian derivation of the first law of black hole thermodynamics in vielbein general relativity  \`a la Wald \cite{wald1993black}. To resolve this puzzle, in \cite{Jacobson:2015uqa} a Lorentz-adjusted notion of diffeomorphism symmetry was introduced through a modification of the Lie derivative. Alternatively, in \cite{DePaoli:2018erh} a Lorentz invariant symplectic potential for vielbein gravity was derived by the addition of appropriate boundary terms, also resolving the puzzle. In this case, the potential turns out to be identical on-shell to the metric Einstein--Hilbert symplectic potential.

	Both of these proposals are encompassed by our framework: the modified Lie derivative is the horizontal projection of the standard Lie derivative (understood as a vector on field-space), and the modified symplectic potential is the horizontal projection of the standard symplectic potential (a 1-form on field space).	
	
	Connections-forms call for Wilson lines. But what is the physical meaning, if any, of a Wilson line in field-space? 
A moment of reflection shows that this Wilson lines provide field-dependent elements of the group of gauge transformations. Most importantly, these are elements which transform covariantly. 
Such an object is precisely what is required for a notion of \lq{}dressing\rq{} \cite{Dirac:1955uv, Lavelle:1995ty, Lavelle:2009zz, bagan2000charges, bagan2000charges2,Capri2005, Attard2018, francoisthesis, Ilderton:2010tf, Lavelle:1994rh}: these are field-dependent gauge transformations that can be combined to charged fields so as to build gauge-invariant `dressed fields'.
Thus far in our work, different notions of dressing seem to correspond to  different connections. 
We will mostly focus on the dressing built out of the SdW connection, and will relate it to important constructions of gauge-invariant fields by different authors, such as Lavelle and McMullan\rq{}s constructions for QCD  \cite{Lavelle:1995ty}---which are  also related to the dressings of  the Gribov--Zwanziger framework \cite{ Zwanziger:1989mf,Capri2005, Vandersickel:2012tz}---and also to those of Vilkovisky \cite{Vilkovisky:1984st, vilkovisky1984gospel}, whose constructions occur in a different, more general  context.
However, due to the non-vanishing of the curvature of the SdW connection, it turns out that dressings can only be defined perturbatively in the non-Abelian theory.
The only notion of nonperturbative dressing that survives in this case is therefore an infinitesimal one, which we argue  corresponds to the horizontal differential  introduced in the first part of the paper.
	
	Lastly we note that in the absence of boundaries, our horizontal covariant symplectic formalism reduces to the standard one of e.g. \cite{WittenCrnkovic, Lee:1990nz, wald1993black, Iyer:1994ys},  thus recovering known results in the literature. 
 However, apart from the fact that physical observers are always contained in bounded regions, where total charges  and currents are  measured, $\varpi$ remains a novel, useful tool also in the absence of boundaries. This is exemplified by e.g. the significant efforts towards a geometrical, gauge-invariant, understanding of  the path integral \cite{DeWitt:1967ub, Vilkovisky:1984st, vilkovisky1984gospel, DeWitt:1995cx, DeWitt_Book, Branchina:2003ek, Pawlowski:2003sk},  and the replacement of Gribov ambiguities by field-space curvature effects. Moreover, new applications of $\varpi$ in the absence of boundaries are not restricted to quantum mechanics;} they include the study of a (spacetime local) $\varpi$ based on charged matter fields and the study of non-Abelian analogues of the Dirac dressing. These and other physical applications are the subject of the second part of the paper.
	 
	 (In appendix \ref{app:DeWitt}, a small dictionary is presented to ease comparison with DeWitt's notation.)

	\part{Mathematical theory}

	\section{Field-space geometry} \label{sec:fieldspace_geo}

	In the next two sections, we will introduce the technical and notational scaffolding for the remainder of the paper. 
	It consists of two pillars:  \textit {the geometry of field-space}---so that we can talk about local gauge transformations in the appropriate framework---and the \textit{geometry of principal fiber bundles}---so that we can talk about a general concept of covariant derivatives.\footnote { Some of the material discussed in this section was already present in \cite{Gomes:2016mwl}.}

	\subsection{Field-space geometrical tools\label{sec_fieldspace}}
	
	Let us ignore some mathematical subtleties and consider the space of field configurations $\Phi=\{\varphi^I(x)\}$ to be a manifold.\footnote{A manifold locally modeled on a Banach space is easy to define (see \cite{Lang} for details on the infinite-dimensional aspects). But here there are intricacies corresponding to the fact that we have sections of infinite-differentiability. To properly define such manifolds, one needs to be more careful, but the end result is what is called an Inverse Limit Hilbert manifold, and it possesses all of the structure we require (see e.g.: \cite{Ebin, Palais, YangMillsSlice, kondracki1983, Mitter:1979un, Michor, fischermarsden} for the different contexts in which these subtleties arise and how they are resolved, and for the validity of the required mathematical theorems in this infinite-dimensional context).} 
	Here, $x\in M$ is a point in a space(time) region and $I$ is a super-index labeling both the field's (finite) types and their components.
	At this level, we are still off-shell, meaning that the field configurations $\varphi\in\Phi$ do not have to satisfy any equations of motion.  
	In the following, a `double-struck' typeface---like in $ \dd$, $\fF$,  $\fLie$, $\fvf$, etc.---will be consistently used for field-space entities.
	
	On $\F$, we introduce the deRham differential $\dd$ \cite{ Crnkovic:1986ex, Crnkovic:1986be, Crnkovic:1987tz}; it should be thought of as the analogue, on $\F$, of the spacetime differential $\d$. 
	A basis of the one-forms on field-space, $\Lambda^1(\F)$, is hence given by $\Big(\dd \varphi^I(x) \Big)$.
	On a functional $f:\F\to \mathbb{R}$ (reals), $\dd$ acts as: 
	\begin{align}
	\dd f = \sum_{I}\int_M \d^n x \left( \frac{\delta f}{\delta  \varphi^{I}(x)} \dd \varphi^{I}(x)\right)  =:\int \frac{\dd f}{\dd \varphi^I} \dd \varphi^I,
	\label{dd}
	\end{align}
	where $\delta/\delta \varphi$ denotes as usual a functional derivative, and the last identity introduces a more homogeneous short-handed notation.
	Higher dimensional (functional) forms are defined by the above formula and antisymmetrization. In particular $ \dd^2 = 0$ (wedge products are left understood).
	
	Functional spacetime-local vector fields on $\F$ are denoted $\fvf \in \mathfrak{X}^1(\F)$. In components, they read
	\be\label{vectors}
	\bb X =  \sum_I \int_M \d^n x \left( \fvf^I(\varphi(x)) \frac{\delta}{\delta \varphi^I(x)}\right) =: \int \fvf^I\frac{\dd}{\dd \varphi^I} \,,
	\ee
	where the introduced notation follows that of \eqref{dd}. 
	When extra emphasis is needed, we will denote by $\bb X_\varphi\in\mathrm T_\varphi \F$ the value of $\bb X$ at $\varphi$.
	In the following, in specific circumstances we will introduce certain spacetime non-local vector fields; we will make clear how locality is relaxed in these occurrences.
	
	Contraction of a vector field with a differential form in $\F$ is denoted with $\fI$, and defined by
	\be
	\fI_\fvf \dd \varphi^I = \fvf^I
	\ee
	and the usual rules of linearity and antisymmetrization. 
	
	We also introduce the functional Lie derivative along $\fvf$ of a generic functional form through the Cartan formula
	\be
	\fLie_\fvf = \fI_\fvf\dd + \dd \fI_\fvf,
	\label{eq_magic}
	\ee
	and denote the Lie bracket between two vector fields with a double-struck notation
	\be
	\fLie_{\lbr \bb X, \bb Y\rbr} = \fLie_{\bb X} \fLie_{\bb Y} - \fLie_{\bb Y} \fLie_{\bb X}  .
	\ee
	
	Finally, on a slightly different note, we conclude by observing that in this formalism the field-space and spacetime differential commute, i.e.
	\be
	\dd \d - \d \dd = 0.
	\label{eq_ddcomm}
	\ee

	\subsection{Field-space of Yang--Mills theory as a principal fiber bundle}
	
	Although a large part of the construction presented here in this context can be generalized to other theories, in this paper we will restrict our attention to Yang-Mills theory with matter (YM).
	
	Consider YM theory with {\it charge group} $G$. Although most of our discussions goes through for any $G$ compact and semi-simple,  in the interest of definiteness, we fix $G$ to $\SU(N)$ or ${\rm U}(1)$.
	Gauge transformations themselves form a group, the \textit{gauge group}
	\be
	g(\cdot) \in \G \cong C^\infty(M, G),
	\ee
	with point-wise composition over $M$.
	To be clear, these are the smooth functions from $M$ to $G$, so each element roughly consists of one choice of group element per point, $g(\cdot):M\to G, \; x\mapsto g(x)$. If there is no risk of confusion with elements of $G$, we will denote the elements of $\G$ simply as $g$.
	Similarly, the Lie algebra of the gauge group is given by
	\be
	\xi(\cdot) \in \fG = C^\infty(M,\fg)
	\ee
	where $\fg={\rm Lie}(G)$.
	
	At this point, we consider $\G$ as the unconstrained group of gauge transformations. That is, the group is unconstrained by fall-off or any other boundary conditions. It is the aim of the framework we introduce in this article to adapt the relevant quantities---charges, symplectic potential, etc.---to abide by these transformations,  in all scenarios. Note that this demand becomes particularly relevant when full gauge covariance is not automatically guaranteed by the standard treatments, i.e. precisely in the presence of boundaries.
	
	Another relevant question is whether {\it all} elements of $\G$ are indeed to be considered `pure gauge' or whether some of them are singled out as `physical' symmetry transformations. E.g. one might expect this to happen for global ${\rm U}(1)$ transformations in electromagnetism. This is a central point of our discussion; we will touch upon this shortly, although what is precisely meant by `physical' and `gauge'  will have to wait until section \ref{ssec:sympl_charges}.

	Back to the general discussion.
	The field-space of YM, $\FYM$, is given by gauge connections $A$ and matter fields $\Psi$,
	\be
	\FYM = \{\varphi =  (A,\Psi) \} .
	\ee
	Gauge connections are $\fg$-valued 1-forms over the spacetime manifold\footnotemark~$M$,%
	\footnotetext{One could also take the (less popular) parametrization of degrees of freedom for YM as $\omega$, a connection on a principal fiber bundle, with base space being spacetime and fiber isomorphic to $G$---the relation between $A$ and $\omega$ requires a section $\sigma:M\rightarrow P$, and is then $A=\sigma^*\omega$. This distinction is important for non-trivial bundles, where $\omega$ exists globally but $A$ only locally.  Although we will stick to the \lq{}physicists\rq{} parametrization, $A$,  our formalism can be readily extended with minimal modifications to the case where $\omega$ is considered as the fundamental variable. \label{ftnt:PFB}}
	\be
	A = A^a_\mu(x) \tau_a\d x^\mu \in \Lambda^1(M, \fg),
	\ee
	where $\fg= \text{Lie}(G)$ and $\{ \tau_a \}_a$ is an orthogonal basis of the latter.
	We take it normalized with respect to the trace in the fundamental representation as\footnote{For $G=\SU(2)$, $\tau_a = -\frac{i}{2}\sigma_a$, with $\sigma_a$ the Pauli matrices.} $\tr(\tau_a \tau_b) = -\frac12 \delta_{ab}$.

	We will consider in the following two types of matter fields: scalar matter fields, $\phi$, which are smooth functions on $M$ valued in $W$, with $W$ the fundamental representation of $G$:
	\be
	\phi = \phi^m(x)|m\rangle \in \mathcal{C}^{\infty}(M,W);
	\ee
	and Dirac spinorial matter fields, $\psi$, which are smooth anticommuting functions on $M$, valued in $\mathbb C^{ 4}\otimes W$:
	\be
	\psi = \psi^{\alpha m}(x)|\alpha\rangle |m\rangle  \in \mathcal{C}_{\rm a}^\infty(M, \C^{4}\otimes W),
	\ee
	where $\alpha$ and $m$ are spinorial and color indices respectively ($W$ is as above)---in this case, the field-space of matter is a supermanifold \cite{DeWitt:1992cy, HenneauxTeitelboim}.
	
	When the spacetime or commutation properties of the fields are not relevant, we will simply denote the matter fields, be they scalar or spinorial, by $\Psi$.

	\begin{figure}[t]
		\begin{center}
			\includegraphics[scale=0.17]{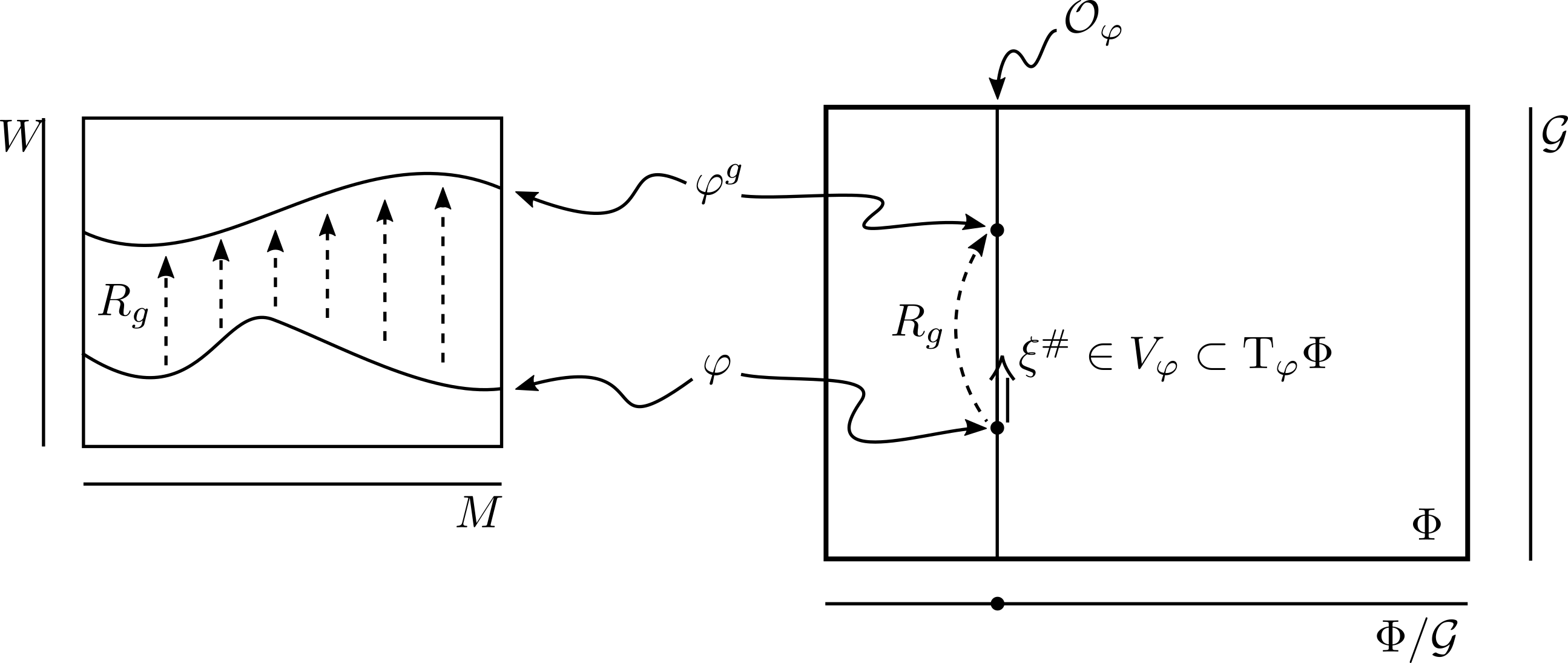}	
			\caption{A pictorial representation of the field-space $\F$ seen as a principal fiber bundle. We have highlighted a configuration $\varphi$, its (gauge-transformed) image under the action of $R_g:\varphi\mapsto\varphi^g$, and its orbit $\mathcal O_\varphi \cong \G$. We have also represented the quotient space of `gauge-invariant configurations' $\F/\G$. On the left hand side of the picture, we have `zoomed into' a representation of $\varphi$ and $\varphi^g$ as sections of a vector bundle over the spacetime region $M$ (here, we are assuming $\varphi$ to be the scalar field $\phi$ valued in $W$). }
			\label{fig1}
		\end{center}
	\end{figure}

	The connection and matter fields transform under the action of the gauge group  and thus gauge transformations induce a natural {\it right} action of $\G$ on $\FYM$:
	\be
	\begin{array}{rccl}
		R :&  \G\times \FYM &\to& \FYM  \\
		&\Big(g(\cdot), \varphi\Big) &\mapsto& R_{g(\cdot)}\varphi=\varphi^g
	\end{array}
	\label{eq_right_action}
	\ee
	where
	\be
	A^g = g^{-1} A g + g^{-1}\d g 
	\quad\text{and}\quad
	\Psi^g = g^{-1}\Psi.
	\label{group_action}
	\ee
	This action `morally' turns $\FYM$ into an {\it infinite-dimensional principal fiber bundle} (PFB)  with base manifold given by the `physical' configurations of the fields, i.e. the space of connections modulo gauge transformations  $\F_\YM/\G$, and fibers isomorphic to the gauge group $\G$. The conditions for  a bona-fide PFB, depicted in figure \ref{fig1}, are not perfectly satisfied by $\G$\rq{}s action on $\F_\YM$. There are in fact several complications, due in particular to the  nature of the quotient of  $\FYM$ by $\G$.
	Although most of the obstructions arising from infinite-dimensionality can be overcome \cite{Ebin, Palais, YangMillsSlice,  Mitter:1979un, kondracki1983}, the foremost technical obstruction to a PFB structure of $\FYM$,  the fact that the gauge orbits provide a foliation of $\FYM$ rather than a fibration, remains. That is because the fibers are only generically, and \textit{not} always, isomorphic to $\G$. In the following, we will find important physical consequences of this fact, an it therefore requires further comments.

\begin{figure}[t]
		\begin{center}
			\includegraphics[scale=0.17]{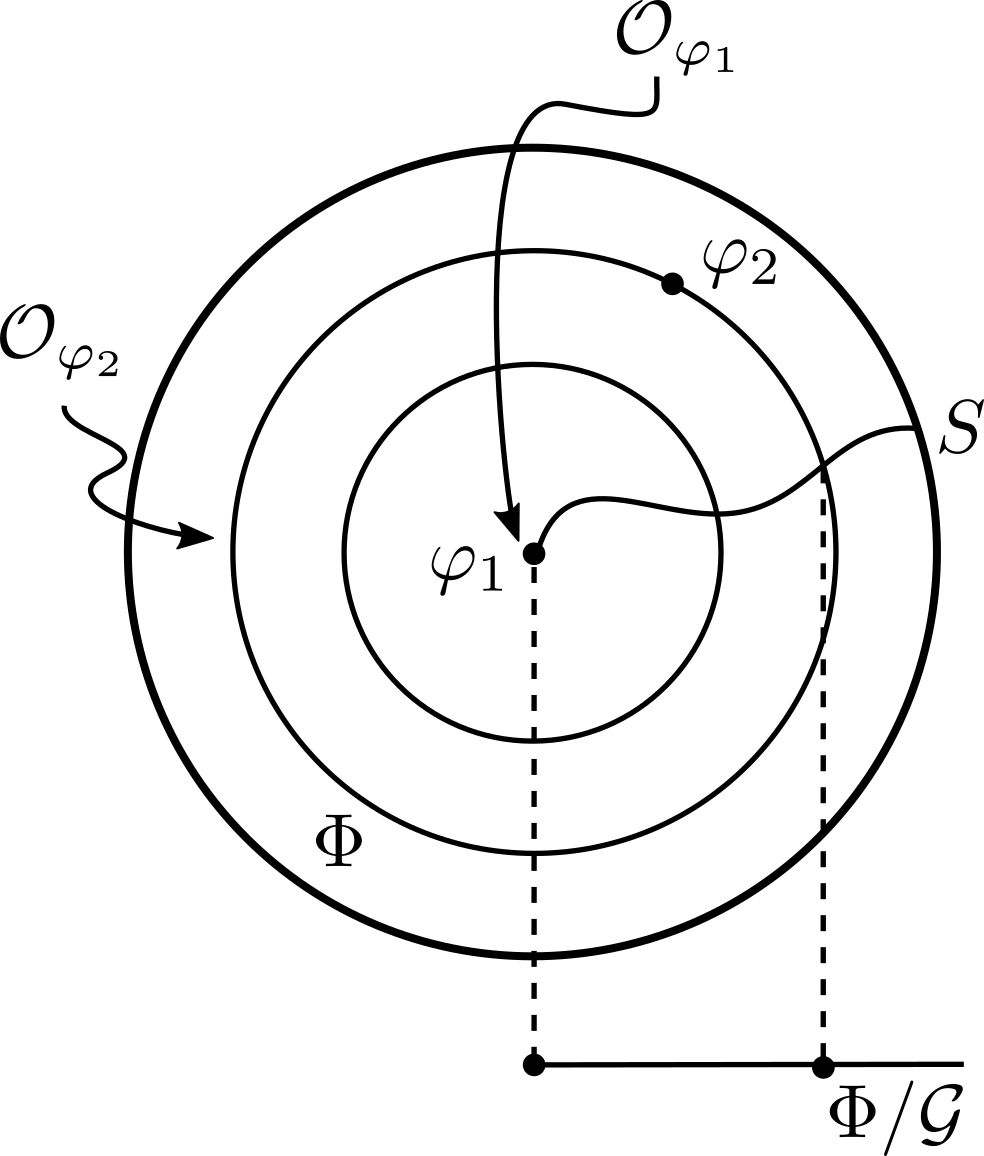}	
			\caption{In this representation $\Phi$ is the page's plane and the orbits are given by concentric circles around $\varphi_1$; $S$ is a section. The field $\varphi_2$ is a generic field in $\F$. The field $\varphi_1$ has a nontrivial stabilizer group: its orbit is reduced to a lower dimensional manifold (in this case just a point). The projection of $\varphi_1$ on  $\F/\G$ sits at a qualitatively different point than that of $\varphi_2$: $\varphi_1$ is part of a lower dimensional stratum (in this case a 0-dimensional one).}
			\label{fig8}
		\end{center}
	\end{figure}

	The orbits of the field configurations $\varphi$, $\mathcal O_\varphi = \{R_{g(\cdot)}\varphi, g(\cdot)\in\G\}$, constitute the fibers of the would-be PFB.
	These orbits are generically isomorphic to $\G$; unless the configuration $\varphi$ is invariant under (conjugacy classes of) subgroups of $\G$.
	A configuration $\varphi$ with this invariance is called {\it reducible}, the subgroups of $\G$ under which it is invariant is referred to as its {\it stabilizer group},\footnotemark~$\mathcal S_\varphi\subset\G$.%
	\footnotetext{In order to obtain an actual PFB structure on $\F$, one can slightly modify the gauge group $\G$ to its `pointed' version $\G_o\subset \G$, where $o\in M$ is fixed (e.g. \cite{Singer:1978dk,Singer:1981xw}). $\G_o$ contains all and only those elements of $\G$ that are the identity at a fixed $x=o$. That is, $g_o\in\G_o$ if an only if $g_o(x=o) = {\rm id}$. This property makes the stabilizer subgroups of $\G_o$ necessarily trivial, while allowing the  (dense) subset $\F_o\subset \F$, constituted by those elements with trivial stabilizer, to have a principal fiber bundle structure. Nevertheless, we discard this option since the stabilizer groups  will play a central role in our discussion, and since they contribute in crucial ways to the topological properties of the quotient $\F/\G$. }
	Since certain configurations \textit{do} have nontrivial stabilizer groups, the symmetry group in question may  give rise to qualitatively different orbits%
	---figure \ref{fig8}. This is the obstruction to the bona-fide PFB structure of $\FYM$, since then
	the quotient space is no longer a manifold, but a patchwork of manifolds of different dimensions called \textit{a stratified manifold}. 
	Indeed, the defining feature of a PFB, that is, its local product structure, can be amended to include these types of group action, because they still have \lq{}slices\rq{}. Slices\footnotemark~have been shown to exist for the spaces of: Yang-Mills potentials, Euclidean metrics, and a certain subset of Lorentzian Einstein metrics \cite{Ebin, Palais, YangMillsSlice, kondracki1983, Mitter:1979un,  isenberg1982slice}.
	The existence of lower strata will be explicitly related to the existence of 
	 `physical' charges  in section \ref{ssec:sympl_charges}. 
	\footnotetext{ Roughly speaking, a slice for the action of a group $\G$ on a manifold $\F$ at a point $\varphi \in \F$ is a manifold $S_\varphi$, transversal to the orbit of $\varphi$,  $\mathcal{O}_\varphi$.  If the stabilizer group $\G_\varphi$ of $\varphi$ is trivial, then $S_\varphi$ can give a local chart for the space $\F/\G$ near $\varphi$. In this case, one can use the slice to parametrize the physically distinct configurations. When the stabilizer groups of $\varphi\in \F$ become non-trivial,  the symmetry group in question may act qualitatively differently on different orbits.
		In that case,  let $\mathcal{N}_{\varphi_o}=\{\varphi\in \F~|~\mathcal{S}_\varphi ~~ \mbox{is conjugate to}~~ \mathcal{S}_{\varphi_o}\}$. Then one can show that $\mathcal{N}_{\varphi}/\G$ is a manifold (since $\mathcal{S}_\varphi$ does {\it not} change dimension). Each such manifold defines a  \lq{}stratum\rq{},  containing  the orbit $\mathcal{O}_\varphi$. The larger the stabilizer group---i.e. the more symmetric the configuration---the smaller the dimension of the stratum; their union forms a concatenation of manifolds of decreasing dimension. See \cite{isenberg1982slice, kondracki1983} for  reviews. \label{footnt:strata}}
	
	 In effect, all of our constructions will appear in the gauge-variant $\F$, not $\F/\G$. This is one of the assets of our formalism, since $\F$ has simple topology and local parametrizations, unlike $\F/\G$. Hence, most of the subtleties distinguishing between an actual PFB and $\F$ are for us immaterial---apart from its generalization to allow for reducible configurations. To summarize: we use $\F$ and require at most the validity of slice theorems.
	
	From the infinitesimal version of the group action on field-space, we can readily define a map from the Lie algebra of the gauge group, $\fG$, into the vector fields on field-space $\fX^1(\FYM)$,
	\be
	\begin{array}{|crcclc|}
		\hline&&&&&\\
		&{}^\# :&  \fG &\to& \fX^1(\FYM)  &\\
		&&\xi &\mapsto& \xi^\# &\\
		&&&&&\\\hline
	\end{array}
	\label{eq_sharpoperatordef}
	\ee
	which associates to an infinitesimal gauge transformation $\xi(\cdot)\in\fG$ the associated flow $\xi^\#\in\fX^1(\FYM)$ on field-space. We will denote $\xi^\#$, $\xi\in\fG$, the {\it fundamental vector fields}.
	
	More explicitly, the action of the flow on functions is defined through
	\be
	{\xi}^\#f   := \fLie_{\xi^\#}f := \frac{\d}{\d t}_{|t=0} R_{\exp(t{\xi})}^*f(A,\Psi) .
	\ee
	From this definition and equations \eqref{eq_right_action} and \eqref{group_action}, it is then immediate to verify that
	\be
	\xi^\# = \int \delta_\xi A \frac{\dd}{\dd A} + \int \delta_\xi \Psi \frac{\dd}{\dd \Psi}
	\label{xi_hash}\ee
	where we adopted the notation of section \ref{sec_fieldspace} for the field-space vectors, and introduced the standard notation for infinitesimal gauge-transformations (along $\xi$),
	\be
	\delta_\xi A = \D \xi := \d \xi + [A,\xi]
	\qquad\text{and}\qquad
	\delta_\xi\Psi = - \xi \Psi,
	\label{eq_infin_gtransf}
	\ee
	with $[\cdot,\cdot] $ the Lie bracket on $\fg = {\rm Lie}(G)$, extended pointwise on $M$ to $\fG$.
	The purpose of the notation $\delta_\xi$ is not only to relate to the usual one for gauge transformations, but also to have a separate notation for the components of the vector field in field-space.

	The meaning of ${\xi}^\# f$ is `the variation of $f$ under the infinitesimal gauge transformation ${\xi}(x)$'.
	In general, the choice of $\xi(\cdot)\in\fG$ can depend on the field configuration,\footnote{Second order of differentiability will be required.} i.e. we will generalize our notation to
	\be
	\boxed{\quad\phantom{\Big|}
		\xi:\FYM\rightarrow\fG.
		\quad}
	\ee
	In this case we speak about {\it field-dependent} gauge transformations.\footnote{Strictly speaking, field-dependent gauge transformations are {\it not} elements of the group of gauge transformation. They are nonetheless natural entities, which technically are the morphisms of the `action groupoid' associated to the action of $\G$ on $\F$ (see e.g. \cite{BaezGroupoids} for a pedestrian account). However, to maintain our presentation simple---but hopefully not confusing!---we will continue to use the loose term `field-dependent gauge transformations'.
	 In the physics literature, the importance of field-dependent extensions of the gauge group has also been stressed by Barnich and collaborators, e.g. \cite{Barnich:2010xq}.}
	The name `fundamental vector fields' is reserved for $\xi$'s which are field {\it independent}.
	
	The transformation taking a given field configuration to a gauge-fixing section is a typical example of field-dependent gauge transformations. 
	{\it Henceforth, we will assume field-dependence to be always non-trivial, unless otherwise stated} (look for the $(\dd\xi=0)$ specifier).
	For example, if $\xi$ is chosen to depend on the matter field $\Psi$ configuration, but not on that of the gauge connection $A$, then $\dd \xi = \int \frac{\dd \xi}{\dd \Psi} \dd \Psi \neq 0$.

	For field-independent $\xi$, it is easy to show that
	\be
	\label{integrable} 
	\lbr {\xi_1^\#},{\xi_2^\#} \rbr={[\xi_1, \xi_2]}^\# \qquad (\dd \xi =0),
	\ee
	while for field-dependent Lie algebra elements, the identity becomes\footnote{This can be seen by writing $\xi = \int \xi^a \tau_a$ with a field- and position-independent basis $\tau_a$ of $\fg$, and field- and position-dependent coefficients $\xi^a$.	}
	\begin{align}
	\lbr \xi_1^\#, \xi_2^\# \rbr = [\xi_1, \xi_2]^\# + \big(\xi_1^\# (\xi_2)\big)^\# - \big(\xi_2^\# (\xi_1)\big)^\#.
	\end{align}
	
	Recall that the double-struck bracket on the left hand side is the canonical Lie bracket between vector fields on $\FYM$, while the Lie bracket on the right hand side is the Lie bracket between elements of $\fg$.

	In the next sections, we will discuss the natural transposition of standard gauge theoretic structures from the spacetime perspective to the field-space one. The easiest way to do this is to rely on the geometric PFB picture presented here, for which such structures are unambiguous and intuitively clear. 
	Exploiting this picture, we will explain the existence and meaning of connection-forms in the field-space context.

	\section{Connection-form on $\FYM$}\label{sec:general_YM}

	\subsection{Basic definitions}
	
	Vector fields $\fvf$ which are tangent to  gauge orbits in $\FYM$ will be called `vertical' and their span at a  $\varphi\in\FYM$ defines a vertical subspace of the tangent space.  In symbols, 
	\be 
	\boxed{\quad\phantom{\Big|}
		\mathrm T_\varphi \FYM \supset {V}_\varphi = \mathrm{Span}\{{\lvf}^\#, {\lvf}\in\fG\}.
		\quad}
	\ee
	Vertical fields represent infinitesimal  gauge transformations. By \eqref{integrable}  they  span integrable distributions of $\mathrm T\F$, and by the Frobenius theorem\footnote{More precisely, by its generalization for infinite-dimensional manifolds, see \cite{Lang, Ebin}.} they span the tangent spaces to the orbits $\mathcal{O}_\varphi=\{ R_{g(\cdot)}\varphi, \, g(\cdot)\in \G\}$. 
	The disjoint union of the vertical tangent spaces is denoted $V\subset {\rm T}\FYM$.

	The vertical subspace of the tangent bundle, $V$, can be complemented with another transversal subspace, $H=\cup_\varphi H_\varphi$, which we call horizontal:%
	\footnote{The conditions for the existence of such direct sums in the case of infinite-dimensions are given in e.g.: \cite{Lang, Michor, Ebin}. } 
	\be
	\boxed{\quad\phantom{\Big|}
		\mathrm T\FYM\simeq {V}\oplus {H}.
		\quad}
	\ee
	Crucially, as in the finite-dimensional case, there is no canonical transversal complement to the vertical subspaces. At each $\varphi$, a choice of $H_\varphi$ corresponds to a choice of a vertical projector, $\hat V_\varphi:\mathrm T_\varphi\FYM\to V_\varphi$, through
	\be
	H_\varphi=\text{ker}(\hat V_\varphi).
	\ee  
	By defining a horizontal subspace, one obtains a (path-dependent) identification between gauge degrees of freedom at different orbits.

	From this perspective, it is natural to introduce on $\FYM$ a functional {\it connection\hyp{}form} $\varpi$ (\textsc{var-pie}) which implements a notion of vertical projection. 
	This works in the following way.
	Mimicking the finite-dimensional case \cite{kobayashivol1}, we define $\varpi$ as a (bosonic) functional 1-form over field-space, valued in the Lie algebra of the gauge group $\fG$,
	\be
	\varpi \in \Lambda^1(\FYM, \fG).
	\ee
	The idea is to use the fact that a one-form naturally contracts with vector fields to define horizontal complements as $\varpi$'s kernel:
	\be
	\boxed{\quad\phantom{\Big|}
		H := {\rm ker}(\varpi) = \{ \fvf \in \mathrm T\FYM \,|\, \fI_\fvf \varpi = 0\}.
		\quad}
	\ee
	
	\begin{figure}
		\begin{center}
			\includegraphics[scale=.17]{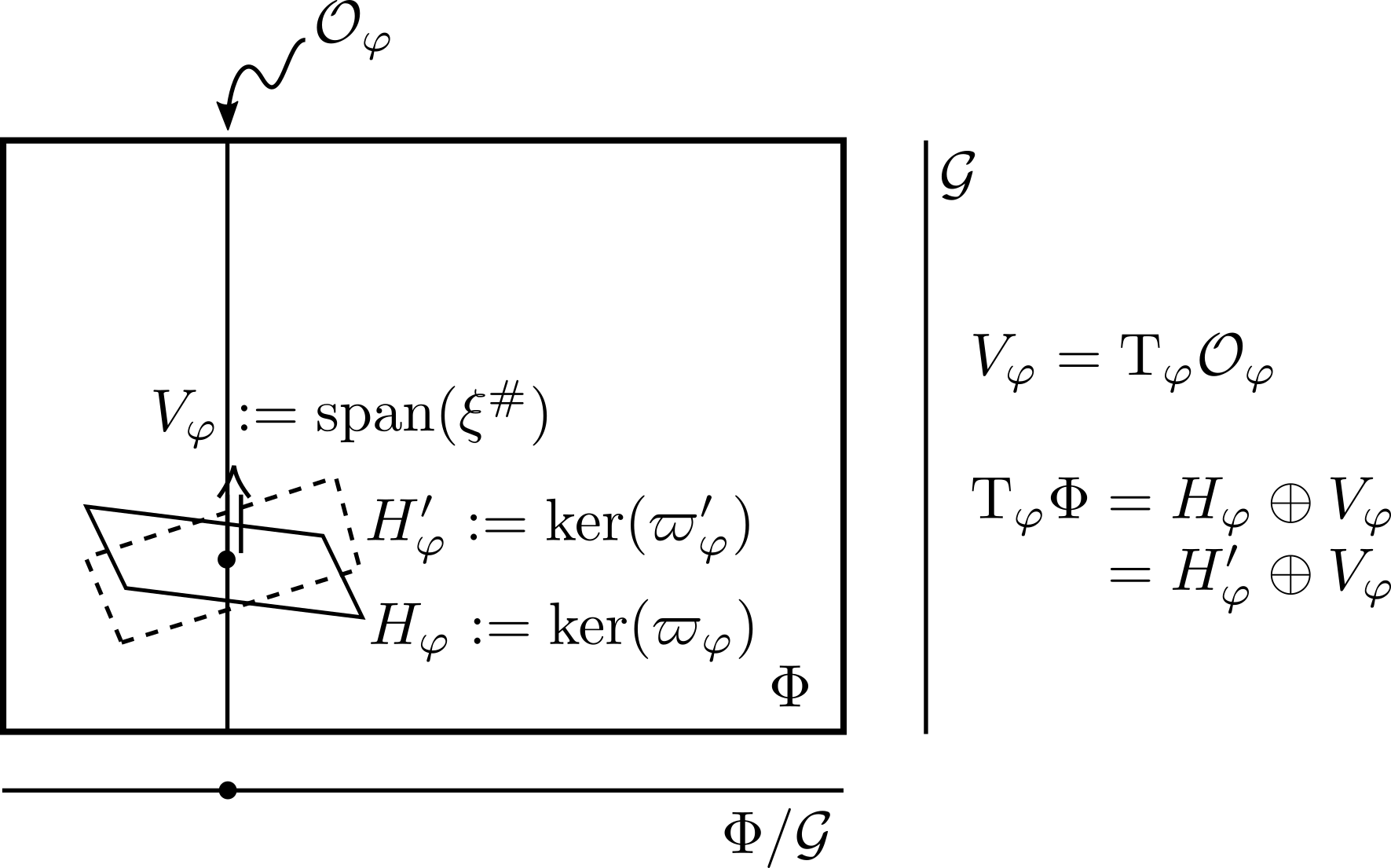}
			\caption{A pictorial representation of the split of ${\rm T}_\varphi \F$ into a vertical subspace $V_\varphi$ spanned by $\{\xi_\varphi^\#, \xi\in\fG\}$ and its horizontal complement $H_\varphi$ defined as the kernel at $\varphi$ of a functional connection $\varpi$. With dotted lines, we represent a different choice of horizontal complement associated to a different choice of $\varpi$.}
			\label{fig2}
		\end{center}
	\end{figure}
	
	In full, gory, detailed components, we have e.g.
	\be
	\varpi = \varpi^a(x)\tau_a 
	=  \int_M \d y \Big( \stackrel{A}{\varpi}\!{}^{a}{}^{\mu}_b(x,y)\dd A^b_\mu(y) +  \stackrel{\psi}{\varpi}\!{}^a{}_{\alpha m}(x,y)\dd \psi^{\alpha m}(y) \Big)\tau_a,
	\label{eq_varpiexplicit}
	\ee
	where in the first equality we have made explicit the fact that $\varpi$ is valued in $\fG$, while in the second we made explicit the differential-form structure on $\FYM$ as well. From \eqref{eq_varpiexplicit}, it is clear why in the following we will work as much as possible in an abstract DeWitt-like notation to keep formulas compact. However, it is important to keep the `multi-layered' structure of $\varpi$  well in mind: once again, {\it $\varpi$ is a  field-space 1-form which takes values in $\fG$}. 
	Moreover, it is crucial to remember that $\varpi$ {\it need not be spacetime local}. How locality is violated will be clarified later on, when we provide explicit examples of $\varpi$'s. The geometric situation is depicted in figure \ref{fig2}.
	
	The gauge character of the vertical subspaces in $\FYM$ requires compatibility of $\varpi$ with gauge transformations. 
	In other words,  in order for $\varpi$ to define an actual connection, it must satisfy the following two fundamental equations, which reflect the facts that  ({\it i}\,) $\varpi$ defines a vertical projector---and hence a horizontal complement to the fibers via its kernel%
	\footnote{In finite dimensions, it is easy to see that property \eqref{varpi_complement} implies that ${\rm ker}\ \varpi$ gives a direct-sum complement to $V$ within ${\rm T}P$ (here $P$ is a finite dimensional PFB and $p\in P$ are its points). That is, as a linear operator $\varpi_p:T_pP\rightarrow \mathfrak{g}$, $\varpi_p$ takes an $m+n$ dimensional space  to an $n$-dimensional one, where $n={\rm dim}(\mathfrak{g})={\rm dim}(V_p)$. Therefore ${\rm dim}({\rm ker}\ \varpi)=m$ and $({\rm ker}\ \varpi)\cap V=0$. In infinite-dimensions the story is more complicated (one must use the Fredholm alternative), but can similarly be resolved under certain conditions, see e.g. \cite{fischermarsden, Ebin, Lang, Palais}. }%
	---and that ({\it ii}\,) $\varpi$ is equivariant, i.e. it transforms `nicely' along the gauge directions. In formulas, for field-independent gauge transformations,
	\begin{subequations}
		\begin{align}
		\fI_{\xi^\#}\varpi & = \xi \label{varpi_complement}\\
		R_{g}^*\varpi &= \Ad_{g^{-1}}\varpi \qquad (\dd \xi =0) \label{varpi_independent}
		\end{align}
	\end{subequations}
	where on the right-hand side of \eqref{varpi_independent} $g(x)$ acts on the $\tau_\alpha$ of formula \eqref{eq_varpiexplicit}.  
	Therefore, the equivariance condition intertwines the action of $\G$ on $\FYM$, the manifold on which $\varpi$ lives, and the action of $\G$ on the `internal' indices of $\varpi$. 
	These are the fundamental equations of $\varpi$, from which all its properties descend.
	
	The infinitesimal version of \eqref{varpi_independent} for field-independent $\xi$ is $\fLie_{\xi^\#}\varpi = [\varpi, \xi ]$. 
	An  equivalent equation can be obtained through Cartan's formula \eqref{eq_magic} and equation \eqref{varpi_complement} with $\dd \xi = 0$:
	\be
	\fI_{\xi^\#} \dd \varpi = [\varpi ,\xi]  \qquad (\dd \xi =0).
	\label{eq_BRST}
	\ee
	The advantage of this equation is that it is pointwise (in field-space) linear in $\xi$, and therefore it must hold for field-dependent $\xi$'s as well. 
	Therefore, using again Cartan's formula and the first connection property \cite{Gomes:2016mwl}, we get 
	\begin{subequations}
		\begin{empheq}[box=\fbox]{alignat=1}
		\phantom{A}\notag\\
		\fI_{\xi^\#}\varpi & = \xi  \label{varpi_compl}\\
		\qquad \fLie_{\xi^\#}\varpi &= [\varpi,\xi] +\dd \xi \qquad \label{varpi_dependent}\\
		\notag
		\end{empheq}\label{eq_fundamental}
	\end{subequations}
	
	The finite version of the latter equation generalizes equation  \eqref{varpi_independent} to field-dependent gauge transformations:
	\be
	R_{g}^*\varpi = \Ad_{g^{-1}}\varpi + g^{-1}\dd g 
	\label{varpi_dependent_finite}
	\ee
	
	Note that equation \eqref{eq_BRST}  can also be written directly as 
	\be
	\fI_{\xi^\#} \Big( \dd \varpi + \tfrac12 [ \varpi, \varpi] \Big) = 0
	\ee
	which is equivalent to saying that  the combination
	\be
	\boxed{\quad\phantom{\Big|}
		\bb F :=  \dd \varpi + \tfrac12 [ \varpi, \varpi]
		\quad}
	\ee
	is purely horizontal. Indeed, as in the finite-dimensional PFB framework, this expression defines \textit{the curvature} of $\varpi$, which, in more invariant terms, would be otherwise defined as  the `horizontal derivative' of $\varpi$, $\bb F = \dd_H \varpi$ (see \cite{kobayashivol1}), which we will introduce shortly.
	
	As a last remark of this section, we note that equation \eqref{eq_BRST} can be written in the alternative form
	\be
	\dd_V \varpi =  -\tfrac12 [\varpi,\varpi],
	\ee 
	where $\dd_V$ indicates that the expression is pulled to the fiber, and is the \lq{}vertical\rq{} complement of the horizontal derivative, which we now introduce. This equation
	was interpreted by two of us in \cite{Gomes:2016mwl} as a generalization of the geometric BRST framework \cite{Bonora1983, thierry1985classical, HenneauxTeitelboim} (in analogy with $\dd_H$, $\dd_V$ is called the `vertical derivative').

	\subsection{Horizontal differentials in $\FYM$}

	Summarizing the previous section, what it means for a vector field in field-space to be vertical is defined intrinsically once the gauge transformation properties of the fields are given. 
	They are by definition tangent to the gauge orbits. 
	The decomposition of an arbitrary vector into its vertical (or gauge) and horizontal parts however is not canonical; it requires an extra ingredient---the connection-form $\varpi$. 
	
	Dually, purely horizontal forms---forms whose contraction with any vertical vector field vanishes---are intrinsically defined. 
	Splits of a generic form into a horizontal and  a vertical part however is not intrinsic, but requires the introduction of $\varpi$.  In formulas, given a field-space 1-form $\alpha\in\Lambda^1(\F)$, its vertical projection is $\hat V\alpha = \fI_{\varpi^\#}\alpha$.
	
	Similarly, the horizontal projection of an exterior derivative (in field-space) is called a horizontal differential, and it is denoted\footnotemark~$\dd_H$.%
	\footnotetext{The reader should not confuse this horizontal differential---which is associated to the {\it gauge} structure of the theory and is in step with the standard finite-dimensional nomenclature---with the notion of horizontal differential appearing in the `variational bi-complex' formalism \cite{anderson}---which is associated to spacetime rather than field-space. Although, in principle, we could have adopted the variational bi-complex formalism to deal with spacetime locality from a field-space perspective, in order to avoid an extra layer of formalism, that route was avoided.}
	Its complement, the vertical differential, is denoted $\dd_V$, and hence $\dd = \dd_H + \dd_V$. 
	Since horizontal planes identify points on neighboring orbits, they define a notion of a parallel transport and the horizontal derivative is nothing but the PFB generalization of the covariant derivative.

	On field-space scalars, the horizontal differential associated to $\varpi$ is\footnote{Note again the similarity with a BRST transformation, where $\delta_\varpi\varphi$ is a formal gauge transformation involving an anticommuting gauge parameter.}
	\be
	\dd_H \varphi := \dd \varphi - \delta_\varpi \varphi,
	\label{horizontal_def}
	\ee
	where $\delta_\xi$ for $\xi\in \fG$ is the standard notation for infinitesimal gauge-trans\-for\-ma\-tions introduced in eq. \eqref{eq_infin_gtransf}. The derivative is  `horizontal' in the sense that for every vertical (i.e. pure gauge) vector field  $\xi^\#$ on $\FYM$
	\be
	\fI_{\xi^\#} \dd_H \varphi  = \delta_\xi \varphi - \delta_{\fI_{\xi^\#}\varpi} \varphi \equiv 0,
	\ee
	thanks to the fundamental property in equation \eqref{varpi_complement}.
	More explicitly,
	\begin{align}
	\boxed{\quad\phantom{\Big|}
		\dd_H A  = \dd A - \D\varpi,
		\quad
		\dd_H \Psi = \dd \Psi + \varpi\Psi
		\quad\text{for which}\quad
		\fI_{\xi^\#} \dd_H A  = 0 = \fI_{\xi^\#} \dd_H \Psi.
		\label{eq_contr=zero}
		\quad}
	\end{align}
	From this,  we now verify (via the fundamental equations for $\varpi$, Cartan's formula, and the identity $\dd^2=0$) that the following covariance properties hold:\footnote{This is analogous to {the following property of the gauge-covariant derivative on spacetime}: $\delta_\xi (\D \Psi ) = -  \xi \D \Psi $  (in contrast to $\delta_\xi (\pp\Psi) \neq - \xi (\pp\Psi)$). }
	\be
	\fLie_{\xi^\#} \dd_H  A = - [\xi, \dd_H A] 
	\qquad\text{and}\qquad
	\fLie_{\xi^\#}\dd_H  \Psi = -\xi \dd_H \Psi. 
	\label{covariance_props}
	\ee
	This can be summarized by saying that $\dd_H\varphi$ are {\it equivariant} 1-forms transforming in the adjoint and fundamental representation respectively.
	Let us first prove this formula for the matter field $\Psi$:
	\be
	\fLie_{\xi^\#} \dd_H \Psi = \fI_{\xi^\#} \dd (\dd \Psi + \varpi \Psi) = \fI_{\xi^\#} (\dd \varpi \Psi - \varpi \dd \Psi) = [\varpi, \xi] \Psi - \xi \dd\Psi + \varpi (-\xi\Psi) = -\xi \dd_H\Psi,
	\ee
	where in the first step we have used  $\dd^2=0$, Cartan's formula \eqref{eq_magic} and equation \eqref{eq_contr=zero}, in the second we have distributed $\dd$, paying attention to the anticommuting properties of the field-space forms, in the third we have used equations \eqref{eq_BRST} and \eqref{eq_infin_gtransf}, again paying attention to the anticommuting properties of forms and interior products, and finally in the fourth we have simply recollected the terms. The calculation for $A$ is similar, and requires also the use of Jacobi identities in $\fg$ as well as the commutation property between $\dd$ and $\d$ (equation \eqref{eq_ddcomm}):
	\begin{align}
	\fLie_{\xi^\#} \dd_H A 
	&= \fI_{\xi^\#} \dd (\dd A - \D \varpi ) = -\fI_{\xi^\#} \dd (\d \varpi +[A, \varpi]) \\
	&= - \d [\varpi,\xi] -\fI_{\xi^\#} [\dd A, \varpi] - \fI_{\xi^\#} [ A, \dd \varpi] \notag\\
	&= \d [\xi,\varpi] - [\d \xi,\varpi] - [[A,\xi],\varpi] + [\dd A ,\xi] - [A,[\varpi,\xi]] \notag\\
	& = [\xi, -\dd A  + \d \varpi + [A,\varpi]] = -[\xi,\dd_H A].\notag
	\end{align}
	
	The horizontal exterior derivative can be extended from field-space scalars to forms as follows: For {\it horizontal equivariant} forms $\lambda\in\Lambda^\bullet(\F,W_\rho)$ transforming in the representation $\rho$ of $\G$, i.e. for field-space forms such that\footnote{We denote $\rho$ also the infinitesimal version of the representation. }
	\be \label{eq:def_equi}
	\fI_{\xi^\#} \lambda = 0 \qquad \text{and}\qquad \fLie_{\xi^\#} \lambda = \rho(\xi)\lambda,
	\ee
	for all $\xi\in\fG$,
	a simple formula analogous to \eqref{horizontal_def} holds:
	\be
	\dd_H \lambda = \dd \lambda - \rho(\varpi)\lambda.
	\label{horizontalequivariant}
	\ee
	In this formula, as an argument of $\rho$, $\varpi$ is seen as a (field-space-)form-valued element of $\fG$---we left implicit the non-commutative differential-form characters of $\varpi$ and $\lambda$. 
	It is then easily checked that $\dd_H\lambda$ is again horizontal and equivariant. Note that, in particular, $\Psi$ and $\dd_H\Psi$ are horizontal and equivariant 0 and 1-forms, respectively. As usual, the case of the horizontal differential of the connection-form, $\varpi$, which defines $\fF$, has to be analyzed separately, e.g. along the lines of the previous section.
	For further details see \cite{Gomes:2016mwl, kobayashivol1}.
	
	Finally, let us notice the general formula
	\be
	\dd_H^2\varphi = -\delta_{\bb F} \varphi,
	\label{eq_dHdH=F}
	\ee
	that is
	\be
	\boxed{\quad\phantom{\Big|}
		\dd_H^2 \Psi = \fF \Psi
		\qquad\text{and}\qquad
		\dd_H^2 A = - \D \fF.
		\label{eq_dHdH=F2}
		\quad}
	\ee
	This  standard identity is most simply proven on a case by case basis: 
	\begin{subequations}
		\begin{align}
		\dd_H^2 \Psi &= \dd (\dd_H \Psi) + \varpi (\dd_H\Psi) = \dd (\varpi \Psi) + \varpi \dd \Psi + \varpi \varpi \Psi = \fF \Psi\\
		\dd_H^2 A & = \dd (\dd_H A) + [\varpi,\dd_H A] = \dd (-\d \varpi - [A, \varpi] ) + [\varpi, (\dd A - \d \varpi - [A,\varpi])] = - D\fF,
		\end{align}
	\end{subequations} 
In the first line we used $\dd^2 = 0$, and equations \eqref{eq:def_equi} and \eqref{horizontalequivariant} together with the fact that for all $\xi$, $\fLie_{\xi^\#}\dd_H\psi = \rho(\xi) \dd_H\psi = - \xi \dd_H \psi$. Further, we used also that $\varpi\varpi \equiv \tfrac12[\varpi,\varpi]$ and the fact that, when `going through' $\varpi$, $\dd$ takes a minus sign. In the second line,
we used  $\fLie_{\xi^\#}\dd_H A = [\dd_H A,\xi]$ (again a rewriting of \eqref{covariance_props}), and then the identities $[\varpi, \d \varpi] = \tfrac12 \d[\varpi,\varpi]$, and $[\dd A ,\varpi] = [\varpi, \dd A]$ where two anticommutation rules intervene to give a global plus sign, as well as the (graded) Jacobi identity%
	\footnote{To prove it, write $\varpi = \varpi_I \dd \varphi^I$, and use the ordinary Jacobi identity for the Lie-algebra valued field-space scalars $\varpi_I$: 
		$$
		0= \Big( [A, [\varpi_I,\varpi_J]] + [ \varpi_I,[ \varpi_J, A]] + [\varpi_J, [A,\varpi_I]] \Big) \dd\varphi^I\dd\varphi^J = [A, [\varpi, \varpi]] + [\varpi, [\varpi, A] - [\varpi, [A,\varpi]  ,
		$$  where for the last term we used $\dd \varphi^I\dd \varphi^J = - \dd \varphi^J \dd \varphi^I$. The last expression is precisely equation \eqref{eq_jacobi2}.
	}
	\be
	[A,[\varpi,\varpi]] + 2 [\varpi, [\varpi, A]] = 0.
	\label{eq_jacobi2}
	\ee

	In section \ref{sec:Noether_hor}, we will see how the horizontal differential $\dd_H$ can be used to introduce a completely gauge-invariant (pre)symplectic geometry on $\FYM$ \cite{Gomes:2016mwl}.

	\subsection{Remarks on section \ref{sec:general_YM}\label{rmks:general_YM}}
	
	\paragraph*{\indent(i) $\varpi$ \textsc{vs.} new degrees of freedom ---}
	First of all,  as remarked in the introduction,  $\varpi$ is {\it not} required to involve new fields.
	It is a one-form living on field-space $\F$. As such its `value' depends on the underlying fields (possibly in a non-local manner), which is  why we may call it \lq{}relational\rq{}; it provides a notion of gauge vs. horizontal splittings \textit{relative} to the underlying fields. 
	This point will become clearer in the next sections.
	
	One might ask whether it is possible at all to build such a $\varpi$ satisfying the fundamental equations \eqref{eq_fundamental} without extending the field-space, and whether this can be done explicitly. 
	We shall answer both these questions affirmatively in the next part of this paper, where we provide various examples of $\varpi$'s built solely out of fields in $\FYM$.

	However, it is certainly true that by extending the field-space via the inclusion of new group-valued degrees of freedom $H\in C^\infty(M,G)$ transforming as $H\mapsto Hg$, a viable $\varpi$ is readily defined by $\varpi=H^{-1} \dd H$ (see the discussion of the `co-rotation principle' in the introduction to section \ref{sec:Singerconnection}). This was indeed the solution implicitly e.g. adopted in \cite{Donnelly:2016auv} (cf. earlier examples \cite{Regge:1974zd, Balachandran:1994up}). However, this not only does not abide to the general philosophical principle of relationalism on which the present approach is based, but it is also unnecessary: the same formal result will be obtained in section \ref{sec:matter} by picking a specific $\varpi$ built out of the matter fields $\Psi$.

	\paragraph*{\indent(ii) Is $\varpi$ unique? ---}
	It is important to note that, although the principal fiber bundle structure invites the introduction of a connection-form, it does {\it not} determine it uniquely. 
	Therefore, the choice of a specific $\varpi$ does require extra input, ultimately equivalent to the choice of a specific horizontal complement to the vertical spaces. 
	In the next section we will argue that a mathematically quite natural---albeit noncanonical---way to determine the horizontal complements exists, and will introduce our $\varpi$'s accordingly.
	In sections \ref{sec:matter} and \ref{sec:dressing}, we will also discuss how the main connections obtained in this way can be interpreted in a physical manner, in terms of  dressings and choices of specific material reference frames.

	\section{Connection-forms and their curvatures from supermetrics}\label{sec:dw_connection}

	In this section, we introduce metrics on the field-space $\F$, or supermetrics, as a means of selecting connection-forms. 
	In short, field-space metrics which are appropriately gauge-compatible can be used to derive a field-space connection $\varpi$ by a demand of orthogonality: as we saw in section \ref{sec:general_YM}, the role of $\varpi$ is to determine a projector onto the vertical subspace $V_\varphi = \mathrm T_\varphi \mathcal O_\varphi\subset \mathrm T_\varphi\F$, and in the presence of a  field-space metric this can be done by orthogonal projection. If the field-space metric is appropriately gauge compatible, the vertical projector indeed transforms as a connection.
	In the following we will give a formal proof of the above statements, and provide an explicit link between the properties of the metric and the curvature of the associated connection.
	
	{\aldo 
	For the decomposition of  $\mathrm T_\varphi \mathcal O_\varphi\subset \mathrm T_\varphi\F$ into horizontal and vertical subspaces to be well-defined, i.e. $V_\varphi\cap H_\varphi = \emptyset$, the supermetric must be {\florian positive definite}. 
	In section \ref{sec:orthogonal}, we will assume that this is the case. However, it will soon be clear that natural choices of supermetrics do not satisfy this hypothesis if {\it spacetime} is equipped with a Lorentzian metric (cf. footnote \ref{footnt:strata}). To circumvent this issue, we will work in spacetime regions $M$ of Euclidean signature or, alternatively, we can work in a ``3+1'' setting where field-space is the space of field configuration on a Cauchy hypersurface $\Sigma$. See also point (\textit{iii}) of \hyperref[rmks:7]{\it Remarks to Section \ref{sec:Singerconnection}} and section \ref{sec:Noether_hor}. 
	From the results of section \ref{sec:Singerconnection}, it will also be clear that the restriction to spaces of Euclidean signature is necessary to obtain a well-posed boundary-value problem for the PDE defining the field-space connection form $\varpi$ from a supermetric.
	}
	
	{\aldo Finally,} in the following sections, we ignore the possibility of  configurations with global symmetries, i.e. with non-trivial stabilizers. This is a simplifying assumption for the time being. All the remaining configurations---those with trivial stabilizers---form a dense subset of the full field-space. Nonetheless, the reducible configurations carry important physical baggage. We will come back to this point soon.

	\subsection{Supermetrics and the functional connection-form}\label{sec:orthogonal}

	A metric $\bb G$ on field-space $\F$, or {\it supermetric}, contracts two field-space tangent vectors  at the same field configuration $\varphi$, e.g. $\bb X, \bb Y \in {\rm T}_\varphi \F$, to return a number. 
	We will consider field-space metrics of the form
	\begin{align}
	\bb G(\bb X, \bb Y) = \int_M \d^d x \, \bb G_{IJ}\big(\varphi(x), x\big)\, \bb X^I\big(\varphi(x), x\big) \bb Y^J\big(\varphi(x), x\big),
	\label{eq_supermetricGG}
	\end{align}
	where, as before, $\bb X = \int_M \d^d x \bb X^I(\varphi; x) \frac{\dd}{\dd \varphi^I(x)}$ indicates a vector.\footnote{We commit a slight abuse of notation: the same notation is used for vectors and vector fields. The distinction should be clear from the context.}
	The domain of integration $M$ is our region of interest, and at this stage can be a Cauchy surface or a region of space(time), and may or may not have boundaries. 
	
	The expression above assumes that $\bb G_{IJ}$ is not only local in  field-space, as any supermetric must be, but also local in spacetime. 
	In fact, we will require more and demand $\bb G_{IJ}$ to be {\it ultralocal} in spacetime, that is we demand that $\bb G(\bb X , \bb Y)$ does not involve any spacetime derivatives of the components $\bb X^I(x)$ (nor of $\varphi$). 
	In  the cases relevant for this article a further simplification is possible: the $\bb G_{IJ}$ can be taken constant throughout $\F$.
	In such cases, we would deal with an ultralocal field-independent supermetric. These types of metrics, if required to be also non-degenerate in field-space, are essentially unique \cite{DeWitt_Book}. We will provide explicit examples of such metrics in the following sections. 
	
	We note that a crucial example of field-dependence  is given in background-independent theories,  such as general relativity, where $\bb G$ depends on the space(time) metric. {In general relativity, $\bb G$ is (one of) the DeWitt supermetrics.}\footnote{\henrique It is only in general relativity (i.e. for spin-2 fields), that there is 1-parameter family of such supermetrics. For a subset of these choices, the emerging field-space supermetric is not positive-definite. \aldo See  the discussion in the introduction to this section.\label{fn:footnt19}} 
	
	Now, from a purely geometrical perspective, demanding  field-independence of the components of $\bb G$ is not a well-defined requirement, since it is a coordinate dependent statement {\aldo (in field space)}. A better demand is that the field-space metric be compatible with the gauge symmetry structure of the theory---which can encompass diffeomorphisms and hence background independence,---i.e. that $\bb G$ be constant along the gauge orbits. { More precisely, demand that the fundamental vector fields $\xi^\#$ be Killing, that is
		\begin{align}
		\fLie_{\xi^\#} \bb G = 0
		\qquad \text{for all} \;
		\xi \in \text{Lie}(\G) \text{\;and\;} \dd \xi = 0
		\label{eq_GG_Killing}
		\end{align}}
	(in fact, a slightly weaker version of this is sufficient for our purposes, see below).
	
	In the following, we will prove that the above requirement allows the construction of a connection via the orthogonality condition sketched in the introduction to this section. 
	Before delving into the proof, let us observe that while a field-space metric determines a connection, the converse is not true: the field-space metric also contains information about the inner product of two vertical vectors and of two horizontal vectors, which is not contained in the connection.
	
	In equations, for a field-space metric $\bb G$, we define the associated connection $\varpi$ through the demand that\footnote{The following equation holds pointwise on $\Phi$, where the vector field $\bb X$ identifies a tangent vector $\bb X_\varphi \in {\rm T}_\varphi \F$. In the main text we have omitted the subscripts.}  for all $\xi \in \text{Lie}(\G)$ and all $\bb X \in \mathfrak{X}^1(\F)$,
	\begin{equation}
	\bb G(\xi^\sharp, \hat H(\bb X)) \equiv \bb G(\xi^\sharp, \bb X - \fI_{\bb X} \varpi^\#) = 0,
	\label{eq4.3}
	\end{equation}
	where as before $\hat H$ stands for the horizontal projection induced by $\bb G$ itself. Formally, this can be solved for $\varpi$ as follows.
	Let $\bb Q_{ab}$ be the pullback to $\fG$ under $\cdot^\#$ of the metric induced from $\bb G$ on the fibers, as expressed in the $\{\tau_a\}_a$ basis:
	\be
	\bb Q_{ab}=\bb G(\tau_a^\#, \tau_b^\#),
	\label{eq_Qab}
	\ee
	and $\bb Q^{ab}$ its inverse.
	Note that we are here committing a slight abuse of notation with respect to the previous sections, since we are assuming that the index $a$ runs not only over a basis of $\fg$, but also over {\henrique (Euclidean)} spacetime points.\footnotemark~In other words, we are assuming here that $\{\tau_a\}$ is the basis of $\fG$ obtained by a point-wise extension of a basis of $\fg$.
	\footnotetext{Hence, the generator $\tau_a$ really stands for $\tau_{a,y}(x)=\delta(x,y)\tau_a  \in \fG$, and has to be contracted with $\xi^a$ which really stands for $\xi^{a,y}=\xi^a(y)$. E.g. in pure YM without matter, $\tau_a^\#$ stands for $\tau_{a,y}^\# =\int \d x \Big(\delta'_{\mu}(x,y)\delta^b_a + \delta(x,y)f_{ca}{}^b A(x)^c_\mu\Big)\frac{\dd}{\dd A(x)^b_{\mu}}$, where $\delta'_{\mu}(x,y)$ is the distribution $\frac{\pp}{\pp x^\mu}\delta(x,y)$. In this way, $\xi^\# = \tau^\#_a \xi^a$ is obtained as $\xi^\# = \tau^\#_{a,y}\xi^{a,y} = \sum_a \int \d y \, \tau^\#_{a,y} \xi^a(y) = \int \d x \, (\D_\mu \xi)^a(x)  \frac{\dd}{\dd A^b_\mu(x)}$.\label{ftnt:tauhash}}

	A couple of comments are in order. 
	First, $\bb Q_{ab}$ does not in general coincide with the (point-wise extensions of the) Killing form in $\fg$.
	And second, $\bb Q_{ab}$ need not be ultralocal, since the lift $\tau_a \mapsto \tau_a^\#$ may contain spacetime derivatives, e.g. when $\varphi$ is the gauge potential. In these cases, $\bb Q_{ab}$ is rather (the spacetime integral of) a bilinear differential operator. 
	If $\bb Q_{ab}$ fails to be ultralocal, the inversion procedure defining $\bb Q^{ab}$ has to be understood in the sense of Green's functions and might be subtle. {\henrique Lastly, {\aldo as explained in the introduction to this section and in footnote \ref{fn:footnt19},} $\bb Q_{ab}$ could have null directions, which would enlarge the kernel of $\varpi$ and have implications for conserved charges.}
	We will have more to say about these items in the study of specific cases performed in the following sections. For the moment, we will keep working formally.

	Now, expanding $\varpi = \varpi^a \tau_a$, equation \ref{eq4.3} can be written as $\bb G(\tau_a^\#, \bb X) = \bb Q_{ab} \fI_{\bb X} \varpi^b$, which is readily inverted as%
	\footnote{ A more precise version of this formula is the following. First  introduce the notation  $\alpha(\cdot):=\cdot^\#$, and denote (in this footnote) the adjoint of $\alpha$ with respect to the ultralocal inner product $\bb G$ by $\alpha^\dagger:{\rm T}\F\rightarrow\fG$. Then,  equation \eqref{eq:varpi_abstract_solution} reads $\varpi=(\alpha^\dagger\circ\alpha)^{-1}\circ\alpha^\dagger$. In the appropriate cases (see the next section), this formula emphasizes the fact that $\alpha^\dagger\circ\alpha$ is an elliptic differential operator provided the supermetric satisfies a positivity property. This abstract expression is more difficult to manipulate than \eqref{eq:varpi_abstract_solution}. See \cite{Singer:1978dk, Singer:1981xw}, and also \cite{gomes_riem}  for the same formula in a different context. } 
	\begin{align}\label{eq:varpi_abstract_solution}
	\varpi = \bb Q^{ab} \bb G (\tau_b^\#, \cdot) \tau_a.
	\end{align}
	Note that $\bb G(\xi^\#, \cdot)$ accepts field-space vectors and hence defines a one-form in field-space. 
	From the last equation, we immediately obtain the first fundamental property, $\fI_{\xi^\#} \varpi = \xi$, equation \eqref{varpi_complement}.
	
	As defined here above, $\varpi$ satisfies the projection property---equation \eqref{varpi_compl}---by construction.
	Let us now see what is required of the field-space metric $\bb G$ to ensure that $\varpi$ also correctly transforms under gauge transformations---equation \eqref{varpi_dependent}---and can therefore be called a connection.
	We claim that $\varpi$ is a connection if and only if
	\begin{align}\label{eq:preservation_of_orth}
	(\fLie_{\xi^\#} \bb G) \big(\eta^\#, \hat H(\bb X)\big) = 0
	\qquad (\dd \xi =0)
	\end{align}
	for all $\xi \in \fG$ with $\dd \xi = 0$, all $\eta\in\fG$ and all vector fields $\bb X$. 
	The notation means that the Lie derivative acts only on the metric components; or, in other words, that one first takes the Lie derivative of the metric, hence obtaining a bilinear operator which is then used to contract $\eta^\#$ and $\hat H(\bb X)$.
	In particular, $\eta^\#$ is a generic vertical vector, and $\hat H(\bb X)$ a generic horizontal vector, and therefore $\bb G(\eta^\#, \hat H(\bb X)) \equiv 0$. 
	The condition \eqref{eq:preservation_of_orth} can thus be read as the requirement that the notion of orthogonality to the fibers provided by the metric be preserved under vertical transport of the metric.  
	Note that this requirement is less stringent than requiring vertical invariance of the full field-space metric, $\fLie_{\xi^\#} \bb G = 0$. This is consistent with the fact that the connection knows only about the metric notion of orthogonality to the gauge orbits.

	To prove our claim, let us Lie derive equation \eqref{eq:varpi_abstract_solution} along a fundamental vector field $\xi^\#$ ($\dd \xi = 0$). Distributing the Lie derivative and using $\lbr\xi^\#, \tau_a^\#\rbr = [\xi, \tau_a]^\#$ yields
	\begin{align}
	\fLie_{\xi^\#} \varpi 
	&= \fLie_{\xi^\#} \big(\bb Q^{ab} \bb G(\tau_b^\#, \cdot) \tau_a \big) \\
	&= - \bb Q^{ac} \fLie_{\xi^\#} \big(\bb G(\tau_c^\#, \tau_d^\#) \big) \bb Q^{db} \bb G(\tau_b^\#, \cdot) \tau_a + \bb Q^{ab} (\fLie_{\xi^\#} \bb G) (\tau_b^\#, \cdot) \tau_a + \bb Q^{ab} \bb G (\lbr\xi^\#, \tau_b^\#\rbr, \cdot) \tau_a \notag\\
	& = \Big(- \bb Q^{ac} (\fLie_{\xi^\#} \bb G)(\tau_c^\#, \tau_d^\#) \bb Q^{db}  \bb G(\tau_b^\#, \cdot) - \bb Q^{ac} \bb G( [\xi, \tau_c]^\#, \tau_d^\#) \bb Q^{db} \bb  G(\tau_b^\#, \cdot)  \notag\\
	& \qquad -\bb Q^{ac} \bb G( \tau_c^\#, [\xi, \tau_d]^\#) \bb Q^{db} \bb  G(\tau_b^\#, \cdot) 
	+ \bb Q^{ab} (\fLie_{\xi^\#} \bb G) (\tau_b^\#, \cdot) + \bb Q^{ab} \bb G ([\xi, \tau_b]^\#, \cdot)\Big) \tau_a .\notag
	\end{align}
	In the first and second term we recognize $\varpi^\# = \tau_d^\# \bb Q^{db} \bb G(\tau_b^\#, \cdot)$.  For the third term, we use that $\bb Q^{ac} \bb G (\tau_c^\#, \eta^\#) = \eta^a$ by construction. Hence,
	\begin{align}
	\fLie_{\xi^\#} \varpi 
	& =  \Big( - \bb Q^{ac} ( \fLie_{\xi^\#} \bb G) (\tau_c^\#, \varpi^\#) - \bb Q^{ac} \bb G( [\xi, \tau_c]^\#, \varpi^\#) - [\xi, \tau_d]^a \bb Q^{db} \bb G(\tau_b^\#, \cdot) \notag\\
	& \qquad + \bb Q^{ab} ( \fLie_{\xi^\#}  \bb G) (\tau_b^\#, \cdot) + \bb Q^{ab} \bb G([\xi, \tau_b]^\#, \cdot) \Big)\tau_a \nonumber\\
	& =  \Big( \bb Q^{ab} ( \fLie_{\xi^\#} \bb G) (\tau_b^\#, \hat H) + \bb Q^{ab} \bb G([\xi, \tau_b]^\#, \hat H) - [\xi, \varpi]^a \Big) \tau_a.
	\end{align}
	For the second line, we have used $({\rm id} - \varpi^\# )= \hat H$. Moreover, the second term in the second line vanishes by construction, and thus we are left with
	\begin{align}
	\fLie_{\xi^\#} \varpi = \bb Q^{ab} ( \fLie_{\xi^\#} \bb G) (\tau_b^\#, \hat H)\tau_a  + [\varpi, \xi] \qquad (\dd \xi = 0).
	\end{align}
	In order for $\varpi$ to be a connection, as we saw in section \ref{sec:general_YM}, it is sufficient for it to satisfy the projection property \eqref{varpi_compl}, i.e. $\fI_{\xi^\#}\varpi = \xi$, together with the equivariance equation \eqref{varpi_independent}, i.e. $\fLie_{\xi^\#} \varpi = [\varpi, \xi]$ for $\dd \xi = 0$. Since the first projection property \eqref{varpi_compl} has already been established, we have thus shown our claim of equation \eqref{eq:preservation_of_orth}: if $(\fLie_{\xi^\#} \bb G)(\hat V, \hat H)= 0$ for all field-independent $\xi$, then $\varpi = \bb Q^{ab}\bb G(\tau_b^\#, \cdot)\tau_a$ is a connection.
	
	To summarize, a field-space metric determines a vertical projector by providing a notion of orthogonality. If gauge transformations preserve orthogonality to the fibers, then the vertical projector gives a connection.

	\subsection{Supermetrics and the curvature of the  functional connection-form\label{sec:curvature}}
	
	A natural question to ask is how the properties of a field-space connection are linked to the properties of the field-space metric that determines it. In particular, one may ask if the curvature of the field-space connection can be calculated directly from the field-space metric in a useful way. The answer is affirmative, as we will now show.
	
	The intuition is the following: $\varpi$ contains information about the horizontal planes, which are the planes orthogonal to the gauge orbits. If those planes can be integrated in the sense of Frobenius theorem to (infinite-dimensional) hypersurfaces, then $\varpi$ is flat. The curvature $\fF$ of $\varpi$ corresponds to the anholonomicity, or non-integrability, of the planes orthogonal to the gauge orbits. We want to obtain that curvature directly from the metric.\footnote{This problem has a finite dimensional analogue in general relativity: the computation of the gravitational Komar charge associated to an infinitesimal diffeomorphism $\eta$ on a codimension two submanifold $S$ of spacetime. If a vector field $\eta$ is tangential to that submanifold, the Komar charge contains precisely the anholonomicity $f$ of the planes orthogonal to that submanifold, schematically $Q_{\rm K} = \int_S g(f, \eta)$ (for details, see e.g. \cite{Donnelly:2016auv}). The Komar charge can be written directly in terms of the spacetime metric $g$ as $Q_{\rm K} = \int_S \ast \d g(\eta)$. The hodge dual picks out the components orthogonal to $S$. This is analogous to equation \eqref{eq:curvature_and_metric}.} 
	
	The resulting relationship between a field-space-metric $\bb G$ and the curvature $\fF$ of the associated $\varpi$ is: 
	\begin{equation}\label{eq:curvature_and_metric}
	\bb G \big((\fI_{\bb Y} \fI_{\bb X} \bb F)^\#, \xi^\#\big) = \fI_{\hat H (\bb Y)} \fI_{\hat H {(\bb X)}} \big(\dd \bb G (\xi^\#)\big)
	\quad \text{for all} \;
	\xi \in \text{Lie}(\G), ~\dd \xi = 0,
	\end{equation}
	and any $\bb X, \bb Y \in \mathfrak{X}^1(\Phi)$.
	On the right hand side, $\bb G(\xi^\#)\equiv\bb G(\xi^\#,\cdot)$ is a one-form on field-space, so $\dd \bb G (\xi^\#)$ is a two-form. By horizontally projecting the dummy vector fields $\bb X, \bb Y$ on the right hand side, we are taking the horizontal-horizontal part of that two-form. Formally solving for $\bb F$, we get
	\begin{align}
	\bb F = \bb Q^{ab} \big(\dd \bb G(\tau_b^\#) \big)_{HH} \tau_a.\label{eq:curv_formula}
	\end{align}
	Note that in these formulas $\dd$ acts on the one-form $\bb G(\xi^\#)$, and---even if $\xi$ is here taken to be field-independent, $\dd \xi=0$,---the operator $\cdot^\#$ generically introduces field-dependence.
	
	To prove the relation \eqref{eq:curvature_and_metric}, we start from its right hand side.
	We have $\fI_{\hat H (\bb X)} \bb G(\xi^\#) = \bb G(\hat H (\bb X), \xi^\#) = 0$ by construction, and, using the fundamental relations of Cartan's calculus---$\fLie_{\bb X} = \dd\,\fI_{\bb X} + \fI_{\bb X}\,\dd$ and $\fLie_{\bb X} \fI_{\bb Y}  = \fI_{\bb Y} \fLie_{\bb X} + \fI_{\lbr\bb X, \bb Y\rbr}$,---we get 
	\begin{align}\label{eq:curvature_and_metric_RHS}
	\fI_{\hat H (\bb Y)} \fI_{\hat H (\bb X)} (\dd \bb G(\xi^\#)) 
	& = \fI_{\hat H(\bb Y)} \big( \fLie_{\hat H (\bb X)} \bb G(\xi^\#) - \dd \fI_{\hat H (\bb X)} \bb G(\xi^\#) \big) \\
	& = \fLie_{\hat H (\bb X)} \big(\fI_{\hat H (\bb Y)} \bb G(\xi^\#) \big)  -\fI_{\lbr \hat H (\bb X), \hat H (\bb Y) \rbr} \bb G(\xi^\#)  \nonumber\\
	& = - \bb G(\lbr\hat H (\bb X), \hat H (\bb Y) \rbr  , \xi^\#) = - \bb G\big(\hat V(\lbr\hat H (\bb X), \hat H (\bb Y) \rbr), \xi^\#\big).\nonumber
	\end{align}
	For the last equality, we used that any vector field may be decomposed as $\bb Z = \hat V(\bb Z) + \hat H(\bb Z)$, and that $\bb G(\hat H (\bb Z), \xi^\#) = 0$. We thus have to show  the standard result that the vertical part of the commutator of two horizontal vector fields is minus the field strength. We have 
	\begin{align}
	\big(\fI_{{\bb Y}} \fI_{{\bb X}} \fF\big)^\# ={}& \big(\fI_{\hat H (\bb Y)} \fI_{\hat H (\bb X)} \fF\big)^\# 
	= \big(\fI_{\hat H (\bb Y)} \fI_{\hat H (\bb X)} \dd \varpi\big)^\#,
	\end{align}
	where we used that $\fF=\dd_H \varpi$. Using the latter again, and the commutation between $\fLie$ and $\fI$, we obtain through a computation analogous to \eqref{eq:curvature_and_metric_RHS} that
	\begin{align}
	\big(\fI_{{\bb Y}} \fI_{{\bb X}} \fF\big)^\# = - \big(\fI_{\lbr\hat H (\bb X), \hat H (\bb Y) \rbr} \varpi\big)^\# = - \hat V\big(\lbr\hat H (\bb X), \hat H (\bb Y) \rbr\big).
	\end{align}
	This and equation \eqref{eq:curvature_and_metric_RHS} prove the sought result, equation \eqref{eq:curvature_and_metric}.


	\subsection{Remarks on section \ref{sec:dw_connection}}
	
	\paragraph*{\indent(i) Supermetrics and connection forms ---}
	Virtually all the connection forms that we are aware of descend from a natural choice of field-space metric (supermetric) in the way described here.
	This way of introducing a connection in $\FYM$ is to the best of our knowledge due to Singer \cite{Singer:1978dk} and to Narasimhan and Ramadas \cite{Narasimhan:1979kf} (see also the early paper by Babelon and Viallet \cite{Babelon:1979wd, Babelon:1980uj}), although it was used implicitly by DeWitt more than a decade earlier \cite{DeWitt:1967ub} 
	There is another type of field-space connection which appeared in the literature, by the hand of Vilkovisky. We will have more to say about it in relation to dressings, in section \ref{sec:dressing}.
	
	\paragraph*{\indent(ii) Origin of $\varpi$'s curvature, $\fF$ --- }
	There are two ways in which the right hand side of  equation \eqref{eq:curvature_and_metric}, and hence $\bb F$, can be non-zero. Firstly, the field-space metric can be explicitly field-dependent. This is not the case for the Yang-Mills metric we will consider in the next section, but it is the case for general relativity, see section \ref{sec:EC}. Secondly, the Lie algebra can act in a field-space dependent way. This is not the case in Abelian gauge theories, but it is the case for non-Abelian ones: in Yang--Mills, $\delta_\xi A= \d \xi + [A,\xi]$ involves the gauge potential $A$, while in general relativity,  $\delta_\xi g = \pounds_\xi g$  involves the metric $g$, { where now $\xi\in\mathfrak{X}^1(M)$ and $\pounds_\xi$ is the Lie derivative along $\xi$.} 
	{Moreover, we note that in both these cases, $\xi^\#$ involves spacetime derivatives of $\xi$, and hence $\bb Q_{ab}$ will turn into a differential operator. 
		In the next section we will introduce some explicit examples of metric and connection forms which are of particular relevance. 
	}

	\section{Singer--DeWitt connection}\label{sec:Singerconnection}

	In this section, we explore specific connection-forms derived from the natural supermetric for Yang-Mills theories, both with and without spatial boundaries.
	We use the term {\it gauge supermetric} for ultralocal field-space metrics contracting fields whose gauge transformation involves a first derivative, such as the gauge vector potential, and the term {\it matter supermetric} for ultralocal field-space metrics contracting fields which transform in the fundamental representation of an internal gauge group, involving no derivatives. 
	Although this distinction is symmetry-group dependent---e.g. under diffeomorphisms all field transformations involve derivatives---here we make it primarily for gauge theories. The reasons for the distinction will become clear later. 
	
	In a theory which contains both gauge vectors and matter fields, such as Yang--Mills  $\FYM=\{(A,\Psi)\}$, we require that the metric $\bb G$ splits these sectors, i.e. that $\bb G$ has no gauge/matter mixed component. This means that we can find connection-forms for each sector separately.  
	Since a gauge-transformation acts equally on all the sectors---they \lq{}co-rotate\rq{} under gauge transformations---a connection-form can be fully determined by the action of the gauge group on a single field-sector.  We will refer to this idea as the {\it co-rotation principle}.  
	
	Because of this principle, it is justified to study the gauge and matter sectors independently.\footnote{ An analogous idea was put forward in the context of the Vilkovisky--DeWitt effective action in \cite{Rebhan1987}.} 
	We will call the connection-form derived from the gauge supermetric a {\it Singer--DeWitt connection} (SdW), and the one derived from the matter supermetric a {\it Higgs connection}, a nomenclature which will be duly justified. %
	
	Having said this, there will be a subtle caveat to this corotation principle--- there are points in $\FYM$ where only some of the fields may have a non-trivial  stabilizer---with interesting physical consequences.

	\subsection{SdW connection without boundaries}
	
	The prime example for an ultralocal field-space metric is the gauge supermetric for Yang-Mills theories.
	In the following, to emphasize the  neglect of matter fields, we introduce the notation $\F_{\rm pYM}$ to indicate the field-space of `pure' Yang--Mills theory. 
	{\henrique In this field-space, our constructions used a positive-definite supermetric. In the Lorentzian case, such an assumption is hard to substantiate. Therefore, we restrict our attention to two cases: (\textit{i}) $M$ is spacetime, but with Euclidean signature, or (\textit{ii}) $M=\Sigma$ represents a (portion) of a spacelike Cauchy hypersurface, in which case: spacetime admits a Lorentzian signature, field-space is understood to be the space of field configurations on $\Sigma$,  and $d$ is the dimension of $\Sigma$ rather than spacetime. For now,  our discussion encompasses both cases, although for definiteness we focus on the latter case. We explain and discuss these options, and their differences, in items (\textit{i-iii})  in the \hyperref[rmks:7]{\it Remarks to Section \ref{sec:Singerconnection}.}}
	

{\new 
	As a starting point before considering more complex situations, consider the case where $\Sigma$ is a spacelike {\it compact} Cauchy surface assumed for now to have no boundary and trivial de Rham cohomology. Here $g_{ij}$ is a {\it fixed} positive-definite metric on $\Sigma$ (it is a background structure, not part of field-space). Notice that $\bb G^{\rm g}$ is independent of $A$.
}

	The gauge supermetric contracts variations of the gauge field, $\bb X = \int \bb X \frac{\dd}{\dd A}\in \mathrm T_A \F_{\rm pYM}$, as in%
	\footnote{For $\bb G^{\rm g}$ to be dimensionless (in units of $\hbar$), it has to be multiplied by $e^{-2}$, where $e$ is the Yang--Mills coupling constant.}
	\begin{equation}\label{metric_YM}
	\bb G^{\rm g} (\bb X, \bb Y)= \int_\Sigma \d^dx\, \sqrt g g^{ij}  \delta_{ab} \bb X_i^a \bb Y_j^b ,
	\qquad 
	{\bb X}, \bb Y \in {\rm T}_A \F_{\rm pYM}.
	\end{equation}

	To illustrate the general features of the SdW connections, let us solve for the connection arising for the  spatial  gauge supermetric introduced above. We have
	\begin{align}
	0 ={}
	\bb G^{\rm g} (\xi^\sharp, \bb X - \fI_{\bb X} \varpi^{\ \#})
	={}& 
	\int \d^d x\,\sqrt g g^{ij} \delta_{ab} \D_i \xi^a\big(\bb X_j^b - \D_j (\fI_{\bb X} \varpi^{b})\big) \nonumber\\
	={}& - \int \d^d x \, \sqrt{g} \delta_{ab}\xi^a \big( \D^i \bb X_{i}^b - \D^i \D_i \fI_{\bb X} \varpi^b \big) ,
	\label{explicit_hor}
	\end{align}
	and, using the arbitrariness of $\xi^a(x)$ and $\bb X^a_i(x)$, we read off 
	\be
	\boxed{\quad\phantom{\Big|}
		\D^2\varpi =  \D^i \dd A_{i},
		\label{varpi_YM}
		\quad}
	\ee
	where $\D^2 := (\D^i \D_i)$ is the gauge-covariant Laplacian. The horizontal vector fields are the kernel of $\varpi$. Contraction with \eqref{varpi_YM} shows that, in this simple case, the horizontal vector fields are  those which are (covariant-)divergence free: $\D^i\bb X_i = 0$, since by definition $\fI_{\bb X} \dd A_i = \bb X_i$ for any vector $\bb X$.
	We see that SdW connections are generically of the form `inverse Laplacian of divergence'.

	In an Abelian pure Yang-Mills theory, $\D = \d$, and the above becomes a Poisson equation on a compact manifold  with trivial de Rham cohomology. It therefore has a unique solution (up to a constant).
	For non-Abelian theories the relevant Laplace operator is field-dependent and the defining equation for $\varpi$ becomes more involved.\footnote{ In \cite{Narasimhan:1979kf}, it is shown that for a space topology of $S^3$,  gauge group $SU(2)$, and appropriate analytic conditions on the Yang-Mills connection, the kernel of the Laplacian is the  stabilizer of $A$, i.e., Lie algebra elements $\xi$ with $D \xi = 0$. Most Yang-Mills connection have a trivial stabilizer, and the Laplacian is invertible on those. We will come back to the reducible configurations later.}
	
	To compute the curvature of $\varpi$, it is most convenient to use equation \eqref{eq:curvature_and_metric}, rather than trying to compute it directly from $\fF = \dd \varpi + \tfrac12 [\varpi,\varpi]$.
	Consider a field-space constant $\xi$, i.e. $\dd \xi = 0$. We have $\D_i \xi^a = \partial_i \xi^a+ f^a{}_{bc} A_i^b\xi^c$ with $f^a{}_{bc}$ the structure constants of $\fg$.
	Then
	\begin{align}
	\dd \bb G^{\rm g}(\xi^\#) ={}& \dd \int \d^d x \sqrt g g^{ij} \delta_{ab} \D_i \xi^a \dd A_j^b
	= - \int \d^d x \sqrt g g^{ij} \xi^c f_{abc} \dd A^a_i \dd A^b_j \qquad (\dd \xi = 0).
	\end{align}
	The horizontal projectors on the right hand side of equation \eqref{eq:curvature_and_metric} have the effect of replacing $\dd$ with $\dd_H$ in the last line. On its left hand side, after an integration by parts, we have
	\be
	\bb G^{\rm g}(\fF^\#, \xi^\#) = - \int \d^d x \sqrt g \xi^a \delta_{ab} \D^i \D_i \fF^b.
	\ee
	Hence, by equating the two and using the cyclicity of the structure constants (i.e. for a compact semisimple Lie algebra), as well as the arbitrariness of $\xi$, we obtain 
	\be
	\boxed{\quad\phantom{\Big|}
		\D^2 \fF =  g^{ij} [\dd_H A_i ,\dd_H A_j ],
		\quad}
	\label{Singercurvature}
	\ee
	or more explicitly, $\D^2 \fF^a =  f^a{}_{bc}  g^{ij} \dd_H A^b_i \dd_H A^c_j $.
	This result for the curvature of the Yang-Mills DeWitt connection was reported by Singer \cite{Singer:1978dk}, in a context where $\Sigma$ is an Euclidean spacetime without boundary (rather than a time slice).

	\subsection{SdW connection in presence of boundaries}

	Consider now the gauge supermetric of equation \eqref{metric_YM} when the region of interest $\Sigma$ has boundaries, $\pp\Sigma\neq\emptyset$. 
	Then, instead of \eqref{explicit_hor}, we obtain
	\begin{align}
	0 ={}&
	\bb G^{\rm g} (\xi^\sharp, \bb X - \fI_{\bb X} \varpi^{\#})\nonumber\\
	={}& - \int_\Sigma \d^d x \, \sqrt{g} \delta_{ab}\xi^a \big( \D^i \bb X_{i}^b - \D^i \D_i \fI_{\bb X} \varpi^b \big) + \int_{\pp \Sigma} \d^{d-1} x\,  \, \delta_{ab}\sqrt g\,  n^j\,   \xi^a \big(\bb X_j^b - \D_j (\fI_{\bb X} \varpi^b)\big),
	\label{explicit_bdy}
	\end{align} 
	where $n^j$ is the outgoing unit normal to the boundary.
	Thus, from the arbitrariness of $\xi^a(x)$, in the bulk {\it and at the boundary}, we find that the appropriate equations defining $\varpi$ are now (omitting Lie-algebra indices): 
	\be
	\boxed{\quad\phantom{\Big|}
		\D^2 \varpi = \D^i \dd A_i
		\quad\text{and}\quad
		n^i\D_i \varpi_{|\pp \Sigma} = n^i\dd A_i{}_{|\pp \Sigma}.
		\quad}
	\label{YM_varpi_boundaries}
	\ee
	In other words, the gauge-covariant Poisson equation for $\varpi$ comes automatically equipped with nonzero gauge-covariant Neumann boundary conditions. Of course, gauge-covariant Neumann boundary conditions are in reality Robin boundary conditions, $(n^i\pp_i \varpi + [n^iA_i,\varpi] - n^i\dd A_i)_{|\pp \Sigma} =0$ (recall that $A$ is fixed in this equation).
	In Abelian theories, this condition automatically guarantees the existence and uniqueness of $\varpi$ within a spacelike $\Sigma$, even when $\pp\Sigma\neq\emptyset$. 
	Moreover, at $A_i^a=0$, the equations are the same as in the Abelian case, modulo the fact that there are $\dim(\fg)$ of them.
	
	Being in the kernel of $\varpi$ as determined by equation \eqref{YM_varpi_boundaries}, the horizontal vectors $\bb X$ in ${\rm T}\F_{\rm pYM}$ (where all fields are understood to be restricted to live in $\Sigma$) satisfy
	\begin{equation}
	\bb X \in H = {\rm ker}\ \varpi \quad
	\text{if and only if~}
	\D^i\bb X^a_i = 0
	\quad\text{and}\quad
	n^i\bb X^a_i{}_{|\pp \Sigma} = 0,
	\end{equation}
	for all Lie algebra components $a$.
	This result could have been deduced directly from demanding $\bb G^{\rm g}(\bb X^h, \D\xi) = 0 $ for all $\xi\in\fG$, which is precisely the horizontality requirement.
	
	In principle, we could add to $\bb G^{\rm g}$ a boundary contribution, $\bb G^{\rm g}_\pp$ represented by an integral over $\pp\Sigma$. 
	Although, for the sake of simplicity, we will not follow this route, we discuss nevertheless what kind of boundary contributions to the supermetric one is allowed to add. We start by demanding that they are ultralocal and that they satisfy the following gluing principle. 
	Let $\bb G^{\rm g}_\Sigma$ be the sum of the bulk contribution \eqref{metric_YM} and of the associated $\bb G^{\rm g}_\pp$, then we demand
	\be
	\bb G^{\rm g}_{\Sigma_1} + \bb G^{\rm g}_{\Sigma_2} = \bb G^{\rm g}_{\Sigma_1\cup\Sigma_2}.
	\ee
	In particular this means that boundary contributions associated to a common boundary must cancel. 
	The natural way to achieve this is through the mismatch in the orientations of the relative boundary integrals.
	Hence, these must depend linearly on $n_i$, the normal to $\pp\Sigma$. The only possible covariant way to contract the indices of $\bb X$, $\bb Y$ and $n_i$ is through the introduction of a derivative, i.e.
	\be
	\bb G^{\rm g}_{\Sigma} (\bb X, \bb Y)= \int_\Sigma \d^d x \sqrt{g} g^{ij} \delta_{ab} \bb X^a_i \bb Y^b_j + \kappa \int_{\pp \Sigma} \d^{d-1}x\sqrt{h} g^{ii'}g^{jj'}  \nabla_{(i}n_{j)} \delta_{ab} \bb X_{i'}^a \bb Y_{j'}^b
	\ee
	where $h$ is the determinant of the induced metric on $\pp\Sigma$, and $\kappa$ is a constant.
	A brief analysis shows that $\kappa$ needs to be a {\it dimensionful} constant.\footnote{Coordinates are dimensionless, and $[g_{ij}]=2$, $[g^{ij}] = -2$, and $[\sqrt{g} ]= d$. Since $g_{ij} = n_i n_j + h_{ij}$, $[n_i]=1$. From the dimensionless-ness of coordinates it follows that $[\nabla_i]=0$ and $[\bb X_i^a] = [A_i^a] = 0$ (this is because $A^a_i\d x^i \tau_a$ is an infinitesimal element of $\fg$, which does not carry any dimensional factor). Thus, $d-2 = [\bb G_{\rm bulk}^{\rm DW}(\bb X,\bb Y)] =  [\bb G^{\rm DW}_\pp(\bb X,\bb Y)] = [\kappa] + (d-1) - 4 + 1$, from which $[\kappa] = 2$. }

	To avoid introducing a new dimensionful constant in the theory, the only available option in pure YM ($d\neq4$) is that $\kappa\propto e^{4/(d-4)}$. In any case, as we said, we avoid the introduction of these boundary terms in order to keep the boundary conditions on horizontal vector as simple as possible.

	\subsection{SdW connection and the composition of regions\label{sec:SdWcomposition}}

	With the SdW connection, we have constructed a relational  notion of horizontal change---i.e. physical change---with respect to the Yang--Mills gauge potential $A$. In this section, we will make the point that this relational notion depends on the choice of region under consideration,  and its geometry.

	Let us explain how this comes about.
	Consider a region $\Sigma=\Sigma_{I}\cup\Sigma_{II}$, with $\Sigma_{I, II}$ embedded manifolds sharing a portion of their boundary, $S=\pp \Sigma_{I}\cap\pp\Sigma_{II}  \neq \emptyset$.
	To each of these regions one can assign its own field-space, Lie-algebra of gauge symmetries, supermetric, and SdW connections. We will denote a restriction to one of the regions by the same subscripts, i.e.  we have a map $\cdot_I: \F\rightarrow \F_I$ and so on. 
	
	From this, one can define the vertical and horizontal projectors $\hat V = \varpi^\#$ and $\hat H = ({\rm id} - \varpi^\#)$, and similarly for $\hat V_{I, II}$ and $\hat H_{I, II}$. Each of these operators acts on field-space vectors intrinsic to either region (i.e. with support restricted to the relevant region, $\Sigma$ or $\Sigma_{I, II}$).
	More concretely, given a vector $\bb X$ supported on $\Sigma$, it can be decomposed into $\bb X = \bb X_{I} + \bb X_{II}$, where { $\bb X_{I, II}$ live respectively on $\Sigma_{I, II}$ understood as intrinsic manifolds with boundary}. Then, say $\hat H_{I}$, acts only on $\bb X_{I}$.
	Note that here $\bb X$ is considered at a fixed configuration $\varphi$ and as such we are only interested in its overall spacetime dependence.
	
	Crucially, although $\bb X = \bb X_{I} + \bb X_{II}$,  the restriction of the horizontal projection is not horizontal:
	\be
	(\hat H(\bb X) )_{I} \notin  H_{I} ,
	\label{violation1}
	\ee

	The reason for this discrepancy is that each relation above violates one of the two horizontality conditions. 
	Consider first equation \eqref{violation1}. Although the restriction $(\hat H(\bb X))_{I}$ of the horizontal vector $\hat H(\bb X)$ to $\Sigma_{I}$ is indeed divergence free in $\Sigma_{I}$, it will generally fail to satisfy the boundary condition $n_{12}\cdot (\hat H(\bb X))_{I}{}_{|S}=0$ at the interface $S=\Sigma_{I}\cap\Sigma_{II}$.
	We can summarize this fact as follows: upon the splitting of a region, a global horizontal vector does not always decompose into two purely horizontal vectors with respect to their regional SdW connections.
		A key word in the previous statements is `purely': although $(\hat H(\bb X))_{I}$ fails to be purely horizontal, generically it still has a non-trivial horizontal component.
	
	Let us show \ref{violation1} in more detail. For conciseness, let us denote in the rest of this subsection
		\begin{align}
		\xi:=  \fI_{\bb X} \varpi, \qquad \xi_I := \fI_{\bb X_I} \varpi_I, \qquad \xi_{II} := \fI_{\bb X_{II}} \varpi_{II}.
		\end{align}
	The vertical projection $\xi^\#_{I}$ within $\Sigma_{I}$ of $\tilde{\bb X}_{I} := (\hat H(\bb X))_{I}$, i.e. $\xi^\#_{I}=\hat V_{I}(\tilde{\bb X}_{I})$, is found by solving the following equations for $\xi_{I}\in(\fG_{I})$, from \eqref{YM_varpi_boundaries}:
	\be
	\D^2 \xi_I= \D_i \tilde{\bb X}_{I}^i \equiv 0 
	\quad \text{and}\quad
	n^i \D_i {\xi_I}_{|\pp\Sigma_I}= n^i \tilde{\bb X}_{Ii}{}_{|\pp\Sigma_I}  ,
	\label{eq513}
	\ee
	the first equation vanishes because the restriction of a horizontal vector is still divergence-free, and it is again a gauge-covariant generalization of a Poisson equation with Neumann boundary conditions.  { For $n^i \tilde{\bb X}_{Ii}{}_{|\pp\Sigma_I}\neq 0$,  the equation will have a non-trivial solution $\xi_I\neq0$.} This shows that 
	$\hat V_I(\tilde{\bb X}_{I}) \neq 0$, i.e. 
	\be
	\label{eq:hor_restriction}
	\tilde{\bb X}_{I}:=(\hat H(\bb X))_{I}\neq\hat H_I(\bb X_I)\neq 0.
	\ee
	
	In addition to the restriction of a horizontal vector not being horizontal, the sum of two horizontal vectors associated to $\Sigma_{I}$ and $\Sigma_{II}$ is not necessarily horizontal within  the whole $\Sigma$: There exist $\bb Y_{I}, \bb Z_{II}$ such that
	\be
	\hat H_{I} (\bb Y_{I}) + \hat H_{II} (\bb Z_{II}) \notin H.
	\label{violation2}
	\ee 
	The reason is that in order to ensure the correct boundary conditions at $S=\pp\Sigma_I\cap\pp\Sigma_{II}$ from within each of $\Sigma_I$ and $\Sigma_{II}$, the total vector field $\bb X = \hat H_I (\bb Y_I) + \hat H_{II} (\bb Z_{II})$ might fail to be divergence free at the surface.

	 We have seen that in general, restriction and projection do not commute. Let us look for a criterion for commutativity. For non-reducible configurations, assuming appropriate analyticity of the vector field $\bb X$, a sufficient and necessary such criterion for $\bb X$ is as follows:
	\begin{align}
		\xi = \xi_I + \xi_{II} \qquad \text{iff} \qquad (\xi_I)_{\vert S} = 	(\xi_{II})_{\vert S}.
	\end{align}
	 In words, the field-space connection for a given vector field coincides with the regional connections for the restricted vector fields if and only if those regional connections match at the shared boundary.
	The LHS implies $\hat V(\bb X) = \hat V_I (\bb X_{I}) + \hat V_{II}(\bb X_{II})$ and hence by completeness also $\hat H(\bb X) = \hat H_I (\bb X_{I}) + \hat H_{II}(\bb X_{II})$, and thus expresses the commutation of restriction and projection. The RHS is a strong condition on the vector field $\bb X$, and will not be satisfied by generic $\bb X$ (see appendix \ref{app:examples_gluing}).
		
	 Let us show the implication. By continuity of solutions, the LHS trivially implies the RHS.
	That the RHS implies the LHS can be seen as follows: The RHS means that $\xi_I + \xi_{II}=:\tilde \xi$ is continuous at $S$. Because $\xi_{I}$ and $\xi_{II}$ satisfy the same Neumann boundary conditions $n^i {D_i} \xi_I {}_{|S} = n^i \bb X_i {}_{|S} = n^i D_i \xi_{II}{}_{|S}$, we get that $\tilde \xi$ is also once continuously differentiable. Using also the Laplace equations for $\xi_{I, II}$, we get that $D^2 \xi_I {}_{|S}= D^2 \xi_{II}{}_{|S}$, which implies together with the previous points that  $\tilde\xi$ is twice continuously differentiable. From the uniqueness of solutions of the equations determining $\varpi(\bb X)$ { and $\varpi_{I, II} (\bb X_{I, II})$,}  it then follows that $\xi = \tilde \xi =\xi_I + \xi_{II}$ as claimed.

	In appendix \ref{app:examples_gluing}, we present two explicit examples illustrating the interplay between horizontality of field-space vectors in the SdW connection and the decomposition of space into regions.\footnote{It is important to reiterate that all of our examples are in the case where the cohomology of all the regions are trivial. For non-trivial cohomology, new \lq{}topological\rq{}  horizontal fields may arise. This is in line with work on the cohomological origin of certain types of charges. For an introduction, see \cite{GiuliniCharge}. } In particular, an explicit example will be provided where a projection of a horizontal field fails to be horizontal, and explain why charges will have the correct composition properties. 
	For simplicity, we will use  the case of electrodynamics. The treatment would go through almost  unaltered for Yang--Mills around the trivial configuration $A=0$. 
	For now, to summarize, we can state: {\it horizontal projections do not commute with restrictions}.

	We conclude this section by reiterating the observation that $\varpi$ provides a relational distinction of what is gauge (i.e. vertical) and what is `physical' (i.e. horizontal), {\it with respect to both a region and  its field content}.

	
	\subsection{Remarks on section \ref{sec:Singerconnection}\label{rmks:7}} 
	
	\paragraph*{\indent(i) On the choice of the super metric $\bb G^{\rm g}$ ---} 
	A rationale for the choice of the gauge supermetric $\bb G^{\rm g}$ is the following (see e.g. \cite{Babelon:1979wd}). 
	In a second-order Lagrangian formalism for the pure Yang--Mills action, field-space is given by the configuration space of the gauge potential $\{A_i^a(x)\}$, and the kinetic term of the Yang--Mills action is obtained by contracting a tangent vector of this space---the velocity $\bb V = \int \dot A^a_i\frac{\dd}{\dd A^a_i}$---with itself: the metric involved in this contraction is precisely $\bb G^{\rm g}$. 
	Therefore this metric also plays  a role in defining the Legendre transform to the Hamiltonian framework, and hence in the definition of the symplectic structure of the theory.

	\paragraph*{\indent(ii) $A_0$ and time-dependent gauge transformations ---}
	Given that the above construction involves a 3+1 splitting, a question remains to be addressed: what is the role of the time component $A_0$ of the Yang-Mills gauge field? Of course, it is a Lagrange multiplier whose dynamical role is to control the vertical (i.e. gauge) motion of $A_i$ during the time evolution. However, the issue is that by excluding $A_0$ from field-space one is suddenly not allowed to perform time-dependent gauge transformations, since they do affect $A_0$. A way out of this problem is to appropriately covariantize $A_0$.
	{
		In the Abelian case, this is done by setting 
		\be
		A_0(t,x) = \lambda(t,x) + \fI_{\bb V(t,x)}\varpi
		\qquad({\rm Abelian})
		\ee 
		where $\varpi$ is the Abelian SdW connection, $\lambda(t,x)$ is a free function (the Lagrange multiplier), and $\bb V$ is the velocity introduced in the previous paragraph. It is easy to see how the second term ensures $A_0$ is gauge transformed appropriately whenever $\xi$ is time dependent.
		In the general non-Abelian case---and where $\varpi$ is also arbitrary---the relevant correction is
		\be
		A_0(t,x) = h\lambda(t,x)h^{-1} - \pp_0 hh^{-1},
		\label{nonAbA0}
		\ee
		where $\lambda$ is now valued in $\fG$ and $h$ is a  $A_i$-dependent element of $\G$ with appropriate transformation properties (`gauge-compatible dressings', in the nomenclature of section \ref{sec:dressing}---cf. this section for technical details on this remark, and formula \eqref{dhh=varpi} for its Abelian limit). 
		By `appropriate transformation property' we mean that under field-dependent gauge transformations of $A_i$, $h$ must transform according to $h\mapsto g^{-1}h$, so that  its derivative transforms as $\pp_0 hh^{-1}\mapsto  g^{-1} (\pp_0 hh^{-1}) g - g^{-1} \pp_0 g $.
		From this, it follows that $A_0 \mapsto g^{-1}A_0 g + g^{-1} \pp_0 g $, as desired.}  In appendix \ref{app:examples_gluing2}, we analyze a simple example in which time-dependent gauge transformations play a central role, while in point {\it (ix)} of the \hyperref[rmks:dressing9]{\it Remarks on section \ref*{sec:dressing}} we discuss a purely infinitesimal version of this `dressing' of $A_0$ which is sufficient to gauge-invariantly deal with field strengths.

	\paragraph*{\indent(iii) Difficulties with a fully spacetime-covariant approach ---}
	So far, we have focused on field spaces associated to regions equipped with a Euclidean spacetime metric. That can be applied to Euclidean field theories, to the configuration space of a Lagrangian field-theory, or to the phase space of the Hamiltonian theory as parametrized by initial data given on a Cauchy surface. We will mostly work with the latter framework in mind---see section \ref{sec:Noether_hor}.
	
	Yet, another option is to work directly in a spacetime covariant fashion. In this case the field-space is the space of histories (with points corresponding to spacetime configurations of the field), and the region $\Sigma$ would be a spacetime region equipped with a Lorentzian metric. 
	This comes, however, with additional complications, both mathematical and physical.
	First of all, to the best of our knowledge, in this field-space no local product structure---no \lq{}slice theorem\rq{}---was ever proven to exist in the Lorentzian case (cf. footnote \ref{footnt:strata}). {\henrique Indeed, no proof exists that we could use the supermetric to define a splitting $\mathrm T\F\simeq H\oplus V$ by orthogonality (without an elliptic operator, the Fredholm alternative cannot be used to prove the decomposition \cite{gomes_riem, Ebin, Palais}).}\footnote{Elliptic equations on closed manifolds have at most a finite-dimensional kernel, and therefore, up to this ambiguity, can be inverted. Hyperbolic equations, on the other hand, can have an infinite-dimensional set of solutions; their inversion is therefore more ambiguous. } In the cases studied in this paper, definition of $H$ involved only elliptic operators, and, when non-trivial,  the finite-dimensionality of $H\cap V$ coincides with the existence of strata.

	Furthermore, if gauge transformations are completely free throughout the boundary of a spacetime region $\Sigma$, given that the corresponding equations for $\varpi$ are hyperbolic in this case (covariant Laplace operators are replaced by covariant d'A\-lem\-ber\-tians), even if we tried to mimic the Euclidean definition of $\varpi$,   the boundary-value problem associated to the PDE defining $\varpi$ would generally be ill-posed. 
	A last difficulty is related to the time-nonlocality of these dressings, as discussed in \cite{Lavelle:1995ty}.
	To avoid confronting all these problems, we ignore the potentially interesting {\aldo Lorentzian} spacetime-covariant case in the rest of this paper.

	\paragraph*{\indent(iv) Gauge transformations in finite regions and the uniqueness of $\varpi$ ---}
	A crucial point that needs to be emphasized concerns the interplay between boundaries and gauge transformations.
	To this purpose, one should notice that had we restricted the  gauge transformations $\xi$ at the boundary---as is customarily done \cite{Regge:1974zd, BeigOMurchadha,Geiller:2017xad}---we would see that $\varpi$ would become less constrained.
	However, {\henrique naively,} this strategy defeats the  purpose of the present treatment, which wants to allow gauge to be treated geometrically in field-space independently of the spacetime points. Allowing non-trivial gauge transformations at the boundary, we obtain a (covariant) Poisson equation and its {\henrique field-dependent, covariant} boundary conditions,  and no further specification is necessary.
	
	{\henrique An alternative, {\new which we will pursue in a forthcoming publication}, is the restriction of \textit{field-space itself}. This is useful if we would to model certain types of subsystems---e.g. isolated subsystems. This topic is picked up in the second remark of \hyperref[rmk:Noether_hor]{\it Remarks on section \ref*{sec:Noether_hor}}.}

	\paragraph*{\indent(v) Choice of $\varpi$ and horizontality  ---}  Horizontal with respect to $\varpi$ describes  relational physics with respect to a field, and moreover this notion can be attributed to any  bounded region within a manifold. Although different choices of $\varpi$ may describe physical processes differently, two given choices will always agree that the process was indeed \lq{}physical\rq{}, and not \lq{}pure gauge\rq{}. In other words, a field-space vector that has a nonzero horizontal part with respect to one choice of $\varpi$ has a non-zero horizontal part with respect to any choice of\footnotemark~$\varpi$.\footnotetext{This remark applies only on regions of $\F$ away from nontrivial stabilizers, or if the stabilizers are identical for the different fields.}

	\paragraph*{\indent(vi) Gluing of regions, horizontality, and `edge modes' ---} 
We showed that given a region $\Sigma$ subdivided into subregions, $\Sigma_{I}$ and $\Sigma_{II}$ divided by the surface $S$, the notions of horizontality and verticality with respect to the SdW connection are generally not preserved by restrictions or gluings. 
For the notions to be preserved, a necessary and sufficient condition is that the regional vertical projections of $\bb X$ match at the boundary. This turned out to be the same as requiring $(\xi_I)_{|S} = (\xi_{II})_{|S}$, where e.g. $\xi_I = \fI_{\bb X_I}\varpi_I$ is the regional projection.

In the next section, we will see  that the role of $\varpi$ is to `covariantize' the presymplectic potential of gauge theories in the presence of boundaries. 
This is precisely the same role which `edge modes' had in \cite{Donnelly:2016auv}.
Therefore, given a field variation $\bb X$ in a general bounded region $\Sigma$, the (Lie-algebra-valued) vertical projection of $\bb X$ at the boundary, {\it $\xi_{|\pp \Sigma}$,  can be understood as the (infinitesimal) edge mode associated to $\bb X$ with respect to the SdW connection of the region $\Sigma$} (note that $\xi$ will depend non-locally on $A$ and $\bb X$ throughout $\Sigma$). 
One of the main points of \cite{Donnelly:2016auv} was to provide a notion of gluing between regions which they called `entangling product'. Roughly speaking, the entangling product consists in `averaging'\footnote{More precisely, the entangling product features a projection on a gauge-invariant subsector. Formally, one can obtain the quotient through a `group averaging' procedure.} over matching values of the edge modes across the common boundary.

To the extent to which the $\xi_{|\pp \Sigma}$'s correspond within our framework to the edge modes of \cite{Donnelly:2016auv}, we see that the matching condition required by the entangling product is restrictive but possesses a precise geometrical meaning. \\

	These last two remarks, being of much physical interest, transition us into the next part of this  article.

\part{Physical applications}

\section{ Noether charges and field-space horizontality}\label{sec:Noether_hor}

In this section we will explain how the horizontal differential introduced above fits into the standard  (spacetime-)covariant symplectic formalism, and we will exemplify by explicitly using the SdW connection. 
We first recall the (spacetime\hyp{})covariant symplectic formalism. 
The treatment provided here, although formal and seemingly general, contains many features particular to Yang--Mills theories---the focus of this paper. We clarify these features below.

\subsection{Gauge-covariant symplectic geometry and Noether charges}\label{sec:general_Noether}

Let $L = \mathscr{L}(\varphi)\d^d x$ be a  Lagrangian (spacetime-){\it density}. We then define the presymplectic potential\footnote{{\it Pre}symplectic means that the symplectic form  $\Omega = \dd \theta$, once integrated on a Cauchy hypersurface, can contain degenerate directions, related to gauge.} $\theta$ implicitly through the field-space derivative of this density: 
\be\label{eq:theta_implicit}
\boxed{\quad\phantom{\Big|}
\dd L = \mathrm{EL}_I(\varphi)\dd\varphi^I  + \d\theta(\varphi),
\quad}
\ee
where $\d$ is the spacetime exterior derivative, and $\mathrm{EL}_I(\varphi)$ are the (densitized) Euler-Lagrange equations for $\varphi^I$. The (pre)symplectic potential,
\be
\theta=\Pi_I\dd\varphi^I\in \Lambda^1(\F)\otimes \Lambda^{d-1}(M),
\label{theta_standard}
\ee
is a field-space one-form and spacetime $d-1$ form. Notice that we used densitized momenta,  which are in the covariant Hamiltonian formalism viewed as functions of the fields and their derivatives, $\Pi_I = \Pi_I(\varphi)\in\Lambda^{d-1}(M)$. Note also that \eqref{eq:theta_implicit} determines $\theta$ only up to a corner ambiguity, $\theta \mapsto \theta + \d \alpha$.

In the following we will make the two central hypotheses 
\be
\fLie_{\xi^\#} L = 0,
\qquad\text{and}\qquad
\fLie_{\xi^\#} \theta = 0 \quad(\dd \xi = 0).
\label{hypothesis}
\ee
The first states that the Lagrangian {\it density} is strictly invariant under (infinitesimal) gauge transformations, and not only up to boundary terms. 
The second one states the same for the symplectic potential, at least for $\xi$'s which are field-independent.
 These demands are quite restrictive, but do apply to the standard Lagrangian and symplectic potential of Yang--Mills theory.

In the following, we will see that our gauge-covariant symplectic geometry---reinterpreted from a spacetime perspective---will make use of the  corner
ambiguity of $\theta$ in a very specific way. This result will {\it not} follow from the definition of a new modified Lagrangian, but from dealing covariantly with gauge symmetries within $\FYM$ itself.

Now, if the two conditions above hold, then
\be
0 = \fLie_{{\lvf}^\#} L = \mathrm{EL}_I\delta_{\lvf}\varphi^I + \d\fI_{\xi^\#}\theta,
\label{eq_LieLagrange}
\ee
and one is led to define the Noether current density%
\footnote{The physicist's vectorial Noether current $J^\mu$ is rather associated to the Hodge dual of $j$: $J_\mu\d x^\mu = \ast j$, or in components $J^\mu = \epsilon^{\mu\nu_1\dots \nu_{d-1}}j_{\nu_1\dots\nu_{d-1}}$, and $\d j = \nabla_\mu J^\mu\d^d x$. Abstractly, it is easier to manipulate currents which are $d-1$ forms.}
$j_{\lvf}$ as (e.g. \cite{Iyer:1994ys}) 
\be
j_{\lvf} := \fI_{{\lvf}^\#} \theta \equiv \theta(\varphi,\delta_{\lvf}\varphi).
\label{eq_def_jxi}
\ee

The on-shell formulas for the presymplectic potential and form are relevant at the level of the so-called `covariant phase space' construction, where the canonical phase space is identified with the space of histories which satisfy the equations of motions, i.e. the subspace of $\FYM$ defined by ${\rm EL}_I = 0$---a condition we signal with $\approx$.
Then, one has
\be
\d j_\xi \approx 0.
\ee
{
Using the arbitrariness of $\xi\in\fG$, one concludes that the Noether current must be of the form
\be
\label{eq:j_is_dQ}
j_\xi = C_a\xi^a + \d Q_\xi \qquad \text{where}\qquad C_a \approx 0.
\ee}
This equation {defines the charge density $Q_\xi$. It also shows the association between gauge symmetries and canonical constraints\footnotemark~{\henrique (see \cite{Lee:1990nz} for the first derivation of these results  in the covariant symplectic formalism)}, the latter being the canonical generator of the relevant gauge symmetries.} To see this relationship, define the presymplectic two-form
\be
\Omega=\dd\theta \in \Lambda^2(\FYM)\otimes\Lambda^{d-1}(M).
\ee 
Then, making use of $\fLie_{\xi^\#}\theta = 0$ (with $\dd\xi=0$), Cartan's formula \eqref{eq_magic}, and the definitions of $j_\xi$ and $\Omega$ above, one readily finds the Hamiltonian flow equation 
\be
\label{Ham_flow}
\fI_{{\lvf}^\#}\Omega = -\dd j_{\lvf} \qquad (\dd \xi =0).
\ee
This formula indicates that the Noether charge $j_\xi$ is the symplectic generator of the (field-independent) symmetry $\xi^\#$ on field-space (notice also that no  equation of motion is required at this level).
\footnotetext{ The canonical constraints are the pullback of $C_a\approx 0$ on a Cauchy surface. }

Furthermore, the two conditions of equation \eqref{hypothesis} directly imply the current algebra equation $\fI_{{\chi}^\#}\fI_{{\lvf}^\#}\Omega = - j_{[\chi,\xi]}$ ($\dd\xi=0=\dd\chi$), or equivalently
\be
\fLie_{\xi^\#} j_\chi = j_{[\xi,\chi]} \qquad (\dd \xi = 0 = \dd \chi) ,
\label{current algebra}
\ee
which follows from
\be
j_{[\xi,\chi]} = \fI_{[\xi,\chi]^\#}\theta = (\fLie_{\xi^\#}\fI_{\chi^\#} - \fI_{\chi^\#}\fLie_{\xi^\#})\theta = \fLie_{\xi^\#} j_\chi
 \qquad (\dd \xi = 0 = \dd \chi).
\ee

 So far, we have dealt with field-independent gauge transformations. Let us now turn to generic {\it field-dependent} gauge transformations. In this case, many of the equations above fail, and have to be modified. 
Their failure stems from the second condition of  \eqref{hypothesis}, which  can be seen by observing that $\dd$ never acts on $\xi^\#$ on {\henrique the first equation, whereas it does on the second one}. 
From this, it follows that the flow equation \eqref{Ham_flow} cannot hold true.
The obstruction can be shown to consist, on-shell, of a pure corner term:
\be
\fLie_{\xi^\#} \theta = \fI_{\xi^\#}\dd \theta + \dd \fI_{\xi^\#}\theta \;\stackrel{\eqref{hypothesis}}{=}\; \fI_{(\dd\xi)^\#}\theta = j_{\dd \xi} \approx \d Q_{\dd \xi}.
\ee

This observation led us to introduce the horizontal symplectic current in \cite{Gomes:2016mwl},
\be
\boxed{\quad\phantom{\Big|}
\theta_H :=\Pi_I\dd_H\varphi^I = \theta - \Pi_I\delta_{\varpi}\varphi^I.
\quad}
\label{defhorizontalpotential}
\ee
Now, this can be written as $\theta_H = \theta - \fI_{\varpi^\#}\theta = \theta - j_\varpi$, and
 if $\theta$ satisfies the conditions of equation \eqref{hypothesis}, then---even for a field-dependent $\xi$ ($\dd \xi \neq 0$)---one has automatically 
\be
\fLie_{\xi^\#}\theta_H  = 0 ,
\ee
since
\begin{align}
\fLie_{\xi^\#}\theta_H 
&= j_{\dd\xi} - \fLie_{\xi^\#} j_\varpi
= j_{\dd\xi} - (\fLie_{\xi^\#} j)_\varpi - j_{[\varpi,\xi] + \dd \xi} = - (\fLie_{\xi^\#} j)_\varpi - j_{[\varpi,\xi]} = 0
\end{align}
{ Recall from \eqref{eq:def_equi} that the last equation means $\theta_H$ is equivariant} (for the trivial representation, since $\theta_H$ has no `open' Lie-algebra indices). { Using equation \eqref{horizontalequivariant} and the fact that $\theta_H$ is in the trivial representation it follows} that
the horizontal presymplectic two-form is automatically $\dd$-exact:
\be
\boxed{\quad\phantom{\Big|}
\Omega_H:=\dd_H\theta_H=\dd\theta_H,
\quad}
\label{horizontalOmega}
\ee
This proves a fortiori that $\Omega_H$ is $\dd$-closed and therefore a viable presymplectic form.

From the above definitions,
\be
\boxed{\quad\phantom{\Big|}
j^H_{\lvf} := \fI_{{\lvf}^\#}\theta_H =  0
\qquad\text{and}\qquad
\fI_{{\lvf}^\#}\Omega_H = 0.
\label{hor_charge_gauge}
\quad}
\ee
These formulas are valid locally on $M$, at the density level, and hold for field-dependent gauge transformations as well.
The message they convey is that such gauge transformations carry {\it no physical charge}, {as we will explain in more detail in the remarks on this section.}

In the next subsection, we make the above formulas explicit in the case of Yang--Mills theory, while in section \ref{ssec:sympl_charges}, we will show that in spite of the general results above, the formalism can still allow for  conserved global charges in some specific circumstances:  in the construction of physically relevant choices of $\varpi$, such as the SdW connection, certain hypothesis we  have here taken for granted will be subtly violated,  providing the loopholes which allow for the  appearance of non-trivial global charges.


\subsection{Yang-Mills theory}\label{sec:YM}

The Yang-Mills Lagrangian is given (in flat spacetime) by\footnote{ Since this Lagrangian is real only up to boundary terms, strictly speaking one should replace in the remainder any term $T$ which involves $\psi$ with $\frac12(T + \bar T)$. The arguments go through unchanged.}
\be
\mathscr L_\YM(A,\psi,\bar\psi) = -\tfrac{1}{4e^2} F_{a\,\mu\nu}F^{a\,\mu\nu} + i \bar\psi \gamma^\mu \D_\mu \psi
\ee
where $F^a_{\mu\nu} = 2\pp_{[\mu}A^a_{\nu]} +f_{bc}{}^a A^b_\mu A^c_\nu $, $f_{bc}{}^a$ are the structure constants of $\fg$ in the $\tau_a$ basis, $\gamma^\mu$ are Dirac's gamma-matrices, $\D_\mu \psi = \pp_\mu \psi + A_\mu^a \tau_a \psi$, and $e$ is the Yang--Mills coupling constant. Also, $\bar\psi = \psi^\dagger \gamma^0$, with the understanding that $\psi$ and $\psi^\dagger$ have to be considered as two independent (complex) variables. The Lie algebra index $a$ is lowered with the Kronecker delta.

In form language, $L_\YM = \mathscr L_\YM \d^d x$ can be written as\footnote{ Formulas in form language are given with the correct signs in even-dimensional spacetimes. We prioritized uncluttered formulas over complete generality. }
\be
L_\YM = -\tfrac{1}{2e^2} (F^a \wedge  \ast F_a) +  (i \bar\psi \gamma_\mu \D\psi) \wedge \ast \d x^\mu 
\ee
where $F = \d A + \tfrac12[A,A]$ and $\ast$ is the Hodge dual. 
It is clear that $\fLie_{\xi^\#} L_\YM = 0$.

Now, we compute the field-space differential of the Lagrangian,\footnotemark %
\begin{align}
\dd L_\YM = & \big( \dd A^a \wedge (- e^{-2}\D \ast F_a - J_a)\big) + \dd \bar\psi \big(i \gamma_\mu \D \psi \wedge \ast \d x^\mu\big) + \big(-i \D \bar\psi \gamma_\mu \wedge * \d x^\mu \big) \dd \psi \notag\\
&  + \d \big(  -e^{-2}\dd A^a\wedge \ast F_a +i \bar\psi \gamma_\mu \dd \psi (\ast \d x^\mu) \big)\label{YM_eom}
\end{align}
\footnotetext{In coordinate notation,
\begin{align}
\dd \mathscr L_\YM 
=& \Big( e^{-2}\D_\mu F^{\mu \nu}_a - J_a^\nu\Big) \dd A_\nu^a + \dd \bar\psi \Big( i \gamma^\mu \D_\mu \psi\Big) +\Big(- i \D_\mu \bar\psi \gamma^\mu\Big)\dd\psi +\pp_\mu\Big( -e^{-2}F^{\mu \nu}_a\dd A_\nu^a   + i \bar\psi \gamma^\mu \dd\psi\Big).\notag
\end{align}
}
and introduce the matter current density,\footnote{In detail, $$ J_a = -\frac{i}{3!}\bar\psi_{\alpha m} (\gamma^\mu)^{\alpha}{}_{\beta} (\tau_a)^m{}_n \psi^{\beta n} \epsilon_{\mu\nu\rho\sigma} \d x^\nu \wedge \d x^\rho \wedge \d x^\sigma. $$
The unusual appearance of $e^2$ in the Gauss law is due to the fact that the physicist's gauge potential is not $A$ but ${\cal A}=e^{-1}A$.}
\be
J_a = - i  \bar\psi \gamma_\mu \tau_a\psi (\ast \d x^\mu)
\ee
which is valued in the dual of the Lie algebra.

From this one finds (using the usual properties of the $\gamma$ matrices)
\be
{\rm EL}_A = - e^{-2} \D \ast F - J,
\qquad 
{\rm EL}_{\bar \psi} = i \gamma^\mu \D_\mu \psi \d^d x,
\qquad 
{\rm EL}_{\psi} = ({\rm EL}_{\bar \psi})^\dagger \gamma^0,
\ee
and
\be
\theta_\YM =  -e^{-2}\dd A^a \wedge  \ast F_a + i \bar\psi \gamma_\mu \dd \psi \ast \d x^\mu.
\ee
From which $\Pi_A = \ast e^{-2}F $, $\Pi_\psi = i \ast \bar\psi \gamma_\mu \d x^\mu$, and $\Pi_{\bar\psi} = 0$. 
Notice that $\theta_\YM$ is invariant under field-independent gauge transformations (as demanded in the previous section), but not under field-dependent ones: 
\be
\boxed{\quad\phantom{\Big|}
\fLie_{\xi^\#} \theta_\YM = -e^{-2}\d (\ast F_a \dd\xi^a ).
\quad}
\ee

The Noether current $j_\xi$ is given by
\begin{align}
j_\xi = \fI_{\xi^\#} \theta_\YM 
& = - e^{-2}\D \xi^a \wedge \ast F_a - i \bar\psi \gamma_\mu  \xi \psi \ast \d x^\mu \notag\\
& = (e^{-2}\D \ast F + J)_a\xi^a - \d( \ast F_a \xi^a)\notag\\
&\approx - e^{-2}\d (\ast F_a \xi^a).
\end{align}
Thus we recognize its pure-boundary character on-shell. The conservation equation $\d j_\xi \approx 0$ trivially follows.

Finally, the presymplectic two-form reads
\begin{align}
\Omega_\YM &= e^{-2}\dd A^a \wedge \ast \D \dd A_a + i \dd \bar\psi \gamma_\mu \dd \psi \ast \d x^\mu 
\end{align}
where $\D \dd A = \d \dd A + [A, \dd A]$.
The Hamiltonian flow equation \eqref{Ham_flow} can be explicitly checked with a simple computation.

Now, introducing a functional connection-form $\varpi$, we can define the horizontal presymplectic potential
\be
\theta_{\YM, H} = - e^{-2}\dd_H A \wedge \ast F + i \bar\psi \gamma_\mu \dd_H \psi \ast \d x^\mu.
\label{eq_thetaYMH}
\ee
Using equation \eqref{eq_contr=zero}, i.e. $\dd_HA=\dd A-D\varpi$ and $\dd_H \psi = \dd \psi +\varpi\psi$, and the Euler-Lagrange equations\footnote{By pulling back on a (portion of a) Cauchy hypersurface, the constraint equations turn out to be sufficient.} one can verify that 
\be \label{eq:theta_H}
\begin{array}{|crlc|}
\hline&&&\\
&\theta_{\YM, H} 
&= \theta_\YM + e^{-2}D \varpi^a \wedge  \ast F_a - \varpi^a J_a &\\
&& = \theta_\YM - (e^{-2}\D \ast F + J)_a \varpi^a + \d (\ast F_a \varpi^a) &\\
&& \approx \theta_\YM +e^{-2} \d ( \ast F_a \varpi^a).&\\
&&&\\\hline
\end{array}
\ee
Thus, we see that the horizontal presymplectic potential is equal, on-shell, to the standard one plus a boundary term involving $\varpi$. { This follows from the formulas $\theta_H = \theta - j_\varpi$ (see text below equation \eqref{defhorizontalpotential}) and $j_\xi \approx \d Q_\xi$.}

 The above expressions allow us to explicitly prove that $\fLie_{\xi^\#} \theta_{\YM, H} = 0$ (of course, this already follows from the  general remarks of the previous section, since $\fLie_{\xi^\#}\theta= 0$ if $\dd\xi=0$): using equation \eqref{covariance_props}, and $\xi^\dagger = - \xi$ from the unitarity of $G$,
\begin{align}
\fLie_{\xi^\#} \theta_{\YM, H} = -e^{-2}[\dd_H A, \xi]^a \wedge\ast F_a - e^{-2}\dd_H A^a \wedge [\ast F_a, \xi]_a \qquad & \nonumber\\
+  i \Big( (\bar \psi \xi) \gamma_\mu \dd_H \psi + \bar\psi \gamma_\mu (-\xi \dd_H \psi) \Big)\ast \d x^\mu & = 0,
\end{align}
where the last equality follows from the proportionality of the Kronecker delta to the Killing form and the ${\rm ad}$-invariance of the latter, i.e.
\be
\langle \xi_1, [\eta, \xi_2] \rangle_{\rm K} = -\langle [\eta, \xi_1], \xi_2 \rangle_{\rm K} ,
\ee
In turn, according to the general arguments of the previous section, the fact that $\fLie_{\xi^\#} \theta_{\YM, H} = 0$ guarantees that the horizontal presymplectic potential $\Omega_H := \dd_H \theta_H$ is $\dd$-exact, equation \eqref{horizontalOmega}.
Nonetheless, we proceed once again to the explicit verification of these claims.
The horizontal presymplectic two-form is given by
\begin{align}
	\Omega_{\YM,H} &=  e^{-2} \dd_H A^a \wedge \ast\dd_H F_a +  i(\dd_H \bar\psi)\gamma_\mu(\dd_H\psi)\ast \d x^\mu + e^{-2} \D\fF\wedge \ast F - \fF^a J^a,
\end{align}
where we used equation \eqref{eq_dHdH=F2} to obtain the last two terms.
On-shell this can be readily recast in the form
\be
\Omega_{\YM,H} \approx
 e^{-2} \dd_H A^a \wedge \ast \dd_H F_a + i (\dd_H \bar\psi )\gamma_\mu (\dd_H \psi) \ast \d x^\mu + e^{-2}\d(  \ast F_a\fF^a   ),
\ee
which emphasizes that it is not enough to replace $\dd$ with $\dd_H$ in $\Omega$ to obtain $\Omega_H$.  This is a consequence of $\dd_H^2 \propto \fF$ (equation \eqref{eq_dHdH=F2}).

A tedious but straightforward calculation\footnote{In \cite{Gomes:2016mwl}, gauge transformations are the inverse of those considered here, hence the difference in sign.} \cite[App. B2]{Gomes:2016mwl} leads to the following alternative form for $\Omega_{\YM,H}$: 
\be
\boxed{\phantom{\Big|}
\Omega_{\YM,H} = \Omega_\YM - \dd\Big( \varpi^a(e^{-2}  \D \ast F  + J)_a -e^{-2} \d (\ast F_a  \varpi^a) \Big) .
	\;\;}
\ee
which, together with equation \eqref{eq:theta_H}, explicitly shows what we had already proven abstractly, i.e. that $\Omega_{\YM,H} = \dd \theta_{\YM,H}$.
Finally, the above formula shows that $\Omega_{\YM,H} \approx \Omega_\YM +e^{-2} \d \dd ( \ast F_a \varpi^a)$, that is
\be
\boxed{\quad\phantom{\Big|}
\Omega_{\YM,H}|_\text{bulk} \approx \Omega_\YM|_\text{bulk}.
\quad}
\ee



\subsection{SdW connection and symplectic charges}\label{ssec:sympl_charges}

In this subsection, we investigate the horizontal Noether currents obtained from the horizontal symplectic potential of the previous subsection.
There, we gave very general arguments for why the horizontal Noether currents associated to a generic gauge transformation must be trivial---equation \eqref{hor_charge_gauge}. Here, revisiting that argument for the SdW connection, we discover that it has to be refined, with physically interesting consequences. It turns out that if the configuration of the Yang-Mills gauge field possesses {\it global} internal symmetries, these are automatically singled out, and their horizontal currents and charges do {\it not} vanish.

The starting point is therefore the horizontal symplectic potential  \eqref{eq:theta_H}, for the SdW connection with boundaries of equation \eqref{YM_varpi_boundaries}.
Recall from section \ref{sec:general_Noether} that the horizontal Noether current $j^H_\xi$ is given in terms of the horizontal symplectic potential $\theta_H$ by\footnote{We remind the reader that this formula is not completely general, but it is correct for Yang--Mills. See the discussion of section \ref{sec:general_Noether}.}
\begin{align}
 j^H_\xi =  \fI_{\xi^\#} \theta_H = \fI_{\xi^\#} \theta - \fI_{(\varpi(\xi^\#))^\#} \theta.
\end{align}
Since $\varpi(\xi^\#) := \fI_{\xi^\#}\varpi =\xi$ by construction, at first glance, one would expect all horizontal Noether currents to vanish, as in equation \eqref{hor_charge_gauge}. 

However, a subtlety we have so far glossed over now becomes important.
Namely, we have always implicitly assumed that for Yang-Mills without matter, the map $\cdot^\#$ from the Lie algebra of gauge transformations to the  vertical vector fields on field-space is an isomorphism. 
The fact is, this is {\it not} always the case, for {\it the map $\cdot^\#$ can have a non-trivial kernel.}

In electromagnetism, this is easy to see: global gauge transformations $\xi(x)=\text{const.}$ leave the gauge field invariant, since in this case $\xi^\#_A \equiv (\d \xi) \frac{\dd}{\dd A} = 0$; global gauge transformations are the only gauge transformations with this property. In pure non-Abelian Yang-Mills theories, whether or not transformations $\xi$'s such that $\xi^\#_A \equiv (\d\xi + [A,\xi]) \frac{\dd}{\dd A}=0$ exist depends on the point in configuration space $A\in\F_\text{pYM}$.
 Configurations for which such $\xi$'s exist are called {\it reducible}. Reducible configurations correspond to the lower dimensional strata of the stratified manifold $\F_{\rm pYM}/\G$ (see discussion after equation \eqref{group_action}).
  In gravity, the analogue consists of metrics with Killing vector fields. For this reason, we will refer generically to infinitesimal gauge transformations $\xi$'s such that $\D\xi=0$ as \lq{}Killing\rq{}, and we will denote them with the letter $\chi$.
 The total number of Killing transformations that a given configuration $A$ admits is always finite and varies from $0$ to $\text{dim}(\fg)$ depending on $A$ itself.\footnote{ Note however that if $A$ has the maximal number of symmetries, $F$ must vanish, because it transforms in the adjoint representation and its stabilizer cannot be $(\text{dim }\fg)$-dimensional unless $F = 0$. For a discussion of global Yang-Mills charges, see \cite{Chrusciel:1987jr}.}

Let us now go back to the horizontal Noether charges. We  do this in presence of fermions, rather than in the pure Yang--Mills theory.
The crucial technical point for this section is that,  by the co-rotation principle (see the beginning of section \ref{sec:Singerconnection}), a $\varpi$ constructed from the gauge field alone---e.g. the SdW connection---is also a valid connection on the full field-space of Yang Mills theory {\it with} matter. Therefore, one can have Killing transformations for $A$ which are not Killing for the matter field, and the connection $\varpi$ will still be transparent to them.  This is true even if the matter is on-shell, although in this case $\chi$ must be Killing for the matter current $J$ (a weaker condition than being Killing for $\psi$). This feature is the source of the results of this section. 

Suppose $A$ is a configuration that admits Killing transformations $\{\xi_n\}$. To not carry too many indices around, let us focus on one particular transformation, denoted simply by $\chi$. Then, the Killing property for $A$ reads $\delta_{\chi} A \equiv \D\chi =0$.  This implies $\fI_{\chi^\#} \varpi = 0$ for the SdW connection, even when $\delta_{\chi} \psi = -\chi \psi \neq 0$.

The horizontal symplectic potential for Yang-Mills theory with fermions is, see equation \eqref{eq_thetaYMH}),
\begin{align}
\theta_{{\rm YM}, H} = - e^{-2}\dd_H A^a \wedge \ast F_a + i \bar\psi \gamma_\mu \dd_H \psi \ast \d x^\mu.
\end{align}
From this, and the above remarks, for a Killing transformation $\chi$ we obtain
\begin{align}
j^H_\chi =  -\chi^a i \bar\psi \gamma_\mu \tau_a * \d x^\mu =  \chi^a J_a
\qquad
(\D \chi = 0).
\end{align}
Thus, the horizontal current for a global transformation is precisely given by the {\it matter current density $J_a$} contracted with $\chi^a$. {\henrique For a linear combination of such Killing directions, we would replace $\chi^a\rightarrow \alpha^n\chi_n^a$.} In electrodynamics, there is a single such direction, $\chi = \text{const.}$ and $j_\chi^H$ is precisely the total current density of electrons. We thus see that {\it the SdW connection with boundary picks out the global charges---when they exist---as the only physical ones}.

Of course, using the Gauss law, $D \ast F_a + J_a \approx 0$ {\it and} the Killing condition $\D \chi_n =0$, the horizontal charge can  be written as
\be
Q_{\chi_n}^H =\int_\Sigma \chi_n^a J_a \approx -e^{-2}\int_{\pp \Sigma} \ast F_a \chi^a_n
\qquad
(\D \chi = 0).
\label{eq:horiz_charge}\ee

\subsection{Remarks on Section \ref{sec:Noether_hor}\label{rmk:Noether_hor}}

\paragraph*{\indent(i)  Horizontal symplectic geometry ---}
Using the covariant field-space derivative in the case of Yang--Mills theory, the symplectic potential becomes completely gauge-invariant, even with respect to field-dependent gauge transformations.
The difference between the standard symplectic potential and the horizontal one is given by a boundary term.  
Despite this modification, it turns out that one is still able to do symplectic geometry, since the symplectic form associated to the horizontal potential is automatically horizontal. In sum, although our formalism does not explicitly refer to boundaries or its degrees of freedom, it  turns out that in the Yang-Mills case it only has an effect on the symplectic structure at the boundary, as promised.
Notice that although the new contributions to the symplectic structure are boundary contributions, their dependence on the field values---in the Singer--DeWitt case---is nonlocal and involves the whole region. Crucially, no information from outside the region of interest is required at any point; i.e. they are still {\it{regional}}.

\paragraph*{\indent(ii)  New gauge charges? ---}
The message conveyed by equations \eqref{hor_charge_gauge} is that  pure gauge transformations carry {\it no physical charge}  with respect to this particular decomposition of vertical/horizontal, or gauge/physical, degrees of freedom even when their support reaches the boundary. Nevertheless, there is still room for  conserved global charges, as we saw in section \ref{ssec:sympl_charges} and will discuss at point {\it (iv)} below. 
 In \cite{Gomes:2016mwl}, it was argued that choices of $\varpi$ represented exactly this: particular decompositions of fields into \lq{}physical with respect to $\varpi$\rq{}---purely horizontal---and \lq{}gauge with respect to $\varpi$\rq{}--- purely vertical. Such charges are relational in the sense that they refer back to a given field, and are written solely in terms of the existing fields. In appendix \ref{app:examples_gluing1} we will look at explicit examples.

 As remarked on item ({\it iv}) of  \hyperref[rmks:7]{\it Remarks on section \ref*{sec:Singerconnection}}, {\florian here we have } made no restrictions on field-space. However,  trying to model specific sorts of subsystems, e.g. isolated subsystems, could change the picture.   Preliminary investigations show that such restrictions would allow for a larger kernel of $\varpi$. Regarding charges,  the qualitative difference between the present approach and other approaches (e.g.: \cite{Donnelly:2016auv})  is insensitive to field-space restrictions: either way,   the charges associated to the construction presented here depend solely on the original field-content of the theory, i.e. on the type of field configurations we allow inside the region. In other approaches, the abstract boundary brings charge contributions that depend on more than the original field-content, e.g. it depends on new boundary fields (edge modes).  

\paragraph*{\indent(iii) Horizontal \textsc{vs.} standard Noether charges ---} 
The standard---i.e. non-ho\-ri\-zon\-tal---Noether charge, given by
\be
Q_\xi = \int_\Sigma j_\xi = \int_\Sigma (-e^{-2}\D\xi^a \wedge \ast F_a + \xi^a J_a)\approx \int_{\pp\Sigma} -e^{-2}\ast F_a \xi^a,
\ee
can coincide, for very particular values of $\xi$, with our expression for the horizontal charge $Q^H_\chi$ in terms of the weighted flux of $\ast F$, given in equation \eqref{eq:horiz_charge}. But 
it is important to realize that the two charges above are conceptually very different and should not be confused. 
First of all, as we have already discussed, Killing transformations, $\chi$ such that $\D\chi=0$, exist  only for certain symmetric gauge-field configurations ---associated to lower dimensional strata of $\FYM$ (see discussion after equation \eqref{group_action})---even there will only appear in finite number. This is in contrast to the standard Noether current of equation \eqref{eq_def_jxi}, which can be written for any gauge parameter (e.g. \cite{Donnelly:2016auv, Iyer:1994ys}).
Second, while both charges can be expressed in the same way as boundary fluxes, the two correspond to very different bulk currents: while the horizontal current \lq{}knows' only about the (charged) matter content of the theory and generates transformations on  the fermionic fields through $\fI_{\chi^\#}\Omega_H = - \dd j^H_\chi$, the standard Noether current is determined by the demand that it is a differentiable symplectic generator of any gauge transformation, off-shell and with respect to $\Omega$, not $\Omega_H$ (differentiability is understood in the sense of Regge and Teitelboim \cite{Regge:1974zd, BeigOMurchadha}).
Ultimately, the horizontal charges appear related to objectively conserved physical quantities, like the total charge in electromagnetism, since $\D\chi=0$ implies that the quantity $J_\chi :=\ast\chi^aJ_a$ satisfies the on-shell flux balance formula $\nabla_\mu J_\chi^\mu\approx0$. More on global charges on point {\it (iv)} below.

\paragraph*{\indent(iv) Reducible configurations and the uniqueness of $\varpi$ ---}
In general, for reducible configurations $A$ (i.e. for those which admit Killing symmetries, see figure \ref{fig8} for the geometric meaning), the  defining equation for $\varpi$ \eqref{YM_varpi_boundaries} a priori admits multiple solutions. These differ from each other by field-space one-forms whose Lie-algebra value is given by some linear combination of the $\chi_n$ (with  spacetime constant factors).  In other words, $\varpi$ is agnostic about the Killing components of a generic $\xi$. This ambiguity can in principle be settled by appealing to the first Noether theorem for global symmetries---which gives automatically $j_\chi = J^a\chi_a$,---but a unified treatment would of course be preferable.
We leave a more detailed study of these issues to future work.

\paragraph*{\indent(v) Horizontal Noether charges \textsc{vs.} the Barnich--Brandt--Henneaux formalism ---}
Our treatment appears to agree in its general conclusions with the results of Barnich, Brandt and Henneaux (see \cite{Barnich:2001jy, Barnich:1994db, Barnich:1994mt}), who claim that physically meaningful charges can only be assigned in the linearized theory around reducible background configurations.\footnote{ A similar conclusion was also reached via non-symplectic methods by DeWitt \cite{DeWitt_Book}.} 
However, important differences have to be kept in mind. First of all, our treatment is essentially kinematical---although one might argue that the dynamics is just hiding in the choice of the supermetric and the requirement of minimal coupling,---while the one by Barnich and Brandt takes heavily into account the dynamics, and their currents are defined as non-trivially conserved objects. Second, in our treatment both the gauge field $A$ and the fermion field $\psi$ can be `large', while in their treatment a split between background and linear fluctuations is necessary to define conserved quantities within the linear theory (in absence of a background, it is not possible to find a generic {\it and} non-trivially conserved charge). It would be interesting to push the two formalisms closer together. The remarks made in \cite{Gomes:2016mwl}  about the relationship of $\varpi$ with the so-called 'geometric' BRST formalism might constitute a valid starting point.

\paragraph*{\indent(vi) Non-trivial cohomology of $\Sigma$ ---}
As we have repeatedly emphasized, our entire paper  implicitly assumes a trivial cohomology for the gauge bundles over $\Sigma$. Deviating from this assumption, several new aspects would have to be considered, many relating to the possibility of closed but not exact forms.  For example, there might be elements for which $\fI_{\bb X}\varpi=0$, for which $\bb X$ is locally of the form $\xi^\#$ (and thus would be usually identified with being \lq\lq{}pure gauge\rq\rq{}), but it is  not globally of that form, due to topological obstructions. This case could produce new, \lq{}topological\rq{} physical charges (see e.g. \cite{GiuliniCharge}). Moreover, our parametrization of field-space through $A$---which requires a choice of a section for the standard (finite-dimensional) $G$-bundle over $\Sigma$---is insufficient; one would require the  parametrization with $\omega$, the PFB connection. The two issues are, of course, related.  See footnote \ref{ftnt:PFB} for more on this last point.

\section{The Higgs connection}\label{sec:matter}

In the previous two sections, we have focused on the Singer--DeWitt connection. 
Derived from a supermetric, it is built solely from the gauge field $A$. 
In this section, we explore which field-space connection one obtains if one focuses on the matter sector of the theory. We call the resulting connection the {\it Higgs connection}.  The reason for this name will be clarified in due time.
As we have reiterated a few times, even connections which are based on only a subset of the fields turn out to be  valid connections on the whole field-space of the theory; this is a consequence of the co-rotation principle (see the very beginning of section \ref{sec:Singerconnection}).

Let us start by denoting by $\F_{\rm mYM}$ the field-space given only by the matter sector of Yang--Mills.
For definiteness, we assume that the matter fields $\Psi\in\F_{\rm mYM}$ transform  in the fundamental representation, i.e. $\Psi \mapsto g^{-1}\Psi$, or infinitesimally $\delta_\xi\Psi = -\xi\Psi$, and that the charge group is $G= {\rm SU}(N)$ (or ${\rm U}(1)$, when specified).
Generalizations to (non-special) unitary groups and other representations are straightforward in principle.
Since the charge group $G$ is unitary, $\xi$ is  antihermitian
\be
\xi^\dagger = - \xi.
\ee
Introducing the basis $\tau_a$ of $\fg$, we see that
\be
\xi = \xi^a(x)\tau_a
\qquad\text{with}\qquad 
\xi^a(x) \in \bb R 
\quad\text{and}\quad
\tau_a^\dagger = -\tau_a.
\ee
Consideration of special unitary groups corresponds to adding a tracelessness condition, $\tr(\tau_a)=0$. 

We will consider matter fields $\Psi$ which under Lorentz transformations transform either as scalars, or Dirac spinors. Other cases can in principle be treated with similar methods. In the following this distinction will be often unimportant, and we will generically omit the Lorentz indices (always assumed to be appropriately contracted). Similarly the distinction between bosons and fermions fields will often be unimportant, provided the right conventions are introduced. We will get to this point briefly.

As usual, we introduce the vector fields $\bb X = \int \bb X^I\tfrac{\dd}{\dd \Psi^I}\in{\rm T}\F_{\rm mYM}$ where $I$ is a multi-index covering the indices $m$ in the $G$ representation vector space $W\cong \bb C^N$ as well as the spatial point $x$ and possibly spinorial indices $\alpha$. Using similar notation as the previous sections, consider the following supermetric on $\F_{\rm mYM}$:
\be
\bb G^{\rm m}(\bb X, \bb Y) = \frac{1}{2} \int_\Sigma \bar{\bb X} \bb Y + \bar{\bb Y} \bb X 
\qquad\text{for}\qquad
\bb X, \bb Y \in {\rm T}_\Psi \F_{\rm mYM}.
\ee
Here, we have introduced the following unified notation for $\Psi$ and the components of $\bb X$ alike\footnote{In our conventions, $(\gamma^0)^\dagger = -\gamma^0$. This is compatible with $[\gamma^\mu,\gamma^\nu]_+ = 2\eta^{\mu\nu}=2(- + + +)$.}
\be
\bar\Psi = 
\begin{cases}
\phi^\dagger & \text{if $\Psi = \phi$ (scalar)}\\
\psi^\dagger\gamma^0 & \text{if $\Psi = \psi$ (Dirac spinor)}\\
\end{cases}
\ee
with $\gamma^\mu$ the Dirac matrices and $\cdot^\dagger$ denoting hermitian conjugation.
The matter supermetric can also be written as 
\be
\bb G^{\rm m}(\bb X, \bb Y) = \int_\Sigma \Re(\bar{\bb X} \bb Y) ,
\ee
with $\Re$  the real part; more explicitly $\bb G^{\rm m}(\bb X, \bb Y)$ is $\tfrac12\int(\bb X^\ast_m \bb Y^m + \bb Y^\ast_m \bb X^m )$ and has dimensions of inverse mass squared in the scalar case, or it is $\frac{1}{2}\int ({\bb X}_{\alpha m}^\ast \gamma^0 \bb Y^{\alpha m} + \bb Y_{\alpha m}^\ast \gamma^0  \bb X^{\alpha m})$ with dimensions of inverse mass in the spinorial one. In both cases, the asterisk stands for complex conjugation of the components.

The supermetric $\bb G^{\rm m}$ is invariant under the gauge flow and can therefore be used to deduce a field-space connection $\varpi$ by orthogonality to the gauge orbits.
Thus, at $\Psi$, the connection $\varpi$ is defined by the following relation that has to hold for any $\xi$ and $\bb X$:
\be
0 = \bb G^{\rm m}\big( \xi^\# , \hat H(\bb X) \big)
= \bb G^{\rm m}\big( \xi^\# , \bb X - (\fI_{\bb X}\varpi)^\# \big) 
=  \int \Re\Big[\big(-\bar{\xi\Psi}\big) \big(\bb X +  (\fI_{\bb X}\varpi)\Psi\big)\Big].  
\ee

Breaking the previous equation into components, and recalling that $\tau_a^\dagger = - \tau_a$, we obtain
\be
0 
= \frac{1}{2} \int \Big[\bar\Psi \tau_a \bb X - \bar{\bb X} \tau_a\Psi + \Big(\bar\Psi \tau_a\tau_b\Psi  + \bar \Psi\tau_b\tau_a \Psi \Big)(\fI_{\bb X}\varpi)^b\Big] \xi^a.  
\ee
From the arbitrariness of $\xi^a(x)$, we deduce a  local equation that must  hold for any $\bb X$:
\begin{align}
 \bar\Psi\tau_a \bb X - \bar{\bb X}\tau_a\Psi + \bar\Psi[\tau_a,\tau_b]_+\Psi (\fI_{\bb X}\varpi)^b = 0
\quad \text{for any} \quad \bb X \in{\rm T}_\Psi \F_{\rm mYM}.
\label{eq157}
\end{align}
Here, $[\tau_a,\tau_b]_+ :=\tau_a \tau_b + \tau_b\tau_a$ is the anticommutator.

Now, since the $\tau_a$ forms a basis of antihermitian matrices, and since $[\tau_a,\tau_b]_+$ is hermitian, we obtain
\be
[\tau_a,\tau_b]_+ = i D_{ab}{}^{c} \tau_c -\tfrac1N d_{ab}\mathbb 1,
\ee
where $\mathbb 1$ is the identity operator on $W\cong \bb C^N$, while $D_{ab}{}^{c}$ and $d_{ab}$ are matrices of real coefficients,\footnotemark~symmetric in the indices $(ab)$. Moreover, if $\{\tau_a\}$ forms an orthogonal basis such that $\tr(\tau_a\tau_b) = -\tfrac12 \delta_{ab}$, as in the previous sections, then $d_{ab}=\delta_{ab}$.%
\footnotetext{Tangentially, we note that $D_{ab}{}^c$ is the tensor entering the expression of the chiral anomaly, e.g. \cite{WeinbergQFT2}.}

Hence, introducing the matrix
\be
\mathcal{D}(\Psi)_{ab} := \bar\Psi [\tau_a,\tau_b]_+ \Psi,
\ee
 the defining equation for $\varpi$, equation \eqref{eq157}, can be written as
\be
 \bar{\bb X} \tau_a \Psi - \bar\Psi\tau_a\bb X
= \mathcal{D}(\Psi)_{ab} (\fI_{\bb X}\varpi)^b.
\label{eq911}
\ee

Suppose for now that $\mathcal D(\Psi)_{ab}$ is invertible---we will come back to this hypothesis shortly---and denote its inverse by
\be
\mathcal E(\Psi)^{ab}= (\mathcal D(\Psi)_{ab})^{-1}.
\ee
Thus, inverting equation \eqref{eq911} and eliminating $\bb X$ (recall that by definition $\fI_{\bb X}\dd \Psi^I = \bb X^I$), we obtain
\be
\boxed{\quad\phantom{\Big|}
\varpi = \mathcal E(\Psi)^{ab} \Big( (\dd \bar\Psi) \tau_a \Psi - \bar\Psi\tau_a (\dd \Psi) \Big)\tau_b.
\label{eq_varpi_matter}
\quad}
\ee

A particularly simple example of this situation is provided by (scalar) $G=\SU(2)$ Yang--Mills theory. 
In this case, $\tau_a = -\frac{i}{2}\sigma_a$, $d_{ab} = \delta_{ab}$ and $D_{ab}{}^c=0$, thus
\be
\varpi_{\rm \SU(2)} 
= \frac{i }{  \bar\Psi\Psi}\Big(   (\dd \bar\Psi)\sigma^a\Psi - \bar\Psi \sigma^a (\dd \Psi)  \Big)\tau_a.
\label{varpiSU2}
\ee

Allowing for non-special unitary groups, another simple example is that of (scalar) quantum electrodynamics (QED).
In this case, $G=\mathrm{U}(1)$, $d_{ab}=0$ and $D_{ab}{}^c$ is equal to 1---the only antihermitian generator of $\rm U(1)$ is $\tau = i$, which is `proportional to the identity'. Hence,
\be
\varpi_\text{QED} = \frac{1}{2\bar\Psi \Psi}\Big((\dd \bar\Psi)\Psi- \bar\Psi(\dd \Psi ) \Big).
\label{varpi_mQED}
\ee

Let us now consider the matter of the invertibility of $\mathcal D(\Psi)_{ab}$.
We focus on $\Psi$ as a scalar
 field,\footnote{If $\Psi$ is a Dirac spinor $\psi$, matters are actually more subtle. Indeed, $(\gamma^0)_{\alpha\beta}$, which defines $\bar\psi$, has null directions. Therefore, if $\psi^{\alpha m}$ is along one such null direction for all $m$, $\mathcal D(\psi)_{ab}$ vanishes, and so does $\bar \psi \psi$. Thus one has to require (at least) that $\bar \psi\psi\neq0$, for $\varpi$ to be well defined at $\psi$. Notice that $\bar\psi\psi$ is a bosonic quantity for which one can define a meaningful vacuum expectation value (vev). Notice also that, since $\gamma^5\gamma^0\gamma^5 = -\gamma^0$ and since the null space of $\gamma^0$ is 2-dimensional, $\bar\psi\psi$ identically vanishes precisely when the Dirac spinor is chiral, i.e. purely right- or left-handed. However, putting aside the points in field-space where $\bar\psi\psi=0$, one reaches the same conclusions as for the scalar field. The relevance of this remark will become clear later. \label{ftnt43}} $\phi$.
Putting aside the obvious case in which $\phi=0$, it is clear that $\mathcal D(\phi)_{ab}$ defines a negative semidefinite metric on $\fG$, which is degenerate precisely in the directions which annihilate $\phi$.
In other words, $\xi^a\mathcal D(\phi)_{ab}\xi^b=0$ if and only if $\xi\phi=0$, i.e. if and only if $\xi\in {\rm Lie}(\mathcal S_\phi)$ where $\mathcal S_\phi \subset \G$ is the  stabilizer of $\phi\in W$.
Indeed, it is clear that in these directions equation \eqref{eq911} reads $0=0$.
Generically, the stabilizer of a vector in $W$ is non trivial. For example, for $G=\SU(N\geq3)$, we have $\mathcal S_\phi \cong  C^\infty(M,S_o)$, with $S_o\cong\SU(N-1)\subset G$ the stabilizer of some non-vanishing reference vector $v_o\in W\cong \bb C^N$.
Note also that unlike the action on the gauge field, the action on the matter field is pointwise. Consequently, the stabilizer for a field configuration $\varphi$ is an {\it infinite} dimensional group. We will come back on this point later.

From this discussion---assuming matter in the fundamental representation and $G$ (special) unitary---we conclude that  {\it $\mathcal D(\phi)_{ab}$ is invertible if and only if $\phi$ does not vanish and the fundamental representation of $G$ is free}, i.e. if and only if $G=\SU(N=2)$ or $G = {\rm U}(1)$.

The fact that $\varpi$ is left undetermined in the directions belonging to the stabilizer should have been expected from the simple fact that a field insensitive to some transformations cannot be used as a reference to measure those very changes. This is made explicit when one tries to write $\varpi$ in an adapted choice of field-space  coordinates.

Let us start by coordinatizing the field $\phi$  as 
\be
\phi = \phi(h,\rho) = \rho h v_o
\qquad
\text{where} 
\;
h \in  C^\infty(\Sigma,G)
\;\text{and}\;
\rho \in  C^\infty(\Sigma,\mathbb R),
\label{eqcoord}
\ee
and $v_o\in V$ is some non-vanishing reference vector in $W\cong \bb C^N$, the fundamental representation of $G=\SU(N)$.
Note that the reference vector $v_o$ is chosen once and for all, i.e. it is not a field-space coordinate, hence 
\be
\dd \phi = (\dd \ln \rho) \phi + (\dd h h^{-1}) \phi .
\ee
Using $h^{-1} =h^\dagger$, as well as the definition of $\mathcal D$ and $\mathcal E$ above---assuming for now the invertibility of $\mathcal D$---we readily obtain through equation \eqref{eq_varpi_matter} the following coordinate expression for the Higgs connection:\footnote{Proof: from equation \eqref{eq_varpi_matter} and the expression for $\dd \phi$,  
\begin{align}
\varpi 
&= \mathcal E(\phi)^{ab} \Big[ \Big( (\dd \ln \rho) \phi +(\dd h h^{-1}) \phi \Big)^\dagger \tau_a \phi - \phi^\dagger\tau_a \Big( (\dd \ln \rho) \phi + (\dd h h^{-1})\phi \Big) \Big]\tau_b \notag\\
& =  -\mathcal E(\phi)^{ab}  \Big( \phi^\dagger [\tau_c  ,\tau_a ]_+ \phi \Big)\;(\dd h h^{-1})^c \tau_b 
 =   -\mathcal E(\phi)^{ab}\mathcal{D}(\phi)_{ca} (\dd h h^{-1})^c \tau_b 
 = - \dd h h^{-1}\notag.
\end{align}
}
\be
\boxed{\quad\phantom{\Big|}
\varpi = - \dd h   h^{-1}.
\quad}
\label{varpim_flat}
\ee
This proves that for free representations, for which $\mathcal D$ is guaranteed to be invertible, $\varpi$ is flat wherever defined, i.e. for those configurations in which $\phi$ does not vanish anywhere. Note that although $\varpi$ can be expressed in coordinates that rely on an arbitrary reference vector $v_0$, $\varpi$ itself is independent of $v_0$. 

Now, let us go back to the problem of understanding what happens when the representation is not free, e.g. for $G=\SU(N\geq3)$. In this case equation \eqref{eqcoord} fixes $h$ only up to an element of ${\cal S}_o$, the stabilizer of $v_o$, i.e.
\be
h \sim hs
\qquad 
s\in {\cal S}_o\cong{\cal C}^\infty(M,\SU(N-1)).
\ee
And therefore, also $\varpi$---if defined {\it through} equation \eqref{varpim_flat}---is defined only up to the following transformations:
\be
\boxed{\quad\phantom{\Big|}
\varpi \sim \varpi - \Ad_{h}\dd s s^{-1}.
\label{upto}
\quad}
\ee
Clearly, $ \Ad_{h}\dd s s^{-1} \in {\rm Lie}({\cal S}_\phi)$ and thus the present ambiguity is consistent with the difficulties in inverting equation \eqref{eq911}.

If we tried to formalize this state of affairs, we would say that $\phi$ is coordinatized by $h\sim hs$ is in the right coset ${\cal S}_o\setminus \G$. But since these cosets are generically not groups themselves, the corresponding expression for $\varpi$ cannot be directly made sense of. 
In other words the Lie bracket of $\fg$ does not close among elements of $\fg/{\rm Lie}(S_o)$, where the quotient is taken in the sense of vector spaces.

Physically, this state of affairs simply means that a connection made out of a field which is stabilized by a subgroup of the gauge transformations can only `detect' and `compensate  for' that part of the transformation that modifies it, while being totally transparent to the part of transformation which stabilizes it.

We will discuss these matters and their physical interpretation in further detail in the remarks below.


\subsection{Remarks on section \ref{sec:matter}\label{rmks:matter}} 

\paragraph*{\indent(i) On the choice of the supermetric $\bb G^{\rm m}$ and chirality ---}
In the case of scalar fields the choice of the supermetric $\bb G^{\rm m}$ can be justified in precise analogy to the gauge supermetric: it is the same supermetric that contracts the quadratic kinetic term appearing in the Lagrangian (there the vectors to be contracted are the velocities $\bb V = \int \dot \phi \frac{\dd }{\dd \phi}$). 
This idea fails in the case of the fermionic action, which is first order. In this case to single out the supermetric $\bb G^{\rm m}$ one has to appeal to the linear nature of the space of the fermion fields, which allows one to identify vectors and configurations. The supermetric is then analogous to the mass term in the Dirac Lagrangian. 
Interestingly, this term is not available for chiral fermions (see footnote \ref{ftnt43}), which therefore need to be addressed with different methods and might hold interesting surprises. An idea in this direction is to use as a supermetric the gradient term of the chiral fermionic Lagrangian, $\bb G\sim \int \Re(\bb X^\dagger \gamma^i\pp_i \bb Y)$; clearly, this supermetric fails to be ultralocal. Is there any viable ultralocal choice for chiral fermions? If not, with what consequences (see e.g. the next point)? We leave these questions to future investigations.

\paragraph*{\indent(ii) Boundaries ---} Crucially, the matter-field connections are solved for {\it locally} in spacetime, and contain no derivatives. This is of course due to the fact that matter fields live in a representation of $G$ and their gauge transformations do not involve derivatives of the group elements.
It also means that the expression for $\varpi$ is completely unaltered in the presence of boundaries.
The horizontal modification of the symplectic structure advocated for in section \ref{sec:Noether_hor}, if based upon the Higgs connection, would then involve boundary-local terms only---this is in contrast to what happens if the Singer--DeWitt connection is used (cf. point {\it (i)} of the \hyperref[rmk:Noether_hor]{\it Remarks on section \ref*{sec:Noether_hor}}).  See also point {\it (viii)} below.

\paragraph*{\indent(iii) Non-existence of $\varpi$ at vanishing matter field configurations ---} The above formulas show that at any $x_o$ where $\phi(x_o)=0$ the connection $\varpi$ is not defined. Moreover, at these points it cannot be uniquely defined by continuity either. 
As explained in footnote \ref{ftnt43}, for Dirac spinors one requires the stronger condition $(\bar\psi \psi)\neq0$ everywhere.
Therefore, there are areas of field-space where the Higgs connection is not defined, implying that it is not the best tool to explore the global features of field-space. Nevertheless, its flatness is a property that makes this connection an appealing tool whenever it is available. 
In particular, this is appealing at a perturbative level around points in field-space where $\phi\neq0$, or $(\bar\psi \psi)\neq0$ everywhere on $\Sigma$. In a quantum parlance one would speak of backgrounds in which the vacuum expectation value (vev) of $\phi$ or $(\bar\psi\psi)$ does not vanish anywhere. 
These are symmetry-broken configurations.

\paragraph*{\indent(iv) Higgs connection, broken phases, and unitary gauge ---}
Perturbatively around a field-space point in which $\phi\neq0$, physics is described by a spontaneously broken Higgs phase. 
Whether this phase can be reached dynamically or not depends on the details of the Hamiltonian, e.g. the shape of the scalar field potential $V(\phi)$, and of the renormalization flow.
Independently of this fact, by perturbing the kinetic term of the (minimally coupled) gauge-invariant Lagrangian around one of these configurations, one readily sees that $\mu_{ab} = - \tfrac12 \mathcal D(\phi)_{ab}$ is the matrix of masses of the gauge vector bosons in the broken phase.
Thus, as well known, the presence of stabilizing directions for $\phi$ in $\mathcal G$ corresponds to the presence of `residual' massless gauge vector bosons, associated to the unbroken gauge symmetries. 
In particular, the variable $h$ introduced above corresponds to the Goldstone mode of the broken symmetry, and the gauge-fixing to which the (flat!) $\varpi=-\dd h h^{-1}$ corresponds to is the celebrated `unitary gauge' \cite{Weinberg:1973ew,WeinbergQFT2}. 

In the next three remarks, we will build up on this interpretation of the Higgs connection in terms of broken Higgs phases, to provide a physical counterpart to the charge `screening' mechanism mathematically described by the vanishing of the horizontal boundary charges---equation \eqref{hor_charge_gauge}.
There, we will also provide a physical explanation for the case in which some charges fail to be screened when the stabilizer $\mathcal S_\phi$ is nontrivial.

\paragraph*{\indent(v) Higgs phases: condensate, screening, and charges ---}
 First, consider the case in which $\phi$ is in a free representation and all boundary charges vanish---equation \eqref{hor_charge_gauge}.
We have emphasized that $\varpi$ is not defined at configurations in which $\phi$ vanishes at some points. On the other hand, configurations in which $\phi$ takes a non-vanishing vev are known as `condensates'. These are  configurations that arise in particular physical conditions (phases) which induce the proliferation of a certain type of particles---these phases are described by states which contain an indefinite number of $\phi$-particles. Since these particles are charged,  they  can provide a physical picture for the screening mechanism manifest in the vanishing of the horizontal Noether charges. 

This is in agreement with the fact that the vector bosons associated to the broken gauge symmetry acquire a non-vanishing mass $\mu$ (see point {\it (iv)}) and as a consequence the interaction they mediate is not long-range in the broken phase (Yukawa potential). 
This affects the physical significance of the Gauss law {\it in the broken phase}, a fact that turns out to be automatically taken into account if one appeals to the `horizontal symplectic charges' (equation \eqref{hor_charge_gauge}) defined via the Higgs connection: in a completely broken phase all gauge bosons are massive (invertibility of $\mathcal D$) and all horizontal charges vanish.

\paragraph*{\indent(vi) Two examples of broken phases --- }
 A prototypical example of a condensation in high energy physics is that of the Higgs field. This is an $\SU(2)$ charged field with non-vanishing vev (in the portion of configuration space relevant to our universe). Using the Higgs field as a reference for the $\SU(2)$ gauge symmetry leads to a flat connection of the form \eqref{varpiSU2} wherever it is defined. Being  flat, this connection can be locally associated to a gauge fixing in those regions of $\F$. Once again, this is indeed the famous unitary gauge.
 
Another beautiful example of  this situation is given by the experimental setup suggested by Susskind to detect the electromagnetic memory effect proposed by Strominger and collaborators \cite{Susskind:2015hpa, strominger2018lectures}. There, the phase of the electromagnetic field is measured with respect to the phase of a superconductor's wave-function. In our language, this would imply the construction of the $G={\rm U}(1)$ Higgs connection out of this wave-function, as in equation \eqref{varpi_mQED}. Note that superconductivity has an effective description in terms of a symmetry breaking mechanism \cite{WeinbergQFT2}. This example emphasizes the relation of $\varpi$ and the choice of a measurement device, or reference frame. We comment further on this perspective at points {\it (viii)} and {\it (ix)}.

\paragraph*{\indent(vii) Non trivial stabilizer, $\mathcal S_\phi \neq \{\rm id\}$ ---}
If the stabilizer of $\phi$ is non-trivial, we have seen that the Higgs connection is mathematically not defined. Nonetheless, we have seen that it is morally valued in the  (coset) vector space $\fG/{\rm Lie}(\mathcal{S}_\phi)$ defined by the equivalence relation $\xi\sim\xi+\sigma$, $\sigma \in {\rm Lie}(S_\phi)$---this does not generally define a Lie sub-algebra of $\fG$.
In the Higgs-phase interpretation of this situation, this means that the symmetry ${\mathcal S}_\phi$ is still unbroken. In turn, this means that the vector bosons associated to this unbroken subgroup of symmetry are still massless: therefore they mediate  long-range interactions whose corresponding charges are not screened by the condensate.
It is then natural to deal with these unbroken symmetries as we did for the standard gauge symmetries in the previous section: we can compensate for them through a Singer--DeWitt connection valued in  ${\rm Lie}(\mathcal S_\phi)$. This would leave open the prospect of global non-vanishing charges associated to the unbroken gauge symmetries while eliminating local, `pure gauge', charges. We postpone the study of combinations of `partial' $\varpi$'s to future work.

\paragraph*{\indent(viii) Higgs connection \textsc{vs.} new boundary degrees of freedom ---}
Taking the horizontal symplectic potential $\theta_H$ with respect to the Higgs connection gives a result formally equivalent to that commonly used when `covariantization' of $\theta$ is obtained through the introduction of new group-valued degrees of freedom, sometimes called `edge modes', as in \cite{Balachandran:1994up,Donnelly:2016auv,Geiller:2017xad,Geiller:2017whh,Speranza:2017gxd}---see point {\it (i)} of \hyperref[rmks:general_YM]{\it Remarks on section \ref*{sec:general_YM}}. 
However, it is important to notice that we get at this result from a very different starting point, since $h$ is here not a new field, but a coordinate on the matter field sector.
In other words, we have shown that our construction bypasses the introduction of group-valued edge modes as new compensating degrees of freedom as in \cite{Donnelly:2016auv}, by obtaining a formally equivalent result in presence of a condensate of more standard matter fields---provided they live in a free representation of the charge group. From this view point, the group-valued edge modes {\it precisely} arise as the Goldstone modes of a spontaneously broken gauge symmetry.\footnote{See also \cite{Attard2018} for another mathematically formal way to relate spontaneous symmetry breaking, gauge-fixings, and dressings.}

\paragraph*{\indent(ix) Higgs connection \textsc{vs.} quantum reference frames ---}
The Higgs connection uses the matter fields already present in the theory as a `phase' reference frame. 
Prototypical examples are given at point {\it (vi)}, where we discussed e.g. how the phase of a superconductor's wave function can be used as a reference frame in the electromagnetic case \cite{Susskind:2015hpa}.
As we have already stated, this translates into the construction of a (matter) connection out of the superconductor's wave-function (equation \eqref{varpi_mQED}).
This suggests an interpretation of  $\varpi$ as a choice of {\it (quantum) reference frame} \cite{aharonov1967:obs, Aharonov1967SSR, Bartlett:2007zz, AharonovBook}.
This interpretation could also explain the inherent ambiguity in the choice of a certain $\varpi$ over another, e.g. of the Higgs connection over the SdW connection, in terms of a choice of {\it experimental} setup:
given the phase the system finds itself in, the observables of interest, and the measurement apparatus of choice, one specific connection-form is singled out.\\

 Suggestively, the first---to the best of our knowledge---discussion of quantum reference frames arose from challenging the notion of superselection rules (SSR) introduced in \cite{wick1952,WWW1970}, and in particular, that of the charge SSR.
In turn, using the algebraic approach to quantum field-theory, the charge SSR can be deduced from the Gauss constraint and the Gauss law at asymptotic boundaries \cite{StrocchiCSR1974, StrocchiErratum}. If made rigorous, this  string of speculative relations would close the circle of ideas presented in this paper: gauge invariance,  Gauss law, non-localities and boundaries are all encompassed in our notion of  field-space connection-forms. Connection-forms are naturally associated to material (quantum) reference frames, and then, from the work of Aharonov and Susskind, these are related to superselection rules, and finally, from the results of Strocchi and Wightman, we get back to the Gauss law again.

\section{A $\varpi$ for the Lorentz gauge symmetry of vielbein gravity}\label{sec:EC}

A case similar to those considered in the previous section is that of Lorentz symmetry in vielbein, or Einstein--Cartan, gravity.
This is a first-order theory in which the spacetime metric field is replaced by a vielbein $e$ and a spin connection $\omega$. These fields are independent and, roughly speaking, canonically conjugate to each other. 
In particular, the vielbein $e$ represents a local inertial frame and its relation to the metric is given by
\be
g_{\mu\nu} = \eta_{IJ} e^I_\mu e^J_\nu,
\ee
where $\eta_{IJ}$ is the flat Minkowski metric, while $\omega$ is the spin connection which on-shell (and in absence of sources of torsion) essentially reduces to the Levi-Civita connection. 

As the contraction to the Minkowski metric above suggests, the indices $I,J,\dots$ carry a representation of the Lorentz group. Under these transformations, the vielbein transforms homogeneously, and the spin-connection transforms---of course---as a connection:
\be
(e,\omega) \mapsto (\Lambda^{-1} e, \Lambda^{-1} \omega \Lambda  + \Lambda^{-1}\d \Lambda ),
\ee
where $\Lambda^I{}_J\in{\rm SO}(1,d)$ is a matrix element of the Lorentz group.

The Einstein--Cartan Lagrangian is
\be
L_{\rm  EC}= \epsilon_{IJKL}e^I\wedge e^J \wedge  F^{KL}[\omega],
\ee
and the associated presymplectic potential
\be
\theta_{\rm  EC} = \epsilon_{IJKL}e^I\wedge e^J \wedge  \dd \omega^{KL}.
\ee
With respect to the Lorentz transformations, $L_{\rm  EC}$ satisfies all the properties required in the analysis of section \ref{sec:general_Noether}. Hence, $\theta_{\rm EC}$ admits a horizontal version with the usual properties. The similarities with Yang--Mills theory are evident.

The goal of this section is to build a Higgs connection for these Lorentz transformations based on the vielbein field. For this, we choose the obvious metric on the space of vielbeins, 
\be
\bb G^e(\bb X, \bb Y) = \int_\Sigma  \sqrt{g_\Sigma} g^{\mu\nu} \eta_{IJ} \bb X^I_\mu \bb Y^J_\nu.
\ee
where, as customary by now, $\bb X_e = \int \bb X^I_\mu(x) \frac{\dd}{\dd e^I_\mu(x)}$, and where we have introduced $g_\Sigma$ to be the determinant of the pull-back of the spacetime metric to the region $\Sigma$ (in this case, $\Sigma$ can actually be chosen to be a spacetime region without further difficulties).

Proceeding as in the previous section, one deduces the following {\it vielbein field-space connection} for the Lorentz symmetry\footnote{Recall the  standard notation $e^{\mu I} := \eta^{IJ}(e^{-1})^\mu_J$.}
\be
\varpi^{IJ} = e^{\mu [I} \dd e^{J]}_\mu.
\label{varpiEC}
\ee
Here, we have slightly  departed from the notation used in the previous sections by denoting directly the matrix elements of $\varpi$ as  elements of $\mathfrak {so}(1,d)$. 
The horizontal field-space derivative of the vielbein becomes essentially the field-space derivative of the metric:
\begin{align}
\dd_H e_\mu^I = \frac12 \dd g_{\mu \nu} e^{\nu I}.
\end{align}
In this equation, $g_{\mu\nu}$ is  just a placeholder for $\eta_{IJ}e^I_\mu e^J_\nu$ and its differential appears simply because $\dd_H e$ must be horizontal, i.e. (very) roughly speaking `the $\dd$ of something gauge-invariant.'

Interestingly, the curvature of $\varpi^{IJ}$does not vanish. It can be obtained from the field-space metric via equation \eqref{eq:curvature_and_metric}, or by using $\dd_H \dd_H e_\mu^I= - \delta_{\fF} e_{\mu}^I = \fF^{IJ} e_{\mu J}$. The result is
\begin{equation}
	\bb F^{IJ} = -\frac14 e^{\mu I}e^{\rho J}g^{\nu \sigma}  \dd g_{\mu \nu}  \dd g_{\rho \sigma} .
\end{equation}

We conclude by providing the horizontal potential for Einstein--Cartan gravity for the vielbein connection (cf. the section \ref{sec:YM} on Yang Mills theory):
\begin{align}
\theta_{{\rm EC}, H} &=  \epsilon_{IJKL}e^I\wedge e^J \wedge  \dd_H  \omega^{KL}\notag \\
& \approx \theta - \d \big(\epsilon_{IJKL} e^I\wedge e^J  e^{\mu K}\wedge \dd e_\mu^L\big).
\label{thetaECH}
\end{align}
where $\approx$ means on-shell of the `Gauss' constraint $\d e^I + \omega^I{}_J \wedge e^J \approx 0$---this is in fact the torsionless constraint that, in absence of fermions, reduces the spin connection $\omega$ to the Levi-Civita connection.

\subsection{Remarks on section \ref{sec:EC}}

\paragraph*{\indent(i) Well-definedness of the vielbein connection ---} As for the Higgs connections defined in the previous sections, also the vielbein connection is only defined when the reference field does not `vanish', or better, when the reference field (here a interpreted as a map $e:{\rm T}_xM\to \bb R^d$) is non-degenerate and thus invertible. The invertibility of the vielbein field is a necessary and sufficient condition (in absence of actual matter degrees of freedom) for Einstein--Cartan gravity to be equivalent to Einstein's metric General Relativity. If the vielbein $e$ is not invertible, the metric $g$ is degenerate, and no spacetime interpretation is available. In a quantum theory of gravity based on these variables, classical spacetimes arise in a broken phase, where the vev of $e$ is nonvanishing.

\paragraph*{\indent(ii) The vielbein connection in the literature ---} 
The vielbein connection constructed here has already been implicitly used in various publications concerned with removing the `redundant' Lorentz degrees of freedom from a vielbein formulation of certain gravitational problems. 
In \cite{Jacobson:2015uqa}, the authors constructed a  modified Lorentz-invariant Lie derivative on spacetime, with the goal of recovering the `usual' first law of black-hole mechanics in the first order formalism, through a construction \`a la Wald \cite{wald1993black, Iyer:1994ys}. This required eliminating polluting contributions coming from the interplay between diffeomorphisms and Lorentz transformations.
 In our language, their Lorentz-Lie derivative ${\cal K}_X$, for  $X\in\mathfrak{X}^1(M)$ an infinitesimal diffeomorphism, implements precisely the action of the horizontal component of the lift of the action of $X$. In formulas, 
\be
{\cal K}_X \varphi \equiv \hat H(X^\#)\varphi = \Big(X^\# - (\fI_{X^\#}\varpi)^\#\Big) \varphi =  \pounds_X \varphi - \big(e^{\mu [I}\pounds_X e^{J]}_\mu \big)^\#\varphi,
\ee
where the last term denotes  the appropriate action on $\varphi$ of the infinitesimal Lorentz transformation $\lambda^{I J} = \fI_{X^\#}\varpi^{IJ} = e^{\mu [I}\pounds_X e_\mu{}^{J]}$---this follows from $\fI_{X^\#}\dd e = \pounds_X e$.

Motivated by the more basic question of the equivalence between the vielbein and Einstein--Hilbert formulations of gravity, the authors of \cite{DePaoli:2018erh} constructed  a Lorentz-invariant presymplectic potential for the Einstein--Cartan action.  Their modification of the symplectic potential is precisely $\theta_{H}$, where horizontality is taken with respect to the vielbein connection of equation \eqref{varpiEC}. This allowed them to solve the problem raised in \cite{Jacobson:2015uqa} without having to introduce the modified Lorentz-Lie derivative ${\cal K}_X$. As far as the Noether charges are concerned, the equivalence of the two constructions follows simply from $\fI_{\hat H(X^\#)}\theta \equiv \fI_{X^\#} \theta_H$.

Finally, the authors of \cite{DePaoli:2018erh} also considered the Holst modification of the Einstein--Cartan action. This is obtained by replacing in all the expressions above $\epsilon_{IJKL}\mapsto P_{IJKL}:= \big( \epsilon_{IJKL} + \gamma^{-1}\eta_{II'}\eta_{KK'} \delta^{I'}_J\delta^{K'}_L\big)$, with $\gamma\in \mathbb R$ the Barbero--Immirzi parameter. This extra term ends up producing an extra boundary contribution to $\theta_H$, i.e. $\gamma^{-1}\d(e^I\wedge \dd e_I)$ whose consequence for the ensuing boundary  theory where studied by \cite{Freidel:2016bxd, Freidel:2015gpa}, albeit in a context where the Lorentz ${\rm SL}(2,\bb C)$ symmetry is partially gauge fixed to give an $\SU(2)$ symmetry.
Here, we showed {\it why} the  constructions of \cite{Jacobson:2015uqa,DePaoli:2018erh} work, and embedded them in a much wider context.
In particular, it would be of interest to understand in this context which consequences ensue from the non-vanishing of the vielbein connection's curvature, $\fF$.

\section{The relation between $\varpi$ and dressings}\label{sec:dressing}

The aim of this section is threefold: first, we use $\varpi$ to construct gauge-invariant combinations of fields, that we will call `dressed'; second, we discuss the relation between these fields an the horizontal symplectic geometry of section \ref{sec:Noether_hor}; and finally, we explain the relationship of the ensuing dressed fields to other constructions of gauge-invariant field combinations already present in the literature (Lavelle and McMullan \cite{Lavelle:1995ty}, Gribov--Zwanziger \cite{vanBaal:1995gg, Capri2005, Vandersickel:2012tz}, and Vilkovisky \cite{Vilkovisky:1984st, vilkovisky1984gospel}). 
We conclude by sketching a proposal which is natural in our framework and new to the best of our knowledge. We name it `historical dressing'.

Our starting point is Dirac's idea that a `bare' (or `Lagrangian' \cite{Lavelle:1995ty}) electron field $\psi$---which is gauge-{\it variant}---has to be {\it dressed} in order to be promoted to a physical field \cite{Dirac:1955uv}.
Dressing means attaching to $\psi$ an electromagnetic cloud whose gauge transformation compensates that of $\psi$. 
This means building composite fields $\hat\psi:=\psi^h$, where $\psi$ is the bare electron and $h=h(A)$ a field-dependent equivariant gauge transformation, i.e. an element of $\G$ such that 
\begin{equation}\label{eq:h_transfo}
R_g^* h(A) = h(A^g) = g^{-1}h(A) \quad \text{for all }g\in\G.
\end{equation} 
If this equation is satisfied, the dressed electron is gauge-invariant: $\hat \psi = \hat{\psi^g}$. 

We call the field-dependent gauge transformations $h$, the {\it dressing factors}. The Dirac dressing factor is 
\be
h_\text{Dirac}(x) := \exp\left( -i\int \frac{\d^3 y}{4\pi} \frac{(\pp^i A_i)(y)}{|x-y|} \right).
\ee
Once a dressing factor is found, it can be used to dress any gauge-variant field, including the gauge potential itself.
An important feature of Dirac's construction is that $h(A)$ does not transform under global (`Killing') gauge transformations, so that $\hat \psi$ is indeed a charged object. 
Generalizations to the non-Abelian case are nontrivial. They are the main subject of this section.%

Recognizing the similarity of $\log h_\text{Dirac}$ and the Abelian SdW connection,\footnotemark~we complement Dirac's idea with the observation that $\varpi$ is precisely an infinitesimal, field-dependent, covariant gauge transformation: it is valued in $\fG$, depends on the field-space point and linearized fields, and transforms covariantly as expressed by equation \eqref{varpi_dependent}. 
Of course, $\varpi$ is a field-space 1-form, and can be turned into an element of $\G$ only once exponentiated {\it and} integrated over a line. 
This leads to our proposal of {\it dressings as field-space Wilson lines}.
\footnotetext{In Dirac's construction it is also important that the electric field created by $h(A)$, see below, is the Coulomb field of the electron. In classical terms, this means that the Poisson bracket of $E$ with $\hat \psi$ is $ \{ E(x) , \hat \psi(y) \} = \frac{-1}{4\pi (x-y)^2}\hat \psi(y)$. Notice that this is a Poisson bracket between gauge-invariant quantities. It is most easily computed in temporal gauge; in Coulomb gauge, in which $h_\text{Dirac}\equiv\rm id$, one needs to first introduce non-local Dirac brackets. The supermetric appearing in the kinetic term of the Lagrangian governs both the definition of canonical momenta and, by Lorentz covariance, also the structure of the Gauss constraint. It is not difficult to convince oneself  that the relationship between the SdW connection and the supermetric is what guarantees  that the Poisson bracket $\{E, \hat \psi\}$ gives the expected result.}

The Wilson line proposal involves an {\it extended object in field-space}---the line $\gamma$. This starts at a reference configuration, $\varphi^\star\equiv\star$, and ends at the configuration of interest, $\varphi$. The reader familiar with background-field methods, or with Vilkovisky and DeWitt's geometric improvements thereof, will not find the dependence on a base point too surprising \cite{Kunstatter:1991kw, DeWitt:1995cx, DeWitt_Book, Branchina:2003ek, Pawlowski:2003sk}. 
The possibility of an actual, full fledged, {\it field-space path} dependence might seem more troubling. We will discuss this in detail.

Before moving to the technical discussion, we spend a word on the `inevitability' of the Wilson-line scenario for dressings. Consider the fundamental formula $\hat \psi = \hat{\psi^g}$: it refers to two field configurations, $\psi$ and $\psi^g$, which correspond to two different points in the field-space $\Phi$.
If $\F$ is generic and hence does not admit a flat supermetric, these two points are a priori incomparable---{\it unless they are transported to the same point along some path in field-space}. This is the meaning of the dressing operation, $\hat\cdot\,:$ a transport operation along a path in field-space.

\subsection{Dressings as Wilson lines}

The Wilson-line dressing requires a path in field-space that ends at the configuration to be dressed, $\varphi$, and starts at a reference configuration configuration fixed once and for all. We label this configuration by a star,
\be
\text{reference: } \quad \varphi^\star \equiv \star.
\ee
Often, $\varphi^\star$ is implicitly chosen to be the configuration of vanishing fields.

\begin{figure}[t]
	\begin{center}
		\includegraphics[scale=.17]{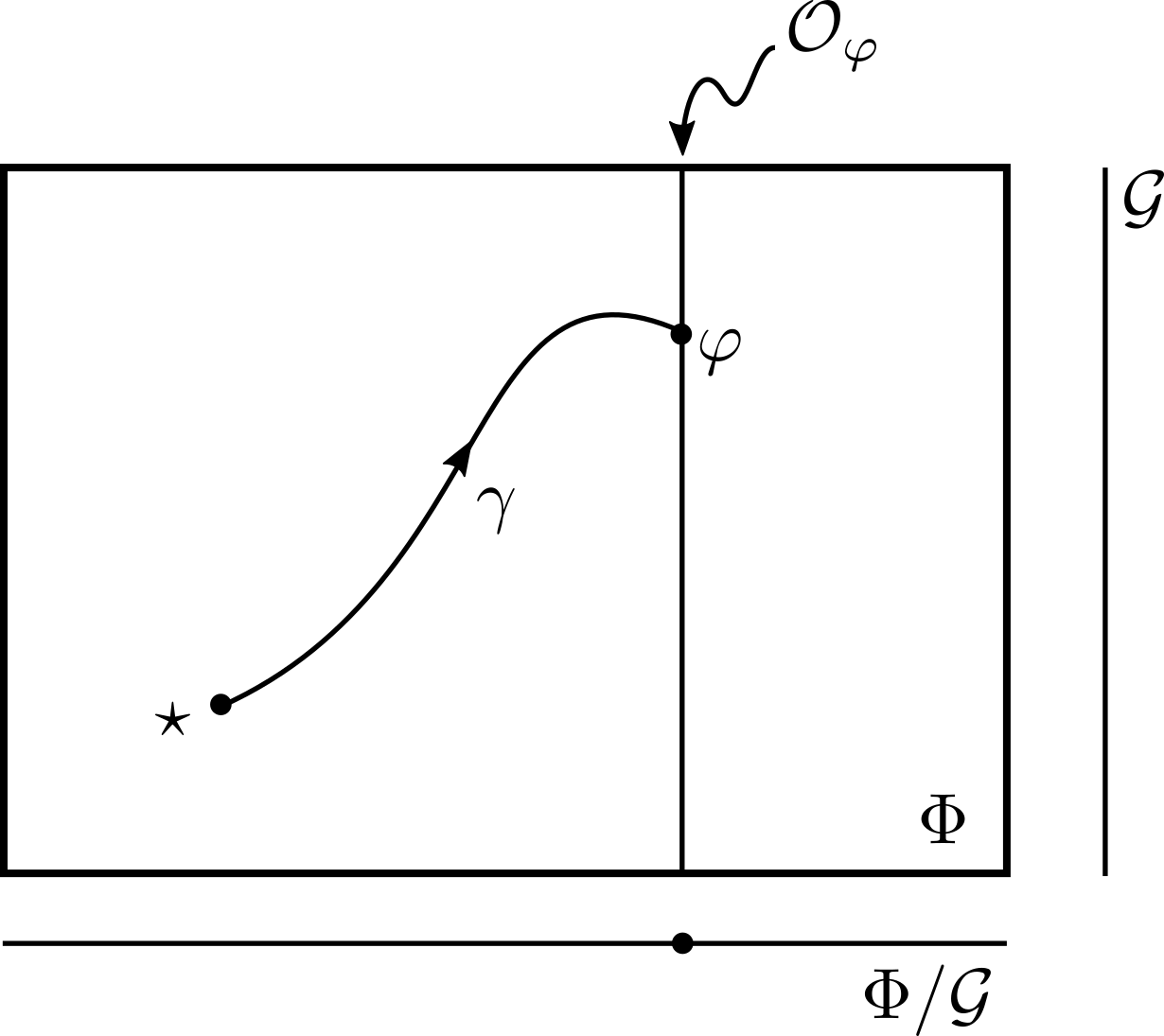}
		\caption{The dressing path $\gamma$ from $\star$ to $\varphi$.}
		\label{fig6}
	\end{center}
\end{figure}

Now, define the field- and {\it path}-dependent gauge transformation $h\in \G$ by 
\be
\label{gstar}
{h}(\gamma;\varphi) = \mathbb{P}\exp \left(\fint_{ \varphi\stackrel{\gamma}{\leftarrow}\star} - \varpi\right) .
\ee
Let us call $h(\gamma;\varphi) $ as the {\it dressing factor of $\varphi$ along $\gamma$}, where $\gamma$ is a path in field-space that starts at $\star$ and ends at $\varphi$---i.e.  $\gamma:[0,1]\to \F$, such that $\gamma(\tau=0)=\star$ and $\gamma(\tau=1)=\varphi$. 
The dressing $h$ is a non-local function on field-space that depends on the chosen path $\gamma$ connecting $\star$ to $\varphi$, figure \ref{fig6}.

The notation  $\mathbb{P}\exp \fint_{ \varphi\stackrel{\gamma}{\leftarrow}\star} $ explores our \lq{}double-struck\rq{} theme for field-space objects; and stands for a path-ordered integral along $\gamma\subset\F$.
As the arrow indicates, composition is from the right.\footnote{The double struck notation does not stand for a multiple integral, but for an integral along a 1-dimensional line embedded in field-space---rather than spacetime. Equivalently, the dressing factor is the solution, evaluated at time $\tau=1$, of the following ODE ($\bb V=\dot \gamma$ is the tangent vector to $\gamma$):
	$$
	\tfrac{\d}{\d \tau} h({\gamma(\tau)})h(\gamma(\tau))^{-1}= - (\fI_{\bb V} \varpi)_{|\gamma(\tau)} 
	\quad \text{and} \quad
	h(\gamma(\tau=))=\rm id.
	$$
	\label{Pexp}
}

Before studying the gauge properties of the so-defined dressing factors, we pause for a moment and consider a couple of simple examples that help elucidate the formalism.

\paragraph*{Example 1: Dirac dressing}
~

We illustrate the general idea by writing the Dirac dressing of the electron as a Wilson line \cite{Gomes:2018shn}. 
Now, in Maxwell theory on flat space, the SdW connection takes the simple form (equation \eqref{YM_varpi_boundaries})
\be
\varpi_{\rm Abelian} = -i \pp^{-2} \pp^i \dd A_i,
\ee
where the only generator $\tau_a = -i$ of the $\rm U(1)$ Lie algebra has been made explicit.

In the Abelian case, the dressing factor is easily computed.
Firstly, since $\bb F_\text{Abelian}=0$---equation \eqref{Singercurvature}---the path dependence of $h(\gamma;A)$ is trivial. 
Therefore we can simply choose $\gamma$ to be the affine path connecting the reference configuration $A^\star=0$ to the target configuration $A$,
\begin{align}
\gamma_\text{affine}(\tau) = (1-\tau)A^\star + \tau A = \tau A, \qquad \tau \in [0,1].
\end{align} 
Secondly, thanks to the Abelian nature of $\fg$, the path ordering can be dispensed of.
Therefore, using the affine structure of field-space to define $\Delta A$ as a tangent vector at $A^\star=0$ with components equal to $A-A^\star=A$, we immediately find 
\be
h_{\rm Abelian}= \exp \big(-\fI_{\Delta A}\varpi_{\rm Abelian}\big) = \exp \big(i\partial^{-2} \partial^i A_i\big).
\label{forAblimit}
\ee
Now, if $\Sigma=\bb R^3$ and fast decaying boundary conditions are fixed at infinity, substituting the Green's function for $\pp^{-2}$ we readily recognize the Dirac dressing
\be
h_{\rm Abelian}(x) = \exp\Big(- i\int   \frac{\d^3 y}{4\pi}  \frac{(\partial^i A_i)(y)}{|x-y|}\Big) = h_\text{Dirac}(x).
\ee

\paragraph*{Example 2: Affine dressing}
~

In the non-Abelian case, $\fF \neq 0$, and the choice of path becomes relevant.
For now, we choose affine paths simply as an illustrative example of the formalism: crucial limitations---to be discussed shortly---make it physically nonviable.
Nonetheless, affine paths allows us to demonstrate that there is a relationship between the Dyson series for the path ordered exponential and a more standard perturbative expansion in powers of the YM coupling constant. 
This will allow us to bridge between our dressing formalism with the one of Lavelle and McMullan \cite{Lavelle:1995ty}.

Let us proceed to the computation. Consider the affine path
\be
A(\tau) \equiv \gamma_{\text{affine},A}(\tau) = \tau A + (1-\tau) A^\star = \tau A \qquad \tau\in[0,1].
\ee
More geometrically, affine paths can be understood as geodesics of the gauge supermetric of equation \eqref{metric_YM}.
From the definition of the dressing factor,
\be
h_\text{affine}(A) =  \bb P \exp\left( \fint_\text{affine} -\varpi\right) = \bb P \exp\left( -\int_0^1 \d \tau \;(\fI_{\bb V_\text{affine}}\varpi)_{|\tau A}\right):
\ee
where  $\bb V_\text{affine}= \dot\gamma_\text{affine}$ is the `velocity' along the affine path:
\be
\bb V_\text{affine} = \int \frac{\d A^a_i(\tau)}{\d \tau}\frac{\dd}{\dd A^a_i} = \int A^a_i \frac{\dd }{\dd A^a_i}
\ee
(we have omitted the spacetime label $x$ and used $A^\star=0$). 
Introducing $\D_\tau := \d + A(\tau)$ and taking $\varpi$ to be the SdW connection (equation \eqref{YM_varpi_boundaries}), we obtain
\be
h_\text{affine}(A) = \bb P \exp \left(-\int_0^1 \d \tau \; \D_\tau^{-2} \D_\tau^i A_i\right).
\ee
This formula requires some clarification on how $\D_\tau$ is supposed to act on $A_i$. By backtracking the formula's origin, one sees that, here, $A_i$ needs be understood as the component of the `velocity' vector $\bb V_i^a$ which---being a tangent vector in $\mathrm T_{A(\tau)}\F_{\rm pYM}$---transforms in the adjoint representation under gauge transformations performed at $A(\tau)\in\F_{\rm pYM}$. Hence,  the gauge-covariant divergence becomes an ordinary divergence:
\be
\D_\tau^iA_i = \delta^{ij}(\pp_iA_j^a + f^a{}_{bc}A(\tau)_i^{b} A_j^c)\tau_a = \delta^{ij}(\pp_i A_j^a + \tau f^{a}{}_{bc} A_i^b A^c_j) \tau_a= \pp^i A_i^a \tau_a,
\ee
where we assumed $g_{ij}=\delta_{ij}$, and again $A^\star = 0$.
This simplification to a standard divergence is entirely due to the choice of an affine path and $A^\star=0$ as a reference configuration. 

To deduce a systematic expansion of the non-Abelian dressing factor around the Abelian Dirac dressing in powers of the Yang--Mills coupling constant $e$, we first notice that the `physicist's' Yang--Mills connection with self-coupling constant $e$ is not $A$, but rather $\mathcal A$ such that
\be
A = e \mathcal A.
\ee
Once expressed in terms of $\mathcal A$, the Yang--Mills Lagrangian `Abelianizes' in the parametric limit $e\to0$.
Now, defining $\epsilon := e\tau \in [0,e]$, the affine dressing factor is readily written in terms of $\cal A$ as
\be
h_\text{affine} = \bb P \exp \int_0^e \d \epsilon \; (- \D_\epsilon^{-2} \pp^i \mathcal A_i),
\ee
where $\D_\epsilon := \d + \epsilon \mathcal A$ is the covariant derivative at coupling $\epsilon$. 

Solving the path order exponential in terms of the Dyson series---a sum of an increasing number of nested integrals,---
\be
h_\text{affine} = {\rm id} 
- \int_0^e \d\epsilon(\D_\epsilon^{-2} \pp^i \mathcal A_i) 
+ \int_0^e\d\epsilon_2\int_0^{\epsilon_2}\d\epsilon_1 (\D_{\epsilon_2}^{-2} \pp^i \mathcal A_i)(\D_{\epsilon_1}^{-2} \pp^i \mathcal A_i) + \cdots,
\ee
and replacing $D_\epsilon^{-2} $ by its Taylor expansion in powers of $\epsilon$, 
\be
(\D_\epsilon^{-2})^a{}_c = \pp^{-2}\delta^a{}_c - \epsilon \pp^{-2} \Big( f^a{}_{bc} (\pp^i \mathcal{A}_i^b) + 2 f^a{}_{bc} \mathcal{A}_i^b \pp^i\Big) \pp^{-2}  + O(\epsilon^2),
\ee
we finally obtain a perturbative expansion of $h_\text{affine}$ in powers of the coupling constant $e$: 
\be
h_\text{affine} = {\rm id} 
- e  \eta
+ \frac{e^{2}}{2}  \Big(\eta^2 -  \pp^{-2} \pp^i[ \mathcal{A}_i,\eta] - \pp^{-2}[\mathcal{A}_i, \pp^i\eta]\Big)   + O(e^3)
\label{haffine_expansion}
\ee
where, for conciseness, $\eta:=\pp^{-2} \pp^i \mathcal A^a_i\tau_a \in \fG$. 

This expansion is structurally similar to that found by Lavelle and McMullan \cite{Lavelle:1995ty} or in the Gribov--Zwanziger framework \cite{Zwanziger:1989mf,Capri2005, Vandersickel:2012tz}.
It also appears related to the Coulomb gauge, since the dressing at a field configuration in that gauge is trivial. We will come back to all these points at the end of the section. Now, we explain the difficulties inherent to the choice of affine dressings.

As we stated at the very beginning, a crucial property that dressing  factors should satisfy is that $h(A^g)= g^{-1}h$.
This is the case for the Dirac dressing above, but not for the affine dressing factors in the non-Abelian theory.

\begin{figure}[t]
	\begin{center}
		\includegraphics[scale=.17]{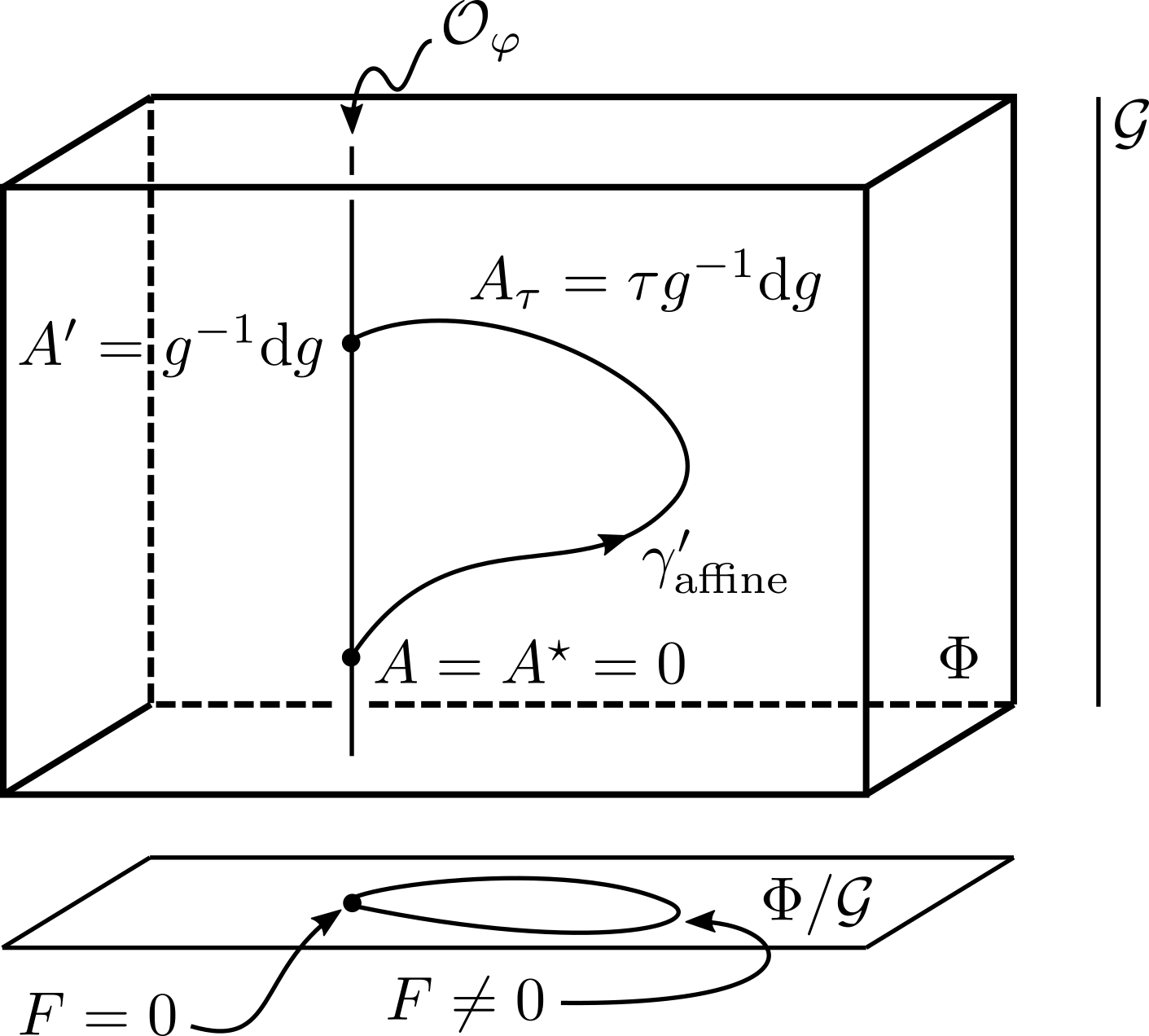}
		\caption{The affine path from $A^\star=0$ to $A'=g^{-1}\d g$ is not vertical to the trivial path $\gamma$ from $A^\star$ to $A=0$: it passes through configurations with nonvanishing curvature.}
		\label{fig5}
	\end{center}
\end{figure}

As a simple counter-example---represented in figure \ref{fig5},---start by considering the following two affine paths, $\gamma$ and $\gamma'$, from $\star\equiv0$ to two gauge-related configurations $A$ and $A'=A^g$.
Choose $A = \star=0 $, such that $\gamma$ is the trivial path $\gamma_\text{aff}(\tau) \equiv \star = 0$, and choose $A' = A^g = g^{-1} \d g $ such that $\gamma'$ is the affine path from $\star$ to $A'$.
From the general discussions of the next section it will be evident that we cannot expect $h(\gamma')$ to be equal to $g^{-1}h(\gamma)=g^{-1}\cdot{\rm id}$, unless  $\gamma'$ is vertical to the trivial path.
Thus, $\gamma'(\tau) = \tau A' = \tau g^{-1}\d g$ would have to be pure gauge for every value of $\tau\in[0,1]$. But in the non-Abelian case it is not, as it can be seen from the fact that $A(\tau)=\tau g^{-1} \d g$ has non-zero Yang--Mills curvature. 
At the perturbative level, the obstruction will not manifest itself before\footnotemark~$O(e^3)$.
\footnotetext{This follows from an application of the non-Abelian Stokes theorem and the fact that $\fF \xrightarrow{e\to0}\fF_\text{Ab}=0$. Moreover, the above perturbative formula can be checked to coincide with the Lavelle--McMullan dressing at least to $O(e^2)$ \cite{Lavelle:1995ty}, whereas the Lavelle--McMullan dressing abides to the covariance property by construction.}We will discuss these facts later on purely geometrical grounds.

In the next section we clarify how and why a manifestly gauge-covariant con\-struction---the Wilson line---can fail to provide an appropriately covariant notion of dressing. The problem lies in what it is meant by `action of a gauge transformation'.

\subsection{Path dependence, gauge dependence, and gauge fixings\label{sec:pathgf}}

In order to unambiguously dress a certain field configuration, we need to fix a choice of path. This choice is relevant whenever $\varpi$ carries curvature, and is encoded in a map
\begin{align}
\Gamma: \F &\rightarrow \{\gamma:[0;1]\to \F, \gamma(0)=\star\} \nonumber\\
\varphi &\mapsto \gamma_\varphi \;\;\text{such that}\;\;\gamma_\varphi(1)=\varphi
\end{align}
from field-space to the set of paths on field-space, which associates to $\varphi$ a path starting at $\star$ and ending at $\varphi$ itself.
A map $\Gamma$ defines a notion of $\Gamma$-dressing, $h_\Gamma(\varphi) := h(\gamma_\varphi, \varphi)$. Affine paths $\Gamma_\text{affine}$ from $A^\star=0$ to $A$ are an example.

To identify the gauge properties of the $\Gamma$-dressing factors, let us first consider what happens to a generic Wilson line $h(\gamma; \varphi)$ if the path is displaced vertically by the action of $R_g$ on $\F$:
\be
h(\gamma;\varphi) \mapsto R_g^\ast h(\gamma;\varphi) = h(R_g\gamma;\varphi^{g(\varphi)})= g(\varphi)^{-1} h(\gamma;\varphi)g(\star),
\label{eq_hgaugetransf}
\ee
where $
\big(R_g\gamma\big)(\tau) := R_{g(\gamma(\tau))} \gamma(\tau)
$
is the gauge-transformed path starting at $R_{g(\star)}\varphi^\star$ and ending at  $\varphi^{g(\varphi)}$.
Unless otherwise stated, $\star$ is henceforth a fully fixed reference configuration. This means that we will consider only field-dependent gauge transformations such that $g(\star) =\rm id$.

Equation \eqref{eq_hgaugetransf} is a standard identity for the gauge transformation of a Wilson line. It follows from the covariance of $\varpi$---equation \eqref{varpi_dependent_finite}---and the definition of $h$---equation \eqref{gstar}. 

The issue with dressings is that in equation  \eqref{eq_hgaugetransf}, the gauge transformation $g$ acts as a vertical diffeomorphism $R_g$ of $\F$,\footnotemark~which generally fails to be compatible with the map $\Gamma$: although $R_g\gamma_\varphi$ does start at $\star$ and ends at $\varphi^{g(\varphi)}$, it will generically fail to be in the image of $\Gamma$, i.e. $R_g\gamma_\varphi \neq \gamma_{\varphi^{g(\varphi)}}$. 
\footnotetext{On standard spacetime Wilson lines, gauge transformations act only `internally'. In contrast, field-space Wilson lines get displaced by the action of a gauge transformation. The difference is that spacetime is the base space of the relevant PFB, while field-space is the total space, on which a field-dependent gauge transformation acts as a generic vertical diffeomorphism $R_{g}$.} 

\begin{figure}[t]
	\begin{center}
		\includegraphics[scale=.17]{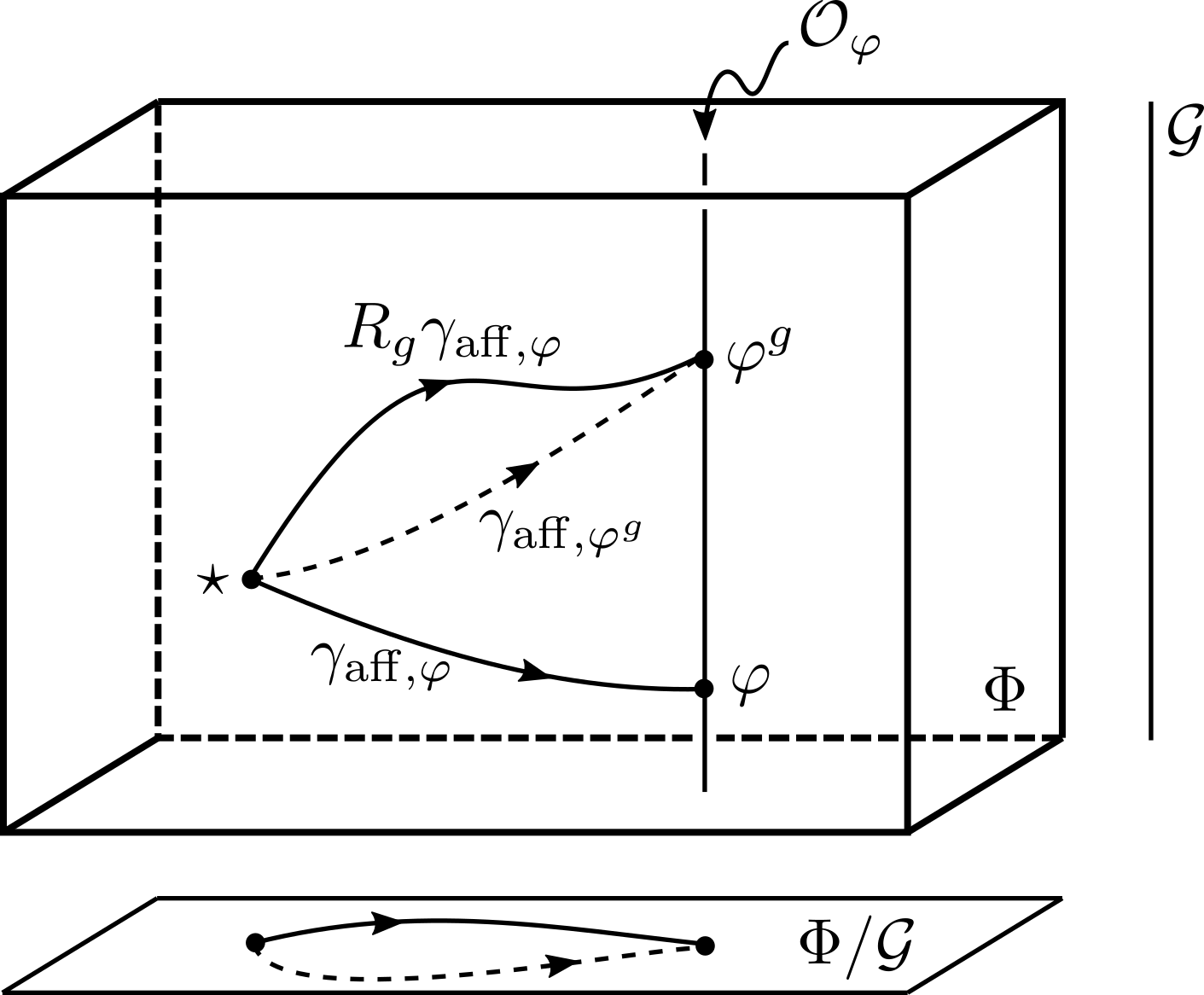}
		\caption{A graphical representation of the failure of gauge-compatibility for $\Gamma_\text{affine}$.}
		\label{fig7}
	\end{center}
\end{figure}

This leads us to the following definition: a choice of paths $\Gamma$ is said to be gauge compatible if the path ending at $\varphi^g$ is always vertical to the path ending at $\varphi$. More formally, $\Gamma$ is {\it gauge compatible} if\footnote{We are here identifying paths that differ by a reparameterization. A change in the parameterization of a path does not affect the dressing factor.}
\be
\text{for any }\varphi\in \F\;\text{and } g\in\G,
\;\text{there exists a }g':\F \to \G
\;\text{such that }
\gamma_{\varphi^{g}}=R_{g'}\gamma_\varphi.
\ee
For $\Gamma$ a gauge compatible choice of paths, it follows that
\be
h_{\Gamma}(\varphi^g)  = R_g^\ast h(\gamma_\varphi;\varphi) = g^{-1}h_\Gamma(\varphi)
\qquad(\Gamma\text{ gauge compatible})
\ee
and thus also that the $\Gamma$-dressed field $\hat \varphi_\Gamma := \varphi^{h_\Gamma(\varphi)}$ is gauge-invariant:
\be
\hat{\varphi^g}_\Gamma =  \hat\varphi_\Gamma \qquad(\Gamma \text{\;gauge compatible}).
\ee
To summarize, gauge compatibility of the choice of paths is sufficient to give the dressing factor the right gauge transformation properties, and thus to construct a dressing. Affine paths for non-Abelian YM are {\it not} gauge compatible---see figure \ref{fig7}.

\subsection{Remarks on section \ref{sec:dressing}}\label{rmks:dressing}

\paragraph*{\indent(i) Dressing of fields \textsc{vs.} dressing of particles --- }
In our approach, based on  the instantaneous configuration space, it is natural to dress matter fields {\it locally in space and instantaneously in time}, possibly with spatially non-local dressing factors. 
The spatial nature of the dressing is particularly evident when we deal with $\varpi$'s associated to finite and bounded regions.  
This is in contrast to the more standard notion of {\it particle} dressing, which deals with the matter fields {\it locally in momentum space}, and the relevant dressing factors are taken to depend on the state of motion of the particle itself, e.g. in \cite{dollard1964asymptotic, Kulish:1970ut, Zwanziger:1973if, Zwanziger:1974jz,  bagan2000charges, bagan2000charges2}  (it is important to notice that the Faddeev-Kulish dressing actually has a dynamical origin).
The difference between the two approaches is deep, and  rooted in the different `philosophical' stances on the nature of quantum matter: either in terms of particles (S-matrix) or in terms of histories of instantaneous configurations of fields (Schr\"odinger path integral).

\paragraph*{\indent(ii) Field-dependent \textsc{vs.} field-independent dressings ---}
The mathematical properties inherent to the notion of `dressing' have been extensively studied even in a context where the dressing factor is not field-dependent, but just a new, extra, `dressing field' \cite{Attard2018, francoisthesis}. 
We have already mentioned this possibility in the context of `edge modes' at points {\it (i)} of \hyperref[rmks:general_YM]{\it Remarks on section \ref*{sec:general_YM}} and {\it (viii)} of \hyperref[rmks:matter]{\it Remarks on section \ref*{sec:matter}}.

\paragraph*{\indent(iii) Dressings \textsc{vs.} gauge fixings \textsc{vs.} Gribov problem --- }
A gauge-fixing corresponds to a smooth section through field-space which intersects every fiber in exactly one point. Such a section defines a smooth dressing, simply by associating to each point of field-space the gauge transformation that transports it to the section. The gauge-invariant dressed field $\hat \varphi$ is then the gauge-fixed field at the intersection of the fiber of $\varphi$ with the section. (We will below describe a dressing-from-gauge-fixing as a Wilson line.)
Conversely, a smooth gauge-compatible dressing factor $h: \F \rightarrow G$ defined throughout $\F$ also defines a preferred smooth section, as its level surface $h = \text{id}$. That the section intersects each orbit in one point follows from the covariance of the dressing factor and the fact that the group acts freely on each fiber.

However, due to well-known topological obstructions, global gauge-sections do not always exist \cite{Gribov:1977wm, Singer:1978dk, Vandersickel:2012tz}; this is the Gribov problem. It follows that if there is a Gribov problem, there is no globally well-defined smooth covariant dressing factor, and gauge-invariant fields cannot be constructed globally in field-space in this manner. To circumvent the Gribov problem, one can either be content to work in a small enough neighborhood of the reference configuration $\star$, or attempt to `piece together' dressing factors for different regions of field-space in a non-smooth way, a difficult task.

 One particularly ambitious attempt for circumventing this topological obstruction  defines this neighborhood as the \lq{}Gribov fundamental domain\rq{}; a region for  which gauge copies do not occur (see \cite{Vandersickel:2012tz} for a review and point {\it (vi)} below). Topological obstructions referring to the non-triviality of the bundle are then transferred to the boundary of this region, which is very difficult to describe in a manner useful for computations \cite{vanBaal:1995gg}.\footnote{ Moreover, one should not underplay the important physical role of such a boundary  in the infinite-dimensional context: it concentrates field-space volume and may be responsible for confinement \cite{Vandersickel:2012tz, Greensite:2004ur}.}

  This modification of the topology of the region under scrutiny elicits an analogy  to the different descriptions of the Aharonov--Bohm effect: one description uses non-trivial topology and zero curvature, the other uses trivial topology and non-zero curvature. This latter description corresponds to our use of the unrestricted $\F$ and of a curvature-full $\varpi$ within it.   $\F$ itself has trivial topology, but admits  a product form over $\F/\G$ if and only if it admits a global section and therefore a connection with zero curvature. Thus the effects of curvature incorporate this aspect of the non-triviality of the principal fiber bundle over $\F/\G$. They provide a manner to deal with the effects of this non-triviality which is distinct from that proposed by the \lq{}fundamental domain\rq{} proposal.

Nonetheless, for many purposes,  working in a small enough neighborhood of the reference configuration $\star$ suffices:  one is often interested in the variations of field functionals around a given background, rather than in the functionals themselves---see section \ref{sec:Noether_hor}. As we saw, no globally well-defined dressing is required to set up an infinitesimal gauge-invariant formalism: replacing generic variations with the horizontal ones constructed from $\varpi$ itself is enough.

Let us make some of these ideas more precise.

\paragraph*{\indent(iv) Dressings \textsc{vs.} horizontal symplectic geometry --- }
If $\fF=0$, as in electrodynamics, the dressing factor turns out to depend only on the `arrival' configuration $A$. We therefore write $h=h(A)$, where $A^\star=0$ is considered fixed once and for all. $h(A)$ is now a smooth $\G$-valued function on $\F$ which defines a gauge-fixing section through the condition $h=\rm id$.
From these considerations, it is clear that
\be
\dd h h^{-1} = -\varpi
\qquad (\fF = 0).
\label{dhh=varpi}
\ee
On the LHS, the differential acts solely on the `arrival' configuration $\varphi$, at which the whole equation is evaluated.\footnote{Despite the similarities, in footnote \ref{Pexp}, $h$ is derived only along the direction of the path $\gamma$, whereas equation \eqref{dhh=varpi} has a general $\dd$.}

Using this equation, the horizontal derivative can be recovered from the dressing:
\be
\dd \hat A = \Ad_{h^{-1}}(\dd_H A  )
\qquad\text{and}\qquad
\dd \hat \Psi = h^{-1} (\dd_H \Psi )
\qquad
(\fF = 0).
\label{eq_kaboom}
\ee
Therefore we conclude that, if $\fF=0$, the horizontal symplectic geometry of section \ref{sec:Noether_hor} is nothing but the symplectic geometry of the dressed fields. {\henrique This equality would also hold at the level of the symplectic potential (and for other Abelian fields, $\varphi$) i.e. $\theta_H(\varphi,\dd \varphi)=\theta(\hat \varphi, \dd \hat \varphi)$ \cite{Gomes:2018shn}.
}

In this sense, {\it  the horizontal derivative can be understood as an infinitesimal dressing}. According to the previous discussion, we conclude that only this infinitesimal version of the dressing survives at the nonperturbative level in non-Abelian YM.

\paragraph*{\indent(v) Choice of paths and dressing /1 Affine paths --- }
Although the notion of dressing is not available at the fully nonperturbative level, it can still be useful in wide portions of field-space. 
In the following remarks, we review how different notions of dressings and paths proposed in the literature fit in our framework and with each other, and finally propose a genuinely different possibility---the {\it historical paths}.

Affine paths arguably constitute the simplest choice of a set of paths in the space of YM potentials. 
They provide for easily computable dressings---see equation \eqref{haffine_expansion}. Unfortunately, as we saw, affine paths are not gauge-compatible beyond the first orders of perturbation theory, and  therefore have to be discarded.

\paragraph*{\indent(vi) Choice of paths and dressing /2 Lavelle--McMullan--Gribov--Zwanziger dressing --- }
Locally in $\F$, a gauge-fixing surface $S_f:=\{\varphi : f(\varphi)=0\}$ can always be chosen. 
Then, field configurations lying on this surface locally parametrize the reduced field-space $\F/\G$ `in a gauge-invariant way', i.e. they are valid representatives of the gauge orbit crossing $S_f$.
These will correspond to the values of the dressed fields $\hat\varphi$ (at least in some neighborhood of the reference configuration). 

To obtain this, we proceed as in section \ref{sec:pathgf}: define the dressing factor $h_f : \F \to \G$ so that it is the identity on $S_f$, $h_{f|S_f}=\rm id$, and satisfies $(\xi^\# h_f) h_f^{-1} = - \xi$ along the vertical directions. This fixes $h$ in the portion of $\F$ over $S_f$ and guarantees its gauge-compatibility, $h_f(\varphi^g) = g^{-1} h_f(\varphi)$. 
Now, one can {\it define} a {\it flat} field-space connection $\varpi_f$ starting from this gauge-fixing adapted choice of dressing factor, simply through equation \eqref{dhh=varpi}: $\varpi_f := -\dd h_f h_f^{-1}$. If $\star\in S_f$, holonomies of $\varpi_f$ from $\star $ to $\varphi$, along any path, will of course give back $h_f(\varphi)$.

The construction of a dressing factor from a gauge-fixing was used by Lavelle and McMullan to construct their notion of dressing \cite{Lavelle:1995ty}. The gauge-fixing condition of their choice was the Coulomb gauge, $f(A) = \pp^iA_i$ (like us, they also worked in a non-manifestly covariant 3+1 framework---see points {\it (ii)} and {\it (iii)} of \hyperref[rmks:7]{\it Remarks on section \ref*{sec:Singerconnection}}).

The Gribov--Zwanziger framework \cite{Zwanziger:1989mf, vanBaal:1995gg, Capri2005} is related as follows. In one incarnation of the framework, one chooses those orbit representatives that correspond to (global) minima along the orbits of the functional $I_A(g)$. 
In our notation, $I_A(h) = ||\Delta (A^h)||^2_{\bb G}$ is the $\bb G$-length of the vector $\Delta (A^h) \in {\rm T}_\star \F_{\rm pYM} $ defined, using the natural affine structure of $\F_{\rm pYM}$, by $\Delta(A):= A - A^\star \equiv A$ .
Since the DeWitt supermetric $\bb G$ is just an identity matrix in the $A$ coordinates over $\F_{\rm pYM}$, it follows that straight lines through $\F_{\rm pYM}$ are geodesics and $|| \Delta A||_{\bb G}$ is the length of the geodesic between $\star$ and $A$. Minimizing it along the orbit ${\cal O}_A$ means selecting the point of ${\cal O}_A$ which is the closest to $\star$. These points, the Gribov--Zwanziger representatives, can be shown to lie on the Coulomb-gauge surface $S_{\pp A}$.
Equivalently, at fixed $A$, the minimization in $h$ of $I_A(h)$ selects a dressing factor that must coincide with the one by Lavelle and McMullan. We call it the Coulomb dressing factor $h_{\pp A}(A)$.

In all these cases, the dependence of the dressing factor on the reference configuration $\star$ is obscured by the construction. 
In contrast, the construction we review next makes this dependence absolutely manifest.

\paragraph*{\indent(vii) Choice of paths and dressing /3 Horizontal Vilkovisky coordinates --- }
In his geometrical approach to the effective action of gauge theories, Vilkovisky introduced a parameterization of field-space in terms of certain Gaussian normal coordinates which we call $\sigma$ \cite{Vilkovisky:1984st, vilkovisky1984gospel} (see also \cite{Kunstatter:1991kw, DeWitt:1995cx, DeWitt_Book}). 
These coordinates are elements of the tangent space at the reference (or background) configuration $\varphi^\star$, i.e. $\sigma \in {\rm T}_\star \F$. The coordinates $\sigma(\varphi,\star)$ of $\varphi$ are defined to be the tangent at $\varphi^*$ of a certain affinely parametrized geodesic $\gamma$ which has $\gamma(\tau=1) = \varphi$. The hardest part of the construction is concocting a {\it notion of field-space parallel transport which is gauge-compatible}.
Gauge compatibility of the parallel transport is understood in the same way as above: a notion of parallel transport is termed gauge compatible if geodetic paths to gauge related configurations are vertical to one another.\footnote{Note that although the gauge supermetric is gauge compatible, the geodesics of its Levi--Civita connection are the affine paths, and not gauge compatible.}

In \cite{Vilkovisky:1984st, vilkovisky1984gospel}, Vilkovisky defined a gauge-compatible  affine connection on $\F$---the Vilkovisky connection { $\Gamma_{\rm Vilk}$}---out of the Levi-Civita connection of $\bb G$ and combinations of the SdW connection, the fundamental vector fields $\tau_a^\#$, and their field-space derivatives.
\footnotetext{It reads
	\be
	\Gamma_{\rm Vilk} = \Gamma_{\rm LC} - \varpi^a \otimes_S \nabla \tau_a^\#  + \frac12 \varpi^a \otimes_S \nabla_{\varpi^\#} {\tau_a^\#} \in \Big( {\mathfrak X}^1\otimes (\Omega^1\otimes_S \Omega^1)\Big)(\F),
	\ee
	where $\nabla$ is the covariant derivative of the Levi-Civita connection $\Gamma_{\rm LC}$, and $\otimes_S$ means that the form-indices are symmetrized, $\alpha_1\otimes_S\alpha_2 = \alpha_1\otimes\alpha_2 + \alpha_2\otimes\alpha_1$, $\alpha_i\in\Omega^1(\F)$.}

\begin{figure}[t]
	\begin{center}
		\includegraphics[scale=.17]{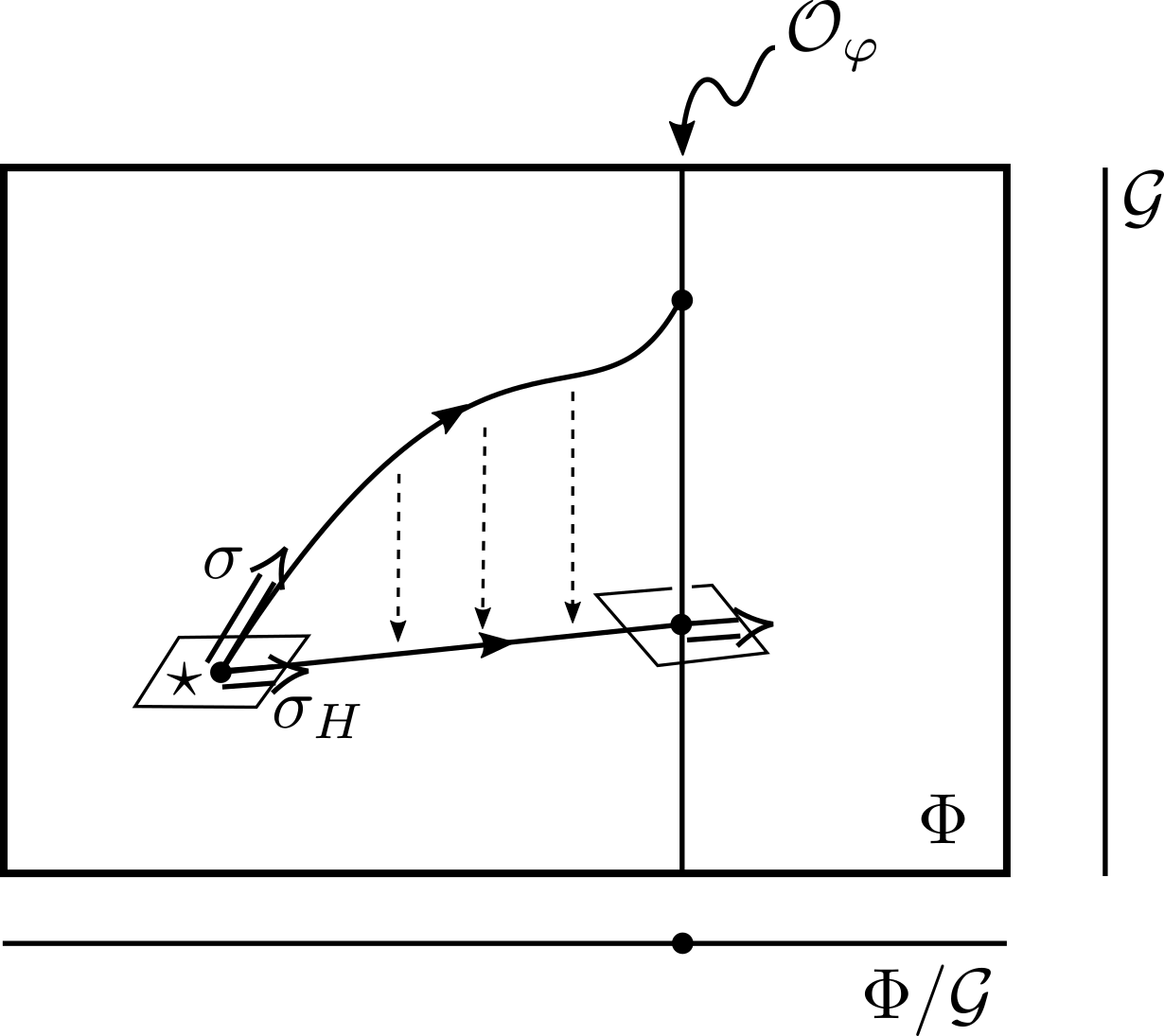}
		\caption{A graphical representation of the Vilkovisky connection and of its properties: (\textit{i}) a Vilkovisky geodesic which has initial horizontal velocity will stay horizontal with respect to the SdW connection, (\textit{ii}) a Vilkovisky geodesic with arbitrary initial velocity $\sigma$ will be vertical to the horizontal Vilkovisky geodesic with initial velocity $\sigma_H$, and (\textit{iii}) horizontal Vilkovisky geodesics are also geodesics with respect to Levi-Civita connection of $\bb G^\text{g}$, and therefore straight lines in the affine coordinates.}
		\label{fig3}
	\end{center}
\end{figure}

At the reference configuration $\star$, where the Vilkovisky coordinates $\sigma$ live, we can use the SdW connection, $\varpi_\star$, to decompose them into their horizontal and vertical components, that is $\sigma_V := \fI_{\sigma}\varpi^\#_\star$ and $\sigma_H := \sigma - \sigma_V$. 
We are now ready to formulate three crucial properties of the Vilkovisky connection---figure \ref{fig3}: (\textit{i}) a Vilkovisky geodesic which has initial horizontal velocity will stay horizontal with respect to the SdW connection, (\textit{ii}) a Vilkovisky geodesic with arbitrary initial velocity $\sigma$ will be vertical to the horizontal Vilkovisky geodesic with initial velocity $\sigma_H$, and (\textit{iii}) horizontal Vilkovisky geodesics are also geodesics with respect to $\Gamma_{\text{LC}}$. Notice that $\Gamma_{\rm LC}$---for $\bb G$ the DeWitt supermetric---vanishes in the standard coordinates.

Therefore, properties (\textit{i}) and (\textit{ii}) express the `gauge compatibility' of the Vilkovisky connection, while (\textit{iii}) provides a relationship to the constructions of the previous paragraph.
In particular, when expressed in the flat affine coordinates centered at $A^\star=0$, the $\sigma_H$ are equal to $\Delta A = A$ for $A$ on the Coulomb section, $\pp A =0$. 
Hence $\sigma_H$,  provides the same kind of notion of dressed fields as the Lavelle--McMullan--Gribov--Zwanziger (LmMGZ) construction. The way Vilkovisky coordinates parameterize points outside the Coulomb section is however more involved.

The bridge between the constructions of Vilkovisky and LmMGZ is provided precisely by our notion of Wilson line dressing. Indeed, we claim that 
\be 
h(\gamma_{\rm Vilk}; A) = h_{\pp A}(A).
\label{V=LmMGZ}
\ee 
We prove---and qualify---this statement in appendix \ref{app:proof}.

Before discussing the next point, we note that the Abelianization phenomenon observed by Vilkovisky \cite{Vilkovisky:1984st} when using Gaussian normal coordinates, was later independently observed in the Lavelle--McMullan framework \cite{Lavelle:2011yc}---where it was also applied to study problems in QCD.

\paragraph*{\indent(viii) Dependence of dressings on the reference configuration $\star$ --- }
The Vilkovisky formulation of the LmMGZ construction highlights the relevance of the reference point $\star$, and so does ours.
In particular, these two formulations make clear that all the burden of gauge covariance is packed into the dependence of all physical quantities on the choice of $A^\star$.
The importance and inevitability of this dependence have been to our knowledge first emphasized by Branchina, Meissner and Veneziano, and then studied in greater detail by Pawlowski, in the context of the  Vilkovisky--DeWitt effective action \cite{Branchina:2003ek, Pawlowski:2003sk} (see also \cite{Wetterich:2017aoy}).
Therefore, what at first might have seemed an unpleasant quirk of our proposal, is in fact a very robust feature of manifestly gauge-invariant objects analyzed from field-space, and it is  consistent with the idea that gauge theories are intrinsically relational \cite{Rovelli:2013fga, MachsBucket}.

We conclude this remark by noting the interesting  possibility of understanding the `Gribov copies' problem \cite{Vandersickel:2012tz} in terms of global geometrical  properties of the Vilkovisky geodesics---such as the appearance of caustics.

\paragraph*{\indent(ix) Choice of paths and dressing /4 Historical paths --- } \label{rmks:dressing9}
A new type of dressing can be constructed from the field history. The SdW connection, as we have defined it, depends on the values of the fields on a Cauchy surface $t = \text{const}$. Any one-parameter family of such data is a path, and a physically motivated choice of path is thus the actual history of the system. We call that choice of paths {\it historical paths}, and the corresponding dressing {\it historical dressing}.

For the framework to be fully gauge-invariant, all paths must share the same starting point (otherwise they would not be directly `comparable' to one another). For the purpose of constructing a dressing, we  may thus think of different configurations as arising from the same initial state, but different preparation procedures. Historical dressings behave well under gauge transformations because of the compatibility of gauge and dynamics. 

We believe that these dressings will be of interest in the quantum mechanical implementation of our framework: the historical paths are what is integrated over in the path integral. Using horizontal lifts of those paths, which amounts to using fields dressed with the historical dressings, in the path integral was suggested in \cite{Narasimhan:1979kf,Gomes:2017jyp}.

\begin{figure}[t]
	\begin{center}
		\includegraphics[scale=.17]{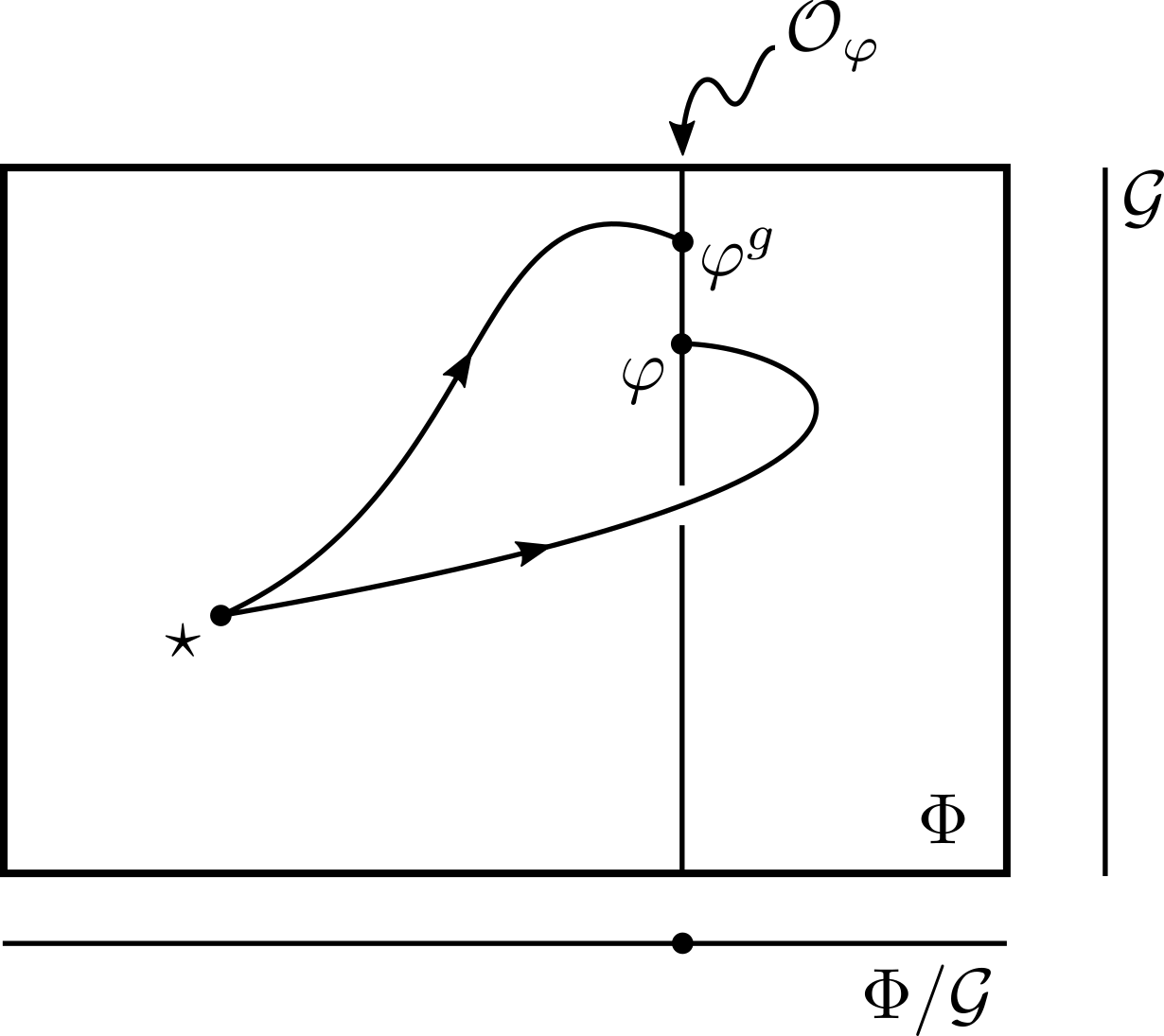}
		\caption{Two different histories connecting the same initial configuration to the same final  physical configuration may lead to different historical dressing factors. The dressed final states may thus differ by a gauge transformation.  In this way, field-space curvature will enter the historical path-integral.}
		\label{fig9}
	\end{center}
\end{figure}

Historical dressings depend on the field history, but dress only the final instantaneous state. Since different preparation procedures may yield the same final state, historical dressings do not associate a unique group element to every instantaneous state. This is in contrast to the other notions of dressing presented here, but also allows historical dressings to sidestep global problems. The situation is illustrated in figure \ref{fig9}.

Infinitesimally, historical dressings could be seen as defining dynamics with respect to the gauge-covariant time derivative $\hat H(\frac{\d}{\d t})$. We point out that this gives a gauge-invariant notion of time evolution on phase space, but without gauge-fixing. This time derivative, with $\varpi$ the SdW connection, has already appeared in the constructions of \cite{Babelon:1980uj}. In a sense, it mimics the gauge compensation properties of the $A_0$ component of the Yang-Mills gauge field: denoting as before by $\bb V$ the field-space vector field with components $\bb V_i = \frac{\d}{\d t} A_i$, the electric field is $E_i = \bb V_i - \D_i A_0$, while, as we know, the gauge-covariant time derivative of $A_i$ reads $\hat H(\bb V)_i = \bb V_i - \D_i (\fI_{\bb V}\varpi)$---that is the standard horizontal projection. See item {\it (ii)} in the \hyperref[rmks:7]{\it Remarks on section \ref*{sec:Singerconnection}} on the `exponentiated' version of this relation.

It would be interesting to further explore the historical dressings in the context of symplectic geometry and charges. If one interprets the dressing factor (or its value on the boundary of the region under consideration) as `edge modes', the historical dressing can be seen to provide non-local (in space), field-dependent dynamics for these edge modes. The variation of the dressing factor, and hence the symplectic geometry, will have a contribution which is non-local in time and involves the curvature of the field-space connection. We leave this direction for future work.

\paragraph*{\indent(x) Higgs dressing --- }
We conclude by a brief remark.
In this section we have focused on SdW dressings. Nonetheless, in the broken phase one should make use of a Higgs dressing \cite{Ilderton:2010tf, Lavelle:1994rh}.
For simplicity, we restrict to the case in which the Higgs connection is well-defined.
Then, we have already noticed that the Higgs connection is flat, and comparing equations \eqref{varpim_flat} and \eqref{dhh=varpi}, it becomes obvious that in this case the Goldstone mode $h$ becomes the dressing factor itself. 
The only subtlety is that $\star$ cannot be the vanishing-field configuration---there the Higgs connection is not even defined!---but has to be taken at $\phi_\star(x) = v_o$.

\subsection*{Acknowledgment}

We would like to thank: Glenn Barnich for enlightening discussions about charges, the Higgs connection, and Vilkovisky's effective action; Wolfgang Wieland and Ali Seraj for early discussions on the Lorentz connection and the regional properties of the SdW connection, respectively; Michele Schiavina, Jordan Fran\c cois, and Simone Speziale for valuable  input on an earlier version of this paper; and Gabriel Herczeg and William Donnelly for feedback which helped us making our exposition clearer.
This research was supported  by Perimeter Institute for Theoretical Physics. Research at Perimeter Institute is supported by the Government of Canada through Industry Canada and by the Province of Ontario through the Ministry of Research and Innovation. FH is supported by a Vanier Canada Graduate Scholarship. HG is supported by The Commonwealth European and International Cambridge Trust.


\appendix
\part*{Appendix}

\section{Examples of a regional $\varpi$ and of the role of $A_0$}\label{app:examples_gluing}

\subsection{Horizontal projections do not commute with regional restrictions \label{app:examples_gluing1}}

We present here an example illustrating the interplay between horizontality of field-space vectors in the SdW connection, and the decomposition of space into regions. We will use for simplicity the case of electrodynamics. The treatment would go through almost  unaltered for Yang--Mills around the trivial configuration $A=0$.

Let us focus on electrodynamics on 4-dimensional Minkowski space, with the field-space restricted by the condition $A_i \rightarrow 0$ fast enough at infinity.\footnote{ Although our formalism does not require explicit boundary conditions, it can also accommodate them. In the case of explicit solutions, it is an assumption which greatly simplifies computations.}
Let $\Sigma = \bb R^3$, and let $\Sigma_{I,II}$ be the lower and upper half spaces $x_3 \le 0, x_3 \ge 0$ respectively. All three regions have associated field-spaces, and there are no boundary conditions at the plane $x_3=0$. Let us equip the field-spaces with the gauge supermetric (\ref{metric_YM}). As before, let us define the horizontal spaces as the orthogonal complement of the fibers under the gauge supermetric. Let $H$, $H_{I,II}$ be the horizontal subspaces of the tangent bundles of the field spaces, and $\varpi$ and $\varpi_{I,II}$ the field-space connections whose kernels are those horizontal spaces. To be horizontal, a vector field on the field-space of the three regions must satisfy
\begin{align}
\bb X  \in {}& H  \qquad & \Leftrightarrow & \qquad \pp^i \bb X_i (x) & ={} & 0 \ \forall x \in \Sigma &&&&\nonumber \\
\bb X_I  \in {}& H_I \qquad & \Leftrightarrow &\qquad \pp^i (\bb X_I)_i (x) & ={} &0 \ \forall x \in \Sigma_I  &\text{ and }& (\bb X_I)_3 (x) \vert_{x_3 = 0} &={}& 0 \nonumber\\
\bb X_{II} \in {}& H_{II} \qquad & \Leftrightarrow & \qquad \pp^i (\bb X_{II})_i (x) & ={} & 0 \ \forall x \in \Sigma_{II}  &\text{ and }& (\bb X_{II})_3 (x) \vert_{x_3 = 0} &={}& 0,
\end{align}
where as before we have used the notation $\bb X = \int \d^3 x \, \bb X_i(x) \frac{\dd}{\dd A_i(x)}$. 
This can be seen from (\ref{YM_varpi_boundaries}) together with the fact that the horizontal vector fields are the kernel of $\varpi$. Notice that on the regions with boundaries $\Sigma_{I, II}$, in addition to the divergence-free condition the vector fields must satisfy a boundary condition to be horizontal (in the whole $\mathbb R^3$ extra boundary conditions are not necessary thanks to the fall-off conditions restricting directly the field-space.).

Let us write $\varpi$ and $\varpi_{I, II}$ by explicitly solving (\ref{YM_varpi_boundaries}). As in section \ref{sec:dressing}, for $\varpi$ we obtain simply 
\begin{align}
\varpi(x) = {}& \pp^{-2} \pp^i \dd A_i = - \int_\Sigma \frac{\d^3 y }{4 \pi} \frac{\pp^i \dd A_i}{|x-y|}.
\end{align}

For the regions $I$ and $II$, while the analogous expression satisfies the bulk equation, it does not satisfy the boundary condition. Hence we must add a solution of the homogeneous Laplace equation implementing the correct Neumann boundary conditions. 
Such a solution is mathematically analogous to the electrostatic potential of an image charge density and boundary charges. 
We obtain, for $x\in\Sigma_I$,
\begin{subequations}
	\begin{align}
	\varpi_I ={}& \varpi_{I,0} + \varpi_{I, \text{img}} + \varpi_{I, \text{s}}\\
	\varpi_{I, 0}(x) ={}& - \int_{\Sigma_I} \frac{\d^3 y }{4 \pi} \frac{\pp^i \dd A_i(y)}{|x-y|}\\
	\varpi_{I, \text{img}}(x) ={}& - \int_{\Sigma_I} \frac{\d^3 y }{4 \pi} \frac{\pp^i \dd A_i(y)}{|x-\overline y|}\\
	\varpi_{I, \text{s}}(x) ={}& - \int_{y_3=0} \frac{d^2 y}{2\pi }\frac{\dd A_3 (y)}{|x-y|}
	\end{align}
\end{subequations}
where, inside the integrals, $\pp_i = \pp/\pp y^i$, and the image charges are located at
\be
\overline{(y_1, y_2, y_3)} = (y_1, y_2, - y_3)
\ee
	The role of $\varpi_{I, 0}$ is to solve the bulk equation for $\varpi_I$, i.e. $\pp^2\varpi = \pp^i\dd A_i$ within $\Sigma_I$. On the other hand $\varpi_{I, \text{img}}$ and $\varpi_{I,s}$ are solutions for the homogeneous (Laplace) equation. The role of $\varpi_{I, \text{img}}$ is to implement  zero Neumann boundary conditions for $\varpi_{I, 0}+\varpi_{I, \text{img}}$ via the method of image charges; whereas the role of  $\varpi_{I, s}$ is to add the appropriate surface charges for $\varpi=\varpi_{I,0} + \varpi_{I, \text{img}} + \varpi_{I, \text{s}}$ to satisfy the nonzero Neumann boundary conditions.
	
	The SdW connection-form $\varpi_{II}$ associated to the region $\Sigma_{II}$ is formally analogous, but the surface charge contribution comes with the opposite sign:
\begin{subequations}
	\begin{align}
	\varpi_{II} ={}& \varpi_{II,0} + \varpi_{II, \text{img}} + \varpi_{II, \text{s}}\\
	\varpi_{II, 0}(x) ={}& - \int_{\Sigma_{II}} \frac{\d^3 y }{4 \pi} \frac{\pp^i \dd A_i(y)}{|x-y|}\\
	\varpi_{II, \text{img}}(x) ={}& - \int_{\Sigma_{II}} \frac{\d^3 y }{4 \pi} \frac{\pp^i \dd A_i(y)}{|x-\overline y|}, \qquad \\
	\varpi_{II, \text{s}}(x) ={}& \int_{y_3=0} \d^2 y\frac{\dd A_3 (y)}{2 \pi|x-y|}.
	\end{align}
\end{subequations}

The sign change can  be seen from the fact that $\pp_3$ is the outgoing normal for $\Sigma_I$, but the ingoing normal for $\Sigma_{II}$. More physically, using again the electrostatic potential analogy, note that a given surface charge density on a plane surface creates opposite normal electric fields on either side: however, the boundary conditions for $\varpi$, which are here analogous to the normal electric fields, come with the same sign on either region, so the surface charge densities which implement them must have opposite signs. If to study a region some auxiliary charges at the boundaries are needed, these compensate each other and disappear when the two regions are combined.

Let $\bb X$ be a field-space vector field on the field-space of $\Sigma$, with components
\begin{align}
\bb X_i(x) = \epsilon_{ijk} \pp^j b^k(x),
\end{align}
where $b^k(x)$  are unspecified, generic functions that fall off sufficiently rapidly as $x \rightarrow \infty$. We immediately see that $\bb X$ cannot be  purely vertical everywhere, since its components are not exact: $\bb X_i \neq \pp_i \xi$. Alternatively, note that $\bb X$ has a non-zero action on a gauge-invariant observable, the magnetic field:
\begin{align}
\bb X (F_{ij}(x)) ={}& 2 \pp_{[i} \epsilon_{j]kl} \pp^k b^l(x) \neq 0.
\end{align}
Since $\pp^i \bb X_i = 0$, the vector field $\bb X$ is purely horizontal on region $\Sigma$. 

Let $\bb X_{I,II}$ be the restrictions of the vector field $\bb X$ to the field-spaces of regions $\Sigma_{I,II}$, explicitly
\begin{align}
\bb X_I = \int_{\Sigma_I} \d^3 x (\epsilon_{ijk}\pp_j b_k(x)) \frac{\dd}{\dd A_i(x)}
\end{align} 
and similarly for $\bb X_{II}$. Now, notice that $\bb X_I$ is {\it not} necessarily horizontal on $\Sigma_I$, since it does not need to satisfy the right boundary conditions. In fact, generally $(\bb X_I)_3 \vert_{x_3 = 0} = (\pp_1 b_2 - \pp_2 b_1)\vert_{x_3= 0} \neq 0$. 
This illustrates the point that the restriction of a horizontal vector field need not be purely horizontal, because what `horizontal' means depends on the shape of the region.  Nonetheless, the restriction \textit{will have} a non-trivial horizontal projection, as per equation \eqref{eq:hor_restriction}.

We now compute the contraction of $\bb X$ with $\varpi$ on the three regions, which we will need to form the horizontal projections. We have
\begin{subequations}
\begin{align}
\varpi(\bb X) ={}& 0\\
\varpi_I(\bb X_I)(x) ={}& - \int_{x_3 = 0} \d^2 y \frac{\pp_1 b_2(y) - \pp_2 b_1(y)}{2 \pi |x-y|}\\
\varpi_{II}(\bb X_{II})(x) ={}& + \int_{x_3 = 0} \d^2 y \frac{\pp_1 b_2(y) - \pp_2 b_1(y)}{2 \pi |x-y|}
\end{align}
\end{subequations}
and thus the horizontal projections are
\begin{subequations}
\begin{align}
\hat H (\bb X) ={}& \bb X\\
\hat H_I (\bb X_I) ={}& \bb X_I - \pp_i \varpi_I(\bb X_I)\\
\hat H_{II} (\bb X_{II}) ={}& \bb X_{II} - \pp_i \varpi_{II}(\bb X_{II}).
\end{align}
\end{subequations}
To summarize, we see explicitly that the restriction of a horizontal vector field need not be the horizontal projection of the restriction $\bb X_I$, and moreover that the horizontal projections of the restrictions $\bb X_I, \bb X_{II}$ do not match at the shared boundary.\\

\subsection{Time dependent gauge transformations and the role of $A_0$  \label{app:examples_gluing2}}

Let us now turn to an example illustrating how time dependent gauge transformations enter  our formalism.  The example is inspired from \cite{Seraj:2017rzw}. Consider a field-space vector field
\begin{align}
\bb X = \int_{\Sigma} \d^3 x \,\pp_i f(t, x) \frac{\dd}{\dd A_i(x)},
\end{align}
where $f(t,x)$ has a nontrivial time dependence.
Is this vector field vertical?  On the one hand, it is a pure gradient, and thus looks like a gauge transformation, but on the other hand, adding a time dependent gradient to $A_i$ changes the electric field.

To resolve the tension, recall that here we are using the instantaneous configurations  $\{A_i (x, t)\}_t$ as field-space coordinates, with the prescription $A_0 = \lambda(t,x) + \varpi(\pp_t A_i)$, with $\lambda(t,x)$ a given function fixed once and for all (see the point {\it (ii)} of \hyperref[rmks:7]{\it Remarks on section \ref*{sec:Singerconnection}}).
Plugging in the SdW connection, we get
\begin{align}
\varpi(\bb X) = \pp^{-2} \pp^i \big(\pp_i f(t,x) \big) = f(t, x).
\end{align}
and therefore the horizontal projection of $\bb X$ vanishes: 
\begin{align}
\hat H (\bb X)_i(t, x) ={}& \bb X_i - \pp_i \varpi = 0.
\end{align}
Now, if $A_i$ undergoes an infinitesimal variation $\bb X_i = \pp_i f$, then $A_0$ changes by $\varpi(\pp_t \bb X) = \pp_t f$, i.e. as if it underwent itself the correct gauge transformation. Hence, the gauge-invariant field strength $F_{\mu \nu}$ does not change under the action of $\hat H(\bb X)$. We conclude that $\bb X$ is purely vertical and, consistently with the interpretation of vertical directions as gauge, does not induce changes in the electromagnetic field. 

Notice that if $f(t,x) = f(t)$ is constant in space, the transformation does not affect $A_i$ in the first place, and therefore neither $A_0$ nor $F_{\mu\nu}$. It is not important that $A_0$, which is in this framework an auxiliary object, does not transform under such `gauge transformations' (see also section \ref{ssec:sympl_charges} for the role of `gauge transformations' which are constant in space).

To conclude,  let us notice that there are two natural alternatives to this treatment. The first uses a covariant phase space approach, in which field-space is coordinatized by $\{A_\mu(t, x)\}$ (for the difficulties this choice runs into see the point {\it (iii)} of \hyperref[rmks:7]{\it Remarks on section \ref*{sec:Singerconnection}}). In this case, $\bb X$ would not be vertical as written since it is missing the appropriate $\int \pp_t f \frac{\dd}{\dd A_0}$ component (in this case the integral are spacetime integrals). The second alternative consists in taking the field-space to be the canonical phase space where the variables $\{A_i, E^i \}$ are taken to be independent. In this case case $\bb X$ would be vertical too.

\section{Proof of equation \eqref{V=LmMGZ}\label{app:proof}}

To prove equation \eqref{V=LmMGZ}---and clarify the hypotheses that go into this statement---we consider the family of Vilkovisky geodesics $\gamma_{{\rm V},s}$ parameterized by the initial velocities $\sigma_s = s\sigma(A) - (1-s)\sigma_H(A)$, $s\in[0,1]$. 

\begin{figure}[t]
	\begin{center}
		\includegraphics[scale=.17]{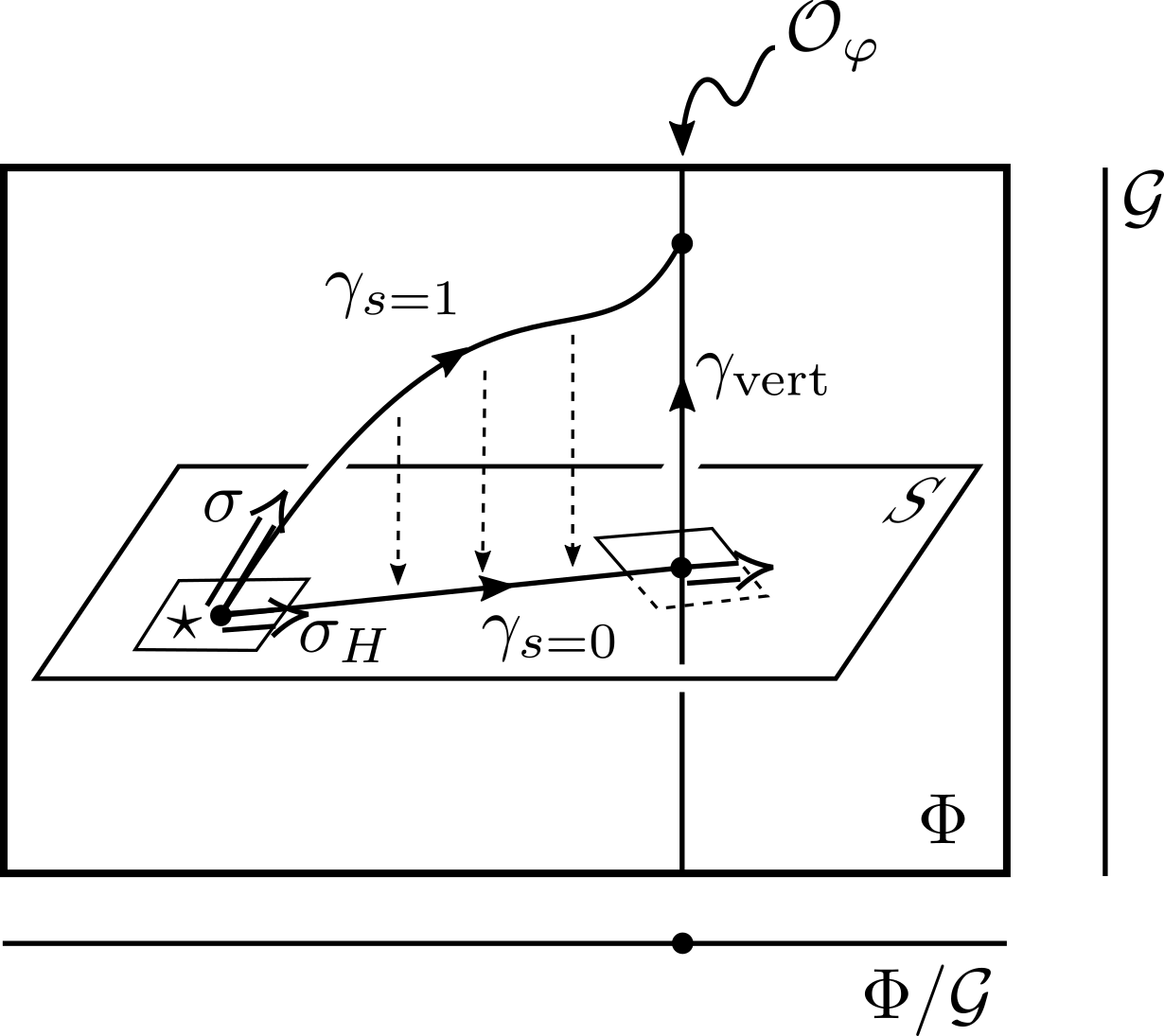}
		\caption{A graphical representation of the Vilkovisky geodesics and some of the properties used in the proof. $S$ is the (affine) surface of connections satisfying the Coulomb condition $\pp A=0$. }
		\label{fig4}
	\end{center}
\end{figure}

We make the hypothesis that the family is smooth. In other words, we suppose that the paths $\gamma_{{\rm V},s=1}$ and $\gamma_{\rm vert}\circ\gamma_{{\rm V},s=0}$ are homotopically equivalent through a family of paths that projects down to $\gamma_{{\rm V,0}}$---here, $\gamma_{\rm vert}$ is a vertical path connecting the arrival points of $\gamma_{{\rm V},0}$ and $\gamma_{{\rm V},1}$ (see figure \ref{fig4}).
There might be global obstructions for configurations far from $A_\star=0$ and from $S_{\pp A}$.

Under the hypothesis, the statement above follows from the non-Abelian Stokes theorem applied to $\gamma_{\rm closed} = \gamma_{{\rm V},1}^{-1}\circ\gamma_{\rm vert}\circ\gamma_{{\rm V},s=0}$. 
A sketch of proof, in a quite loose notation, is the following.
Denoting by $\mathbb S\exp$ a `surface ordered integral', we see that
$h(\gamma_\text{closed},\star)  = \mathbb S \exp \fint_C \fF_{IJ} = \rm id$, where the last result follows from the fact that the tangent plan to the surface $C= \cup_s\gamma_{{\rm V},s}$ has one vertical direction, while $\fF$ is purely horizontal. The conclusion is now reached by observing that $h(\gamma_{{\rm V},0} = \rm id)$ is a consequence of property (\textit{ii}) and that the SdW holonomy $\bb P\exp \left(\fint_{{\rm vert}}-\varpi\right)$ along $\gamma_{\rm vert}$ is  equal to the Coulomb dressing factor $h_{\pp A}(A)$ by property (\textit{iii}).

\newpage
\section{A short guide to DeWitt's notation\label{app:DeWitt}}
\begin{center}
\begin{tabular}{|c|c|c|c|}
\hline
{\sffamily Name} & {\sffamily Our notation} & {\sffamily DeWitt's} & {\sffamily Coordinate expression}\\\hline
supermetric$^*$ & $\bb G(\cdot,\cdot)$ & $\gamma_{ij}$ & eq. \eqref{eq_supermetricGG}\\
vector field & $\bb X$ & $V^i$ & equation \eqref{vectors}\\
fundamental vect. f.  & $\xi^\#=\tau^\#_a\xi^a$ & $Q^i_\alpha \xi^\alpha$ & eq. \eqref{xi_hash} \& fn. \ref{ftnt:tauhash}\\
-- & $\tau_a^\#$ & $Q_\alpha^i $ & fn. \ref{ftnt:tauhash}\\
-- & $\bb Q_{ab}=\bb G(\tau_a^\#,\tau_b^\#)$ & $\mathfrak{F}_{\alpha\beta}=-Q^i_\alpha \gamma_{ij} Q^j_\beta$ & eq. \eqref{eq_Qab}\\
-- & $ \bb Q^{ab} $& $\mathfrak {G}^{\alpha\beta}$ &eq. \eqref{eq_Qab}\\
 connection-form$^\dagger$ & $\varpi$ & -- & eq. \eqref{eq_varpiexplicit} \\ 
SdW connection-form & $\varpi = \bb Q^{ab}\bb G(\tau_a^\#,\cdot)\tau_b$ & $\omega^\alpha_i= \gamma_{ij} Q^j_{\beta}\mathfrak{G}^{\beta\alpha}$ & eq. \eqref{eq:varpi_abstract_solution}\\
vertical projector & $\hat V = \varpi^\#$ & $Q^i_\alpha\varpi^\alpha_j$ & --\\
horizontal projector & $\hat H $ & $\Pi^i{}_j$ & -- \\\hline
\end{tabular}
\end{center}

\noindent$^*$ DeWitt's $Q^i_\alpha$ is often denoted $R^i_\alpha$ in the literature.\\
$^\dagger$ In DeWitt's case the symbol is most often used in relation to the SdW connection. When only the projection property is of interest (and not its covariance), it is denoted $P^\alpha_i$.\\



\begin{thebibliography}{10}
\providecommand{\url}[1]{\texttt{#1}}
\providecommand{\urlprefix}{URL }
\newcommand{\eprint}[1]{\href{http://arxiv.org/abs/#1}{\tt arxiv:#1}}

\bibitem{Rovelli:2013fga}
{Rovelli},
\newblock 2014
  \href{http://dx.doi.org/10.1007/s10701-013-9768-7}{\textit{{Why
  Gauge?}}}
\newblock Found. Phys.
\newblock 44(1) 91 [\eprint{1308.5599}]
  

\bibitem{MachsBucket}
{Barbour} \protect\BIBand{} {Pfiste} (Eds.) 1995.
\newblock \textit{Mach's Principle: From Newton's Bucket to Quantum Gravity}.
\newblock
\newblock Birkhauser

\bibitem{Ebin}
{Ebin}, 1970.
\newblock \textit{The Manifold of Riemannian Metrics}.
\newblock Symp. Pure Math., AMS,
\newblock 11,15

\bibitem{Palais}
{Palais}, 1961.
\newblock \textit{On the existence of slices for the actions of non-compact
  groups}.
\newblock Ann. of Math.
\newblock 73 295

\bibitem{YangMillsSlice}
{Wilkins}, 1989.
\newblock \textit{Slice Theorems in Gauge Theory}.
\newblock Proc. Royal Ir. Acad. A: Math. and  Phys. Sciences
\newblock 89A(1) 13

\bibitem{kondracki1983}
{Kondracki} \protect\BIBand{} {Rogulski}, 1983.
\newblock \textit{On the Stratification of the Orbit Space for the Action of
  Automorphisms on Connections. On Conjugacy Classes of Closed Subgroups. On
  the Notion of Stratification}.
\newblock Preprint: Instytut Matematyczny.
\newblock Inst., Acad.

\bibitem{Mitter:1979un}
{Mitter} \protect\BIBand{} {Viallet},
\newblock 1981
  \href{http://dx.doi.org/10.1007/BF01209307}{\textit{{On
  the Bundle of Connections and the Gauge Orbit Manifold in {Yang-Mills}
  Theory}}}.
\newblock Commun. Math. Phys.
\newblock 79 457

\bibitem{Singer:1978dk}
{Singer},
\newblock 1978
  \href{http://dx.doi.org/10.1007/BF01609471}{\textit{{Some
  Remarks on the Gribov Ambiguity}}}.
\newblock Commun. Math. Phys.
\newblock 60 7

\bibitem{Singer:1981xw}
{Singer},
\newblock 1981
  \href{http://dx.doi.org/10.1088/0031-8949/24/5/002}{\textit{{The
  Geometry of the Orbit Space for Nonabelian Gauge Theories.
  (Talk)}}}.
\newblock Phys. Scripta
\newblock 24 817

\bibitem{isenberg1982slice}
{Isenberg} \protect\BIBand{} {Marsden}, 1982.
\newblock \textit{A slice theorem for the space of solutions of Einstein's
  equations}.
\newblock Phys. Rep.
\newblock 89(2) 179

\bibitem{Gribov:1977wm}
{Gribov},
\newblock 1978
  \href{http://dx.doi.org/10.1016/0550-3213(78)90175-X}{\textit{{Quantization
  of Nonabelian Gauge Theories}}}.
\newblock Nucl. Phys. B139

\bibitem{Vandersickel:2012tz}
{Vandersickel} \protect\BIBand{} {Zwanziger},
\newblock 2012
  \href{http://dx.doi.org/10.1016/j.physrep.2012.07.003}{\textit{{The
  Gribov problem and QCD dynamics}}}.
\newblock Phys. Rept.
\newblock 520 175 [\eprint{1202.1491}]
  

\bibitem{kobayashivol1}
{Kobayashi} \protect\BIBand{} {Nomizu}, 1963.
\newblock \textit{Foundations of differential geometry. {V}ol {I}}.
\newblock
\newblock Interscience Publishers, a division of John Wiley \& Sons, New
  York-Lond on

\bibitem{Gomes:2016mwl}
{Gomes} \protect\BIBand{} {Riello},
\newblock 2017
  \href{http://dx.doi.org/10.1007/JHEP05(2017)017}{\textit{{The
  observer’s ghost: notes on a field space
  connection}}}.
\newblock JHEP
\newblock 05 017 [\eprint{1608.08226}]
  

\bibitem{Thierry-MiegJMP}
{Thierry-Mieg},
\newblock 1980
  \href{http://dx.doi.org/10.1063/1.524385}{\textit{Geometrical
  reinterpretation of Faddeev-Popov ghost particles and BRS
  transformations}}.
\newblock J. Math. Phys.
\newblock 21(12) 2834

\bibitem{Bonora1983}
{Bonora} \protect\BIBand{} {Cotta-Ramusino},
\newblock 1983
  \href{http://dx.doi.org/10.1007/BF01208267}{\textit{{Some
  remarks on BRS transformations, anomalies and the cohomology of the Lie
  algebra of the group of gauge
  transformations}}}.
\newblock Comm. Math. Phys.
\newblock 87(4) 589

\bibitem{Donnelly:2014fua}
{Donnelly} \protect\BIBand{} {Wall},
\newblock 2015
  \href{http://dx.doi.org/10.1103/PhysRevLett.114.111603}{\textit{{Entanglement
  entropy of electromagnetic edge modes}}}.
\newblock Phys. Rev. Lett.
\newblock 114(11) 111603 [\eprint{1412.1895}]
  

\bibitem{Wadia}
{Wadia} \protect\BIBand{} {Yoneya},
\newblock 1977
  \href{http://dx.doi.org/https://doi.org/10.1016/0370-2693(77)90010-7}{\textit{The
  role of surface variables in the vacuum structure of Yang-Mills
  theory}}.
\newblock Phys. Lett. B
\newblock 66(4) 341

\bibitem{Regge:1974zd}
{Regge} \protect\BIBand{} {Teitelboim},
\newblock 1974
  \href{http://dx.doi.org/10.1016/0003-4916(74)90404-7}{\textit{{Role
  of Surface Integrals in the Hamiltonian Formulation of General
  Relativity}}}.
\newblock Annals Phys.
\newblock 88 286

\bibitem{Henneaux:2018gfi}
{Henneaux} \protect\BIBand{} {Troessaert},
\newblock 2018
  \href{http://dx.doi.org/10.1007/JHEP05(2018)137}{\textit{{Asymptotic
  symmetries of electromagnetism at spatial
  infinity}}}.
\newblock JHEP
\newblock 05 137 [\eprint{1803.10194}]
  

\bibitem{Carlip:2004mn}
{Carlip},
\newblock 2005
  \href{http://dx.doi.org/10.1088/0264-9381/22/7/007}{\textit{{Horizon
  constraints and black hole entropy}}}.
\newblock Class. Quant. Grav.
\newblock 22 1303 [\eprint{hep-th/0408123}]
  

\bibitem{Balachandran:1994up}
{Balachandran}, {Chandar}, \protect\BIBand{} {Momen},
\newblock 1996
  \href{http://dx.doi.org/10.1016/0550-3213(95)00622-2}{\textit{{Edge
  states in gravity and black hole
  physics}}}.
\newblock Nucl. Phys.
\newblock B461 581 [\eprint{gr-qc/9412019}]
  

\bibitem{Donnelly:2016auv}
{Donnelly} \protect\BIBand{} {Freidel},
\newblock 2016
  \href{http://dx.doi.org/10.1007/JHEP09(2016)102}{\textit{{Local
  subsystems in gauge theory and gravity}}}.
\newblock JHEP
\newblock 09 102 [\eprint{1601.04744}]
  

\bibitem{Narasimhan:1979kf}
{Narasimhan} \protect\BIBand{} {Ramadas},
\newblock 1979
  \href{http://dx.doi.org/10.1007/BF01221361}{\textit{{Geometry of SU(2) gauge fields}}}.
\newblock Commun. Math. Phys.
\newblock 67 121

\bibitem{Babelon:1979wd}
{Babelon} \protect\BIBand{} {Viallet},
\newblock 1979
  \href{http://dx.doi.org/10.1016/0370-2693(79)90589-6}{\textit{{The
  Geometrical Interpretation of the {Faddeev-Popov}
  Determinant}}}.
\newblock Phys. Lett.
\newblock 85B 246

\bibitem{Babelon:1980uj}
{Babelon} \protect\BIBand{} {Viallet},
\newblock 1981
  \href{http://dx.doi.org/10.1007/BF01208272}{\textit{{On
  the Riemannian Geometry of the Configuration Space of Gauge
  Theories}}}.
\newblock Commun. Math. Phys.
\newblock 81 515

\bibitem{AsoreyMitter}
{Asorey} \protect\BIBand{} {Mitter},
\newblock 1981
  \href{http://dx.doi.org/10.1007/BF01213595}{\textit{{Regularized,
  Continuum {Yang-Mills} Process and {Feynman-Kac} Functional
  Integral}}}.
\newblock Commun. Math. Phys.
\newblock 80 43

\bibitem{DeWitt_Book}
{DeWitt}, 2003.
\newblock \textit{The Global Approach to Quantum Field Theory, Vol. 1}, volume
  114 of \textit{International Series of Monographs in Physics, 114}.
\newblock
\newblock Clarendon Press, Oxford
  
\bibitem{Gomes:2018shn}
{Gomes} \protect\BIBand{} {Riello},
\newblock 2018
\href{https://doi.org/10.1103/PhysRevD.98.025013}{\textit{{Unified geometric framework for boundary charges and particle dressings}}}.
\newblock Phys. Rev.
\newblock D98 025031 [\eprint{1804.01919}]

\bibitem{Lee:1990nz}
{Lee} \protect\BIBand{} {Wald},
\newblock 1990
  \href{http://dx.doi.org/10.1063/1.528801}{\textit{{Local
  symmetries and constraints}}}.
\newblock J. Math. Phys.
\newblock 31 725

\bibitem{WeinbergQFT2}
{Weinberg}, 2005.
\newblock \textit{The Quantum Theory of Fields. Volume 2. Modern Applications}.
\newblock
\newblock Cambridge Univ. Press

\bibitem{Weinberg:1973ew}
{Weinberg},
\newblock 1973
  \href{http://dx.doi.org/10.1103/PhysRevD.7.1068}{\textit{{General
  Theory of Broken Local Symmetries}}}.
\newblock Phys. Rev.
\newblock D7 1068

\bibitem{Jacobson:2015uqa}
{Jacobson} \protect\BIBand{} {Mohd},
\newblock 2015
  \href{http://dx.doi.org/10.1103/PhysRevD.92.124010}{\textit{{Black
  hole entropy and Lorentz-diffeomorphism Noether
  charge}}}.
\newblock Phys. Rev.
\newblock D92 124010 [\eprint{1507.01054}]
  

\bibitem{wald1993black}
{Wald}, 
\newblock 1993
\href{https://doi.org/10.1103/PhysRevD.48.R3427}{{Black hole entropy is the Noether charge}}.
\newblock Phys. Rev.
\newblock D48 R3427(R) [\eprint{gr-qc/9307038}]


  

\bibitem{DePaoli:2018erh}
{De~Paoli} \protect\BIBand{} {Speziale}, 2018
\href{https://doi.org/10.1007/JHEP07(2018)040}{\textit{{A gauge-invariant symplectic potential for tetrad general
  relativity}}}.
  \newblock JHEP 2018:40  [\eprint{1804.09685}]
  

  

\bibitem{Dirac:1955uv}
{Dirac},
\newblock 1955
  \href{http://dx.doi.org/10.1139/p55-081}{\textit{{Gauge
  invariant formulation of quantum
  electrodynamics}}}.
\newblock Can. J. Phys.
\newblock 33 650

\bibitem{Lavelle:1995ty}
{Lavelle} \protect\BIBand{} {McMullan},
\newblock 1997
  \href{http://dx.doi.org/10.1016/S0370-1573(96)00019-1}{\textit{{Constituent
  quarks from QCD}}}.
\newblock Phys. Rept.
\newblock 279 1 [\eprint{hep-ph/9509344}]
  

\bibitem{Lavelle:2009zz}
{Lavelle}, {Heinzl}, {Ilderton}, {Langfeld}, \protect\BIBand{} {McMullan},
  2009.
\newblock \textit{{Infra-red problems and a response}}.
\newblock PoS
\newblock QCD-TNT09 023

\bibitem{bagan2000charges}
{Bagan}, {Lavelle}, \protect\BIBand{} {McMullan}, 2000.
\newblock \textit{Charges from dressed matter: construction}.
\newblock Ann. Phys.
\newblock 282(2) 471

\bibitem{bagan2000charges2}
{Bagan}, {Lavelle}, \protect\BIBand{} {McMullan}, 2000.
\newblock \textit{Charges from dressed matter: physics and renormalisation}.
\newblock Ann. Phys.
\newblock 282(2) 503

\bibitem{Capri2005}
{Capri}, {Dudal}, {Gracey}, {Lemes}, {Sobreiro}, {Sorella}, \protect\BIBand{}
  {Verschelde},
\newblock 2005
  \href{http://dx.doi.org/10.1103/PhysRevD.72.105016}{\textit{{A
  Study of the gauge invariant, nonlocal mass operator $\text{Tr} \int \d^4 x F_{\mu \nu} (D^2)^{-1} F_{\mu \nu}$ in Yang-Mills
  theories}}}.
\newblock Phys. Rev.
\newblock D72 105016
  [\eprint{hep-th/0510240}] 

\bibitem{Attard2018}
{Attard}, {Fran{\c{c}}ois}, {Lazzarini}, \protect\BIBand{} {Masson}, 2018.
\newblock \textit{The Dressing Field Method of Gauge Symmetry Reduction, a
  Review with Examples}, pp. 377--415.
\newblock
\newblock
\newblock Springer International Publishing, Cham
  \href{https://doi.org/10.1007/978-3-319-64813-2_13}{ISBN 978-3-319-64813-2}
\newblock [\eprint{1702.02753}]

\bibitem{francoisthesis}
{Fran\c{c}ois}, 2014.
\newblock \href{http://www.theses.fr/2014AIXM4037}{\textit{Reduction of gauge symmetries: a new geometrical approach}}.
\newblock Ph.D. thesis,
\newblock Aix-Marseille University

\bibitem{Ilderton:2010tf}
{Ilderton}, {Lavelle}, \protect\BIBand{} {McMullan},
\newblock 2010
  \href{http://dx.doi.org/10.1088/1751-8113/43/31/312002}{\textit{{Symmetry
  Breaking, Conformal Geometry and Gauge
  Invariance}}}.
\newblock J. Phys.
\newblock A43 312002 [\eprint{1002.1170}]
  

\bibitem{Lavelle:1994rh}
{Lavelle} \protect\BIBand{} {McMullan},
\newblock 1995
  \href{http://dx.doi.org/10.1016/0370-2693(95)00046-N}{\textit{{Observables
  and gauge fixing in spontaneously broken gauge
  theories}}}.
\newblock Phys. Lett.
\newblock B347 89 [\eprint{hep-th/9412145}]
  

\bibitem{Zwanziger:1989mf}
{Zwanziger},
\newblock 1989
  \href{http://dx.doi.org/10.1016/0550-3213(89)90122-3}{\textit{{Local
  and Renormalizable Action From the Gribov
  Horizon}}}.
\newblock Nucl. Phys.
\newblock B323 513

\bibitem{Vilkovisky:1984st}
{Vilkovisky},
\newblock 1984
  \href{http://dx.doi.org/10.1016/0550-3213(84)90228-1}{\textit{{The
  Unique Effective Action in Quantum Field
  Theory}}}.
\newblock Nucl. Phys.
\newblock B234 125

\bibitem{vilkovisky1984gospel}
{Vilkovisky}, 1984.
\newblock \textit{The gospel according to DeWitt}.
\newblock
\newblock In \textit{Quantum theory of gravity. Essays in honor of the 60th
  birthday of Bryce S. DeWitt}

\bibitem{WittenCrnkovic}
{{Crnkovic}} \protect\BIBand{} {{Witten}}, 1987.
\newblock \textit{{Covariant description of canonical formalism in geometrical
  theories.} In `Three Hundred Years of Gravitation.'}, chapter~12, pp.
  676--684.
\newblock
\newblock Cambridge

\bibitem{Iyer:1994ys}
{Iyer} \protect\BIBand{} {Wald},
\newblock 1994
  \href{http://dx.doi.org/10.1103/PhysRevD.50.846}{\textit{{Some
  properties of Noether charge and a proposal for dynamical black hole
  entropy}}}.
\newblock Phys. Rev.
\newblock D50 846 [\eprint{gr-qc/9403028}]  
  

\bibitem{DeWitt:1967ub}
{DeWitt},
\newblock 1967
  \href{http://dx.doi.org/10.1103/PhysRev.162.1195}{\textit{{Quantum
  Theory of Gravity. 2. The Manifestly Covariant
  Theory}}}.
\newblock Phys. Rev. 162 1195.
\newblock [,298(1967)]

\bibitem{DeWitt:1995cx}
{DeWitt} \protect\BIBand{} {Molina-Paris}, 1995.
\newblock \textit{{Gauge theory without ghosts}}.
\newblock In \textit{{Physics. Proceedings, 2nd International A.D. Sakharov
  Conference, Moscow, Russia, May 20-24, 1996}},
\newblock pp. 396--408
  [\eprint{hep-th/9511109}] 

\bibitem{Branchina:2003ek}
{Branchina}, {Meissner}, \protect\BIBand{} {Veneziano},
\newblock 2003
  \href{http://dx.doi.org/10.1016/j.physletb.2003.09.020}{\textit{{The
  Price of an exact, gauge invariant RG flow
  equation}}}.
\newblock Phys. Lett.
\newblock B574 319 [\eprint{hep-th/0309234}]
  

\bibitem{Pawlowski:2003sk}
{Pawlowski}, 2003.
\newblock \textit{{Geometrical effective action and Wilsonian flows}}
   [\eprint{hep-th/0310018}]
  

\bibitem{Lang}
{Lang}, 1999.
\newblock \textit{{Fundamentals of Differential Geometry}}, volume 191 of
  \textit{Graduate Texts in Mathematics}.
\newblock
\newblock
\newblock Springer New York, New York, NY
  \href{http://link.springer.com/10.1007/978-1-4612-0541-8}{ISBN
  978-1-4612-6810-9}
  
  \bibitem{Michor}
{Kriegl} \protect\BIBand{} {Michor}, 1997.
\newblock \textit{{The Convenient Setting of Global Analysis}}.
\newblock
\newblock Mathematical Surveys and Monographs Vol. 53, American Mathematical Soc.

\bibitem{fischermarsden}
{Fischer} \protect\BIBand{} {Marsden}, 1979.
\newblock \textit{The initial value problem and the dynamical formulation of
  general relativity.}
\newblock
\newblock In S.W. Hawking and W. Israel (Eds.) \textit{General relativity : an Einstein centenary survey.}
  Cambridge Univ. Press, pp. 138-211

\bibitem{Crnkovic:1986ex}
{Crnkovic} \protect\BIBand{} {Witten}, 1986.
\newblock \textit{{Covariant description of canonical formalism in geometrical theories}}
In S.W. Hawking, S.W. and W. Israel (Eds.) \textit{Three hundred years of gravitation}, 676-684

\bibitem{Crnkovic:1986be}
{Crnkovic},
\newblock 1987
  \href{http://dx.doi.org/10.1016/0550-3213(87)90221-5}{\textit{{Symplectic
  Geometry and (Super)poincare Algebra in Geometrical
  Theories}}}.
\newblock Nucl. Phys.
\newblock B288 419

\bibitem{Crnkovic:1987tz}
{Crnkovic},
\newblock 1988
  \href{http://dx.doi.org/10.1088/0264-9381/5/12/008}{\textit{{Symplectic
  Geometry of the Covariant Phase Space}}}.
\newblock Class. Quant. Grav.
\newblock 5 1557

\bibitem{anderson}
{Anderson}, 1992.
\newblock \textit{Introduction to the variational bicomplex}.
\newblock In M.~Gotay, J.~Marsden, \protect\BIBand{} V.~Moncrief (Eds.)
  \textit{Mathematical Aspects of Classical Field Theory, Comptemporary
  Mathematics Vol 132}.
\newblock AMS

\bibitem{DeWitt:1992cy}
{DeWitt}, 2012.
\newblock \textit{{Supermanifolds}}.
\newblock Cambridge Monographs on Mathematical Physics.
\newblock
\newblock Cambridge Univ. Press, Cambridge, UK
  \href{http://www.cambridge.org/mw/academic/subjects/physics/theoretical-physics-and-mathematical-physics/supermanifolds-2nd-edition?format=AR}{ISBN
  9781139240512, 9780521423779}

\bibitem{HenneauxTeitelboim}
{Henneaux and Teitelboim}, 1994.
\newblock \textit{Quantization of Gauge Systems}.
\newblock
\newblock Princeton Univ. Press

\bibitem{BaezGroupoids}
{Baez},
\newblock 2007
   \href{http://math.ucr.edu/home/baez/week249.html}{This Week's Finds in Mathematical Physics (Week 249)}.
   \texttt{http://math.ucr.edu/home/baez/week249.html}


\bibitem{Barnich:2010xq}
{Barnich},
\newblock 2010
  \href{http://dx.doi.org/10.1063/1.3527427}{\textit{{A
  Note on gauge systems from the point of view of Lie
  algebroids}}}.
\newblock AIP Conf. Proc.
\newblock 1307(1) 7 [\eprint{1010.0899}]
  



\bibitem{thierry1985classical}
{Thierry-Mieg}, 1985.
\newblock \textit{Classical geometrical interpretation of ghost fields and
  anomalies in Yang-Mills theory and quantum gravity}.
\newblock
\newblock In \textit{Proceedings of a symposium on anomalies geometry topology}

\bibitem{gomes_riem}
{Gomes}, 2011.
\newblock \textit{Classical Gauge Theory in Riem}.
\newblock J. Math. Phys. 52, 082501
\newblock [\eprint{0807.4405v7}]

\bibitem{Rebhan1987}
{Rebhan},
\newblock 1987
  \href{http://dx.doi.org/https://doi.org/10.1016/0550-3213(87)90241-0}{\textit{The
  Vilkovisky-DeWitt effective action and its application to Yang-Mills
  theories}}.
\newblock Nucl. Phys. B
\newblock 288 832

\bibitem{GiuliniCharge}
{Giulini},
\newblock 2018
  \href{http://dx.doi.org/10.1007/978-1-4939-7708-6_12}{\textit{{Matter
  from Space}}}.
\newblock Einstein Stud.
\newblock 14 363 [\eprint{0910.2574}]
  

\bibitem{BeigOMurchadha}
{Beig} \protect\BIBand{} {OMurchadha},
\newblock 1987
  \href{http://dx.doi.org/https://doi.org/10.1016/0003-4916(87)90037-6}{\textit{The
  Poincar\'e group as the symmetry group of canonical general
  relativity}}.
\newblock Ann. Phys.
\newblock 174(2) 463

\bibitem{Geiller:2017xad}
{Geiller},
\newblock 2017
  \href{http://dx.doi.org/10.1016/j.nuclphysb.2017.09.010}{\textit{{Edge
  modes and corner ambiguities in 3d Chern–Simons theory and
  gravity}}}.
\newblock Nucl. Phys.
\newblock B924 312 [\eprint{1703.04748}]
  

\bibitem{Chrusciel:1987jr}
{Chrusciel} \protect\BIBand{} {Kondracki},
\newblock 1987
  \href{http://dx.doi.org/10.1103/PhysRevD.36.1874}{\textit{{Some
  Global Charges in Classical {Yang-Mills}
  Theory}}}.
\newblock Phys. Rev.
\newblock D36 1874

\bibitem{Barnich:2001jy}
{Barnich} \protect\BIBand{} {Brandt},
\newblock 2002
  \href{http://dx.doi.org/10.1016/S0550-3213(02)00251-1}{\textit{{Covariant
  theory of asymptotic symmetries, conservation laws and central
  charges}}}.
\newblock Nucl. Phys.
\newblock B633 3 [\eprint{hep-th/0111246}]
  

\bibitem{Barnich:1994db}
{Barnich}, {Brandt}, \protect\BIBand{} {Henneaux},
\newblock 1995
  \href{http://dx.doi.org/10.1007/BF02099464}{\textit{{Local
  BRST cohomology in the antifield formalism. 1. General
  theorems}}}.
\newblock Commun. Math. Phys.
\newblock 174 57 [\eprint{hep-th/9405109}]
  

\bibitem{Barnich:1994mt}
{Barnich}, {Brandt}, \protect\BIBand{} {Henneaux},
\newblock 1995
  \href{http://dx.doi.org/10.1007/BF02099465}{\textit{{Local
  BRST cohomology in the antifield formalism. II. Application to Yang-Mills
  theory}}}.
\newblock Commun. Math. Phys.
\newblock 174 93 [\eprint{hep-th/9405194}]
  

\bibitem{Susskind:2015hpa}
{Susskind}, 2015.
\newblock \textit{{Electromagnetic Memory}} 
  [\eprint{1507.02584}] 

\bibitem{strominger2018lectures}
{Strominger}, 2018.
\newblock \textit{Lectures on the infrared structure of gravity and gauge
  theory}.
\newblock Princeton Univ. Press
\newblock [\eprint{1703.05448}]



\bibitem{Geiller:2017whh}
{Geiller},
\newblock 2018
  \href{http://dx.doi.org/10.1007/JHEP02(2018)029}{\textit{{Lorentz-diffeomorphism
  edge modes in 3d gravity}}}.
\newblock JHEP
\newblock 02 029 [\eprint{1712.05269}]
  

\bibitem{Speranza:2017gxd}
{Speranza},
\newblock 2018
  \href{http://dx.doi.org/10.1007/JHEP02(2018)021}{\textit{{Local
  phase space and edge modes for diffeomorphism-invariant
  theories}}}.
\newblock JHEP
\newblock 02 021 [\eprint{1706.05061}]
  

\bibitem{aharonov1967:obs}
{Aharonov} \protect\BIBand{} {Susskind}, 1967.
\newblock \textit{Observability of the sign change of spinors under 2 $\pi$
  rotations}.
\newblock Phys. Rev.
\newblock 158(5) 1237

\bibitem{Aharonov1967SSR}
{Aharonov} \protect\BIBand{} {Susskind},
\newblock 1967
  \href{http://dx.doi.org/10.1103/PhysRev.155.1428}{\textit{Charge
  Superselection Rule}}.
\newblock Phys. Rev.
\newblock 155 1428

\bibitem{Bartlett:2007zz}
{Bartlett}, {Rudolph}, \protect\BIBand{} {Spekkens},
\newblock 2007
  \href{http://dx.doi.org/10.1103/RevModPhys.79.555}{\textit{{Reference
  frames, superselection rules, and quantum
  information}}}.
\newblock Rev. Mod. Phys.
\newblock 79 555

\bibitem{AharonovBook}
{Yakir~Aharonov} and {Daniel~Rohrlich}, 2008.
\newblock \textit{Quantum Paradoxes: Quantum Theory for the Perplexed}.
\newblock
\newblock John Wiley and Sons

\bibitem{wick1952}
{Wick}, {Wightman}, \protect\BIBand{} {Wigner}, 1952.
\newblock \textit{The intrinsic parity of elementary particles}.
\newblock Phys. Rev.
\newblock 88(1) 101

\bibitem{WWW1970}
{Wick}, {Wightman}, \protect\BIBand{} {Wigner},
\newblock 1970
  \href{http://dx.doi.org/10.1103/PhysRevD.1.3267}{\textit{Superselection
  Rule for Charge}}.
\newblock Phys. Rev. D
\newblock 1 3267

\bibitem{StrocchiCSR1974}
{Strocchi} \protect\BIBand{} {Wightman},
\newblock 1974
  \href{http://dx.doi.org/10.1063/1.1666601}{\textit{Proof
  of the charge superselection rule in local relativistic quantum field
  theory}}.
\newblock J. Math. Phys.
\newblock 15(12) 2198
  

\bibitem{StrocchiErratum}
{Strocchi} \protect\BIBand{} {Wightman},
\newblock 1976
  \href{http://dx.doi.org/10.1063/1.522818}{\textit{Erratum:
  Proof of the charge superselection rule in local relativistic quantum field
  theory}}.
\newblock J. Math. Phys.
\newblock 17(10) 1930
  

\bibitem{Freidel:2016bxd}
{Freidel}, {Perez}, \protect\BIBand{} {Pranzetti},
\newblock 2017
  \href{http://dx.doi.org/10.1103/PhysRevD.95.106002}{\textit{{Loop
  gravity string}}}.
\newblock Phys. Rev.
\newblock D95(10) 106002
  [\eprint{1611.03668}] 

\bibitem{Freidel:2015gpa}
{Freidel} \protect\BIBand{} {Perez}, 2015.
\newblock \textit{{Quantum gravity at the corner}}
   [\eprint{1507.02573}]
  

\bibitem{vanBaal:1995gg}
{van Baal}, 1995.
\newblock \textit{{Global issues in gauge fixing}}.
\newblock In \textit{{Nonperturbative approaches to quantum chromodynamics.
  Proceedings, International Workshop, Trento, Italy, July 10-29, 1995}},
\newblock pp. 0004--23
  [\eprint{hep-th/9511119}] 

\bibitem{Kunstatter:1991kw}
{Kunstatter},
\newblock 1992
  \href{http://dx.doi.org/10.1088/0264-9381/9/S/009}{\textit{{The
  Path integral for gauge theories: A Geometrical
  approach}}}.
\newblock Class. Quant. Grav.
\newblock 9 S157

\bibitem{dollard1964asymptotic}
{Dollard}, 1964.
\newblock \textit{Asymptotic convergence and the Coulomb interaction}.
\newblock J. Math. Phys.
\newblock 5(6) 729

\bibitem{Kulish:1970ut}
{Kulish} \protect\BIBand{} {Faddeev},
\newblock 1970
  \href{http://dx.doi.org/10.1007/BF01066485}{\textit{{Asymptotic
  conditions and infrared divergences in quantum
  electrodynamics}}}.
\newblock Theor. Math. Phys. 4 745.
\newblock [Teor. Mat. Fiz.4,153(1970)]

\bibitem{Zwanziger:1973if}
{Zwanziger},
\newblock 1973
  \href{http://dx.doi.org/10.1103/PhysRevD.7.1082}{\textit{{Reduction
  formulas for charged particles and coherent states in quantum
  electrodynamics}}}.
\newblock Phys. Rev.
\newblock D7 1082

\bibitem{Zwanziger:1974jz}
{Zwanziger},
\newblock 1975
  \href{http://dx.doi.org/10.1103/PhysRevD.11.3481}{\textit{{Scattering
  Theory for Quantum Electrodynamics. 1. Infrared Renormalization and
  Asymptotic Fields}}}.
\newblock Phys. Rev.
\newblock D11 3481

\bibitem{Greensite:2004ur}
{Greensite}, {Olejnik}, \protect\BIBand{} {Zwanziger},
\newblock 2005
  \href{http://dx.doi.org/10.1088/1126-6708/2005/05/070}{\textit{{Center
  vortices and the Gribov horizon}}}.
\newblock JHEP
\newblock 05 070 [\eprint{hep-lat/0407032}]
  

\bibitem{Lavelle:2011yc}
{Lavelle}, {McMullan}, \protect\BIBand{} {Sharma},
\newblock 2012
  \href{http://dx.doi.org/10.1103/PhysRevD.85.045013}{\textit{{The
  Factorisation of glue and mass terms in SU(N) gauge
  theories}}}.
\newblock Phys. Rev.
\newblock D85 045013 [\eprint{1110.1574}]
  

\bibitem{Wetterich:2017aoy}
{Wetterich},
\newblock 2018
  \href{http://dx.doi.org/10.1016/j.nuclphysb.2018.07.002}{\textit{{Gauge-invariant
  fields and flow equations for Yang-Mills
  theories}}}.
\newblock Nucl. Phys.
\newblock B934 265 [\eprint{1710.02494}]
  

\bibitem{Gomes:2017jyp}
{Gomes},
\newblock 2017
  \href{http://dx.doi.org/10.1088/1361-6382/aa8cf9}{\textit{{Quantum
  gravity in timeless configuration space}}}.
\newblock Class. Quant. Grav.
\newblock 34(23) 235004 [\eprint{1706.08875}]
  

\bibitem{Seraj:2017rzw}
{Seraj} \protect\BIBand{} {Van~den Bleeken},
\newblock 2017
  \href{http://dx.doi.org/10.1007/JHEP08(2017)127}{\textit{{Strolling
  along gauge theory vacua}}}.
\newblock JHEP
\newblock 08 127 [\eprint{1707.00006}]
  

\end{thebibliography}


\end{document}